%% file: TrackJetUEPaper.tex
\documentclass[aps,twocolumn,prd,showpacs,preprintnumbers]{revtex4-1}

\usepackage[hyperindex,breaklinks]{hyperref} 

\usepackage{graphicx}
\usepackage{subfigure}
\usepackage{atlasphysics}
\usepackage{multirow}
\usepackage{placeins}

\def\thedate{\today}

\def\Pythia{P{\sc{ythia}}}
\def\PythiaS{P{\sc{ythia~6}}}
\def\PythiaE{P{\sc{ythia~8}}}

\def\Herwigpp{{Herwig\raise0.2ex\hbox{\scriptsize\kern-0.5pt+\kern-0.5pt+}}}
\def\geant{{\uppercase{g}eant4}}

\def\antikt{anti-$k_{t}$}

\def\GeV{{\rm GeV}}
\def\MeV{{\rm MeV}}
\def\TeV{{\rm TeV}}

\def\jetEtaMax{1.5}
\def\etaJet{\ensuremath{\eta^{{\mathrm{jet}}}}}
\def\etaTrack{\ensuremath{\eta^{{\mathrm {track}}}}}
\def\phiJet{\ensuremath{\phi^{\mathrm{jet}}}}
\def\pp{\ensuremath{pp}}
\def\ppbar{\ensuremath{p\overline{p}}}
\def\Obs{\ensuremath{\cal{O}}}
\def\Nev{\ensuremath{N_{{\mathrm{ev}}}}}
\def\ptJet{\ensuremath{p_{\mathrm{T}}^{{\mathrm{jet}}}}}
\def\ptTrack{\ensuremath{p_{\mathrm{T}}^{{\mathrm{track}}}}}
\def\ptExt{\ensuremath{p_{\mathrm T}^{{\mathrm{ext}}}}}
\def\sumpt{\ensuremath{\Sigma p_{\mathrm T}}}
\def\ntrks{\ensuremath{N_{\rm ch}}}
\def\ptave{\ensuremath{\overline{{p}}_{\mathrm T}}}

\def\ltap{\raisebox{-.4ex}{\rlap{$\sim$}} \raisebox{.4ex}{$<$}}
\def\gtap{\raisebox{-.4ex}{\rlap{$\sim$}} \raisebox{.4ex}{$>$}}
\def\phiPart{\ensuremath{\phi^{\rm particle}}}
\def\simge{
    \mathrel{\rlap{\raise 0.511ex
        \hbox{$>$}}{\lower 0.511ex \hbox{$\sim$}}}}
\def\simle{
    \mathrel{\rlap{\raise 0.511ex
        \hbox{$<$}}{\lower 0.511ex \hbox{$\sim$}}}}

\usepackage{preprintcover}
\PreprintCoverPaperTitle{Underlying event characteristics and their dependence on jet size 
of charged-particle jet events in $ pp $ collisions 
at $\sqrt{s}=7\ \TeV$ with the ATLAS detector}
\PreprintIdNumber{CERN-PH-EP-2012-148}
\PreprintCoverAbstract{Distributions sensitive to the underlying event are studied in events containing
one or more charged-particle jets produced in \pp\ collisions at 
$\sqrt{s}=7$~TeV with the ATLAS detector at the Large Hadron Collider (LHC). 
These measurements reflect $ 800 \mu \rm{b}^{-1} $ of data taken during 2010.
Jets are reconstructed using the \antikt\ algorithm with radius parameter $R$ varying between $0.2$ and 1.0. 
Distributions of the charged-particle multiplicity, the scalar sum of the transverse momentum of charged particles, 
and the average charged-particle \pt\ are measured as functions of
\ptJet\ in regions transverse to and opposite the leading jet for $4\ \GeV < \ptJet < 100\ \GeV$. 
In addition, the $R$-dependence of the mean values of these observables is studied. 
In the transverse region, both the multiplicity and the scalar sum of the
transverse momentum at fixed \ptJet\ vary significantly with $R$, 
while the average charged-particle transverse momentum has a minimal dependence on $R$.
Predictions from several Monte Carlo tunes have been compared to the data;
the predictions from \PythiaS, based on tunes that have been determined using LHC data, 
show reasonable agreement with the data, including the dependence on $R$. 
Comparisons with other generators indicate that additional tuning of soft-QCD parameters is necessary for these generators.  
The measurements presented here provide a testing ground for further development of the Monte Carlo models.
} 
\PreprintJournalName{Physical Review D}

\begin{document}

\title{Underlying event characteristics and their dependence on jet size 
of charged-particle jet events in $ pp $ collisions
at $ \sqrt{s} = 7~\TeV $ with the ATLAS detector}
\author{The ATLAS Collaboration}
\date{\thedate}
\begin{abstract}
Distributions sensitive to the underlying event are studied in events containing
one or more charged-particle jets produced in \pp\ collisions at 
$\sqrt{s}=7$~TeV with the ATLAS detector at the Large Hadron Collider (LHC). 
These measurements reflect $ 800 \mu \rm{b}^{-1} $ of data taken during 2010.
Jets are reconstructed using the \antikt\ algorithm with radius parameter $R$ varying between $0.2$ and 1.0. 
Distributions of the charged-particle multiplicity, the scalar sum of the transverse momentum of charged particles, 
and the average charged-particle \pt\ are measured as functions of
\ptJet\ in regions transverse to and opposite the leading jet for $4\ \GeV < \ptJet < 100\ \GeV$. 
In addition, the $R$-dependence of the mean values of these observables is studied. 
In the transverse region, both the multiplicity and the scalar sum of the
transverse momentum at fixed \ptJet\ vary significantly with $R$, 
while the average charged-particle transverse momentum has a minimal dependence on $R$.
Predictions from several Monte Carlo tunes have been compared to the data;
the predictions from \PythiaS, based on tunes that have been determined using LHC data, 
show reasonable agreement with the data, including the dependence on $R$. 
Comparisons with other generators indicate that additional tuning of soft-QCD parameters is necessary for these generators.  
The measurements presented here provide a testing ground for further development of the Monte Carlo models.
\end{abstract}

\maketitle

\clearpage

\section{Introduction and Overview}
\label{sec:introduction}
Quantum Chromodynamics (QCD)~\cite{Gross:1973id,Politzer:1973fx}
predicts the cross sections for the production of objects with large transverse momentum ($\pt$)
~\footnote{ATLAS uses a right-handed coordinate system with its origin at the nominal interaction point (IP) in the center of the detector and the $z$-axis along the beam pipe. 
The $x$-axis points from the IP to the center of the LHC ring, and the $y$ axis points upward. 
Cylindrical coordinates $(r,\phi)$ are used in the transverse plane, $\phi$ being the azimuthal angle around the beam pipe. 
The pseudorapidity is defined in terms of the polar angle $\theta$ as $\eta=-\ln\tan(\theta/2)$. 
The transverse momentum $ \pt $ is defined as $ \pt \equiv \sqrt{E^{2}-m^{2}} \sin ( \theta ) $, where $ E $ is the total energy and m is the mass.} 
in hadronic collisions. 
Such calculations are performed by factorizing the interaction into a hard scattering process that can be calculated
perturbatively and a set of soft processes that must be described phenomenologically.  
The high-$\pt$ jet production cross section is calculated~\cite{ellis-stirling-webber} by convolving the 
matrix elements for the scattering of two initial-state partons (quarks and gluons), 
with the corresponding parton distribution functions (PDF), to produce a partonic final state.
To predict the momentum spectrum of the final particles, additional effects must be considered.
The outgoing partons fragment into jets of hadrons.
The beam remnants also hadronize and the spectator partons in the proton
can also interact, leading to multiple parton interactions (MPI).
QCD radiation from the initial- and final-state partons occurs, leading
to additional jets and to an increase in the ambient energy.  These effects
vary with the momentum transfer of the hard parton scattering.  
Some of these processes take place at an energy scale where the QCD coupling
constant is large and perturbation theory cannot be used.
They must therefore be described using QCD-motivated phenomenological models, implemented in Monte Carlo (MC) event generators.
In general, these models contain a number of free parameters with values that must be obtained by fitting to experimental data.

For a single proton-proton ($pp$) collision, the Underlying Event (UE) is defined to be any hadronic activity not
associated with the jets or leptons produced in the hard scattering process.
In practice, because color fields connect all the strongly interacting partons in the proton-proton event, 
no unambiguous assignment of particles to the hard scattering partons or UE is possible. 
Instead, distributions that are sensitive to UE modeling are constructed
from the tracks that are far from the direction of the products of the hard scatter.
This direction is approximated by the direction of the highest-\pt\ (leading) object in the event.  

Measurements of observables sensitive to the UE characteristics in \ppbar\ collisions
were performed at the Tevatron by CDF using jet events at center-of-mass energies
$\sqrt{s}=630$~GeV~ and $\sqrt{s}=1.8$~TeV~\cite{Acosta:2004wqa} ,
and using Drell-Yan and jet events at $\sqrt{s}=1.96$~TeV~\cite{Aaltonen:2010rm}.
Underlying event observables have been measured in \pp\ collisions with $\sqrt{s}=900$~GeV and $\sqrt{s}=7$~TeV 
from the distribution of charged particles in the region transverse to
leading charged particles and leading charged-particle jets by 
CMS~\cite{Khachatryan:2010pv,Chatrchyan:2011id} and in the regions transverse
to and away from leading charged particles by ATLAS~\cite{Aad:2010fh} and ALICE~\cite{ALICE:2011ac}.
A complementary analysis by ATLAS studied UE properties using both neutral and charged particles~\cite{Aad:2011qe}.

This paper extends previous studies of the UE by measuring the charged-particle multiplicity and transverse-momentum
density in the transverse and away regions with respect to leading charged-particle jets 
reconstructed  using the \antikt~\cite{antikt} algorithm and varying the radius parameter, $R$, of that algorithm between $0.2$ and $1.0$.
We can regard the leading charged particle as a charged-particle jet with $ R \approx 0 $.
As long as $R$ remains below a characteristic value determined by the momentum transfer ($ q^{2} $) of the hard scatter,
the $ \pt $ of charged-particle jets formed with larger $ R $ are better indicators of the hard scatter energy.
Such charged-particle jets are reconstructed with high efficiency at
low jet transverse momentum, \ptJet, and can therefore be used to study the behavior of the UE
in the transition region between soft QCD interactions and hard partonic scattering.
At larger \ptJet, charged-particle jets provide a complement to calorimeter-based measurements, with results that
are independent of calorimeter calibrations, selections and uncertainties. 
Variations in the mean values of these UE observables with $R$ provide additional information
on the interplay between the perturbative and non-perturbative components of QCD-inspired MC models.

This paper is organized as follows.
The ATLAS detector is described briefly in Section~\ref{sec:detector}.
The QCD Monte Carlo models used in this analysis
are discussed in Section~\ref{sec:samples}.
The variables sensitive to UE activity are defined in 
Section~\ref{sec:definitions}.
The event and object selections are presented in 
Section~\ref{sec:reconstruction}. 
Section~\ref{sec:analysis} contains a description of the
analysis. In Section~\ref{sec:systematics}, the treatment of systematic 
uncertainties is discussed. Fully corrected data distributions are
presented in Section~\ref{sec:results} and the results are compared to 
the predictions of several Monte Carlo generators.
Section~\ref{sec:rdependence} discusses the dependence of the UE
observables on $R$.
Conclusions are provided in Section~\ref{sec:conclusion}.
\section{The ATLAS Detector}

\label{sec:detector}
The ATLAS detector is described in detail in 
Ref.~\cite{ATLAS}.  The subsystems
relevant for this analysis are the inner detector (ID) and the trigger system.

The ID is used to measure the trajectories and momenta of charged particles.
It consists of three detectors: a pixel detector, a silicon strip tracker
(SCT) and a transition radiation straw-tube tracker (TRT).  
These
detectors are located
inside a solenoid  that provides a 2~T axial magnetic field,
and have full coverage in the azimuthal angle 
$\phi$  over the pseudorapidity range
$|\eta |<2.5$
The ID barrel (end-cap) typically has 3 (2 $\times$ 3) pixel layers, 
4 (2 $\times$ 9) layers of double-sided silicon strip modules, and 
73 (2 $\times$ 160) layers of TRT straw tubes.
A track traversing the
barrel typically has 11 silicon hits (3 pixel clusters, and 
8 strip clusters), and more
than 30 straw-tube hits.
Typical position resolutions are 10, 17 and 130$\,\mathrm{\mu m}$ for the 
$r$-$\phi$ coordinate of the pixel detector, SCT, and TRT respectively.
The resolution of the second measured coordinate is 115$\,\mathrm{\mu m}$
for the pixel detector and 580$\,\mathrm{\mu m}$ for the SCT.

The ATLAS trigger consists of three levels of event selection: 
Level-1 (L1), Level-2 (L2), and Event Filter. 
The trigger relevant for this  analysis is
the single-arm L1 minimum bias trigger\cite{ATLAS-CONF-2010-068}, which uses information from
the Beam Pickup Timing devices
(BPTX) and the Minimum Bias Trigger Scintillators (MBTS).
The BPTX stations are composed
of electrostatic button pickup detectors attached to the 
beam pipe at $\pm 175$~m from the center of the
ATLAS detector. 
The coincidence of the BPTX signal between the two sides of the detector is 
used to determine when bunches collide in the center of the detector.
The MBTS consists of 32 scintillation counters located
at pseudorapidities  $2.09 <|\eta| < 3.84$.  
The trigger requires a single hit above threshold in the MBTS together
with a BPTX coincidence.

\section{Monte Carlo Samples}
\label{sec:samples}
MC event samples are used to
compute detector acceptance and reconstruction efficiencies,
determine background contributions, unfold the measurements for
detector effects,
and estimate systematic uncertainties on the final results.
The samples used here are the same as those used in Ref.~\cite{PhysRevD.84.054001}.
The baseline Monte Carlo event generator used to determine acceptance and efficiencies
is  \PythiaS.4.21~\cite{Sjostrand:2006za}  with the
ATLAS tune~{\sc AMBT1}, which uses the MRST2007LO* PDF~\cite{Sherstnev:2007nd}.
This tune was derived using the measured properties of
minimum bias events~\cite{Aad:2010ir}. 
Generated events are simulated using the
ATLAS detector  simulation~\cite{2010arXiv1005.4568T},
which is based on the
\geant\ toolkit~\cite{Agostinelli:2002hh}.
The simulated events are processed
using the same software as the data.
Several other  simulated samples are used
to assess systematic uncertainties on the detector response: 
\PythiaS\ using the Perugia~2010
tune~\cite{Sjostrand:2006za} (CTEQ5L PDF~\cite{Pumplin:2002vw}),
\PythiaS\  with the ATLAS MC09 tune~\cite{ATL-PHYS-PUB-2010-002} 
(MRST2007LO* PDF) and \PythiaE.145 with tune 4C~\cite{Sjostrand:2007gs}
(CTEQ6L1 PDF).

In the past year, significant work has been done to improve
agreement between the MC generators and LHC data by
tuning the parameters of the phenomenological models used to
describe soft QCD processes~\cite{ATL-PHYS-PUB-2011-014,
field-arXiv:1010.3558v1}.
UE data, after applying corrections for detector effects, 
are compared to recent tunes of
several MC event generators.  
Samples with high statistics were produced using the
\PythiaS.4.21,
\PythiaE.145, and 
\Herwigpp~2.5.1~\cite{Gieseke:2011na} generators.
Several different \PythiaS\
samples were generated with different UE tunes.
AUET2B~\cite{AUET2B} is an ATLAS tune that uses
the \pt-ordered parton shower, interleaved initial-state radiation
(ISR) and MPI which has been tuned to UE
data from CDF and ATLAS. This employs the MRST2007LO**~\cite{Sherstnev:2007nd} PDF. 
The CMS tune Z1~\cite{field-arXiv:1010.3558v1} is very 
similar, but is fitted to CMS UE
data, and uses CTEQ5L~PDF. 
The Perugia~2011 tune~\cite{Skands:2010ak}
uses similar settings,
with the UE parameters constrained by minimum bias and UE data from CDF and ATLAS.
Perugia~2011~NOCR is a tune of the \pt-ordered
model which does not employ color reconnection.
The \PythiaE\ generator includes many new features over \PythiaS,
such as fully interleaved \pT-ordered
MPI+ISR+FSR evolution, and a different model of hard diffraction.
The default author tune~4C~\cite{Corke:2010yf} with CTEQ6L1 
PDF~\cite{Pumplin:2005rh} is used.
The recently released \Herwigpp~2.5.1 version with tune UE7-2 used here 
has angle-ordered parton showers, 
employs MRST2007LO* PDF and has an improved model of color reconnection.

\section{Definition of Variables}
\label{sec:definitions}
After reconstructing jets from charged tracks in the event,
$\ptJet$ refers to the transverse momentum ($\pt$) of the jet with the highest $\pt$.
The variation of the UE properties with $\ptJet$ is measured in this analysis.
In addition, the dependence of these properties on $R$  is studied.

\begin{figure}[tbh]
\includegraphics[width=0.48\textwidth]{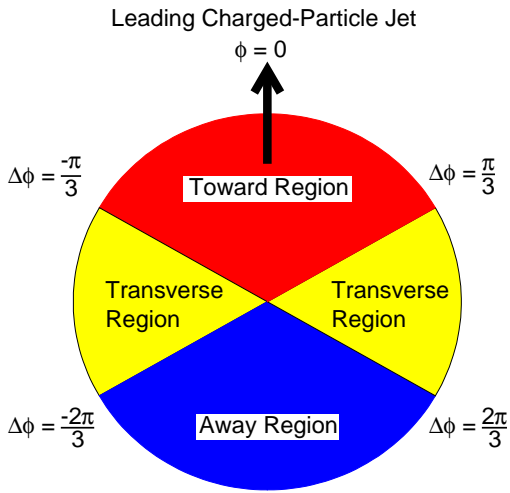}
\caption{Definition of the transverse and away regions with respect to the
leading jet.}
\label{fig:regions}
\end{figure}

Particles are defined to be in the {\it transverse} region if their
azimuthal angle differs from that of the leading jet
($|\Delta \phi | \equiv | \phiPart - \phiJet | $) by between
$\pi/3$ and $2\pi/3$ radians, as shown in Figure~\ref{fig:regions}.  
This region is most sensitive to the UE.
Particles with $|\Delta \phi|>2\pi/3$ are defined to be in the {\it away} region. 
This region is likely to contain the fragmentation products
of the sub-leading parton produced in the hard scattering. 

Three observables sensitive to UE activity 
are studied in the transverse and away regions:
\begin{itemize}
\item $\ntrks$: the number of tracks in the region
\item $\Sigma \pT$: the scalar sum of the transverse momentum of the
tracks in the region
\item $\ptave$: the average \pt\ of the tracks 
in the region ($\ptave \equiv \sumpt / \ntrks $)
\end{itemize}

\section{Event Selection and Reconstruction}
\label{sec:reconstruction}
The events used in this analysis were collected with the ATLAS detector
at a center-of-mass energy $\sqrt{s}=7$~TeV during early 2010.  The
data sample, event selection and reconstruction are identical to those
used to measure the cross section and fragmentation functions
of jets reconstructed from tracks~\cite{PhysRevD.84.054001}.  
Events are required to have passed the L1 minimum bias trigger which was highly prescaled.
The sample represents an integrated luminosity of $~800\ \mu {\rm b}^{-1}$ after the trigger prescale.
The average number of collisions per bunch crossing, $\mu$, varied throughout the data-taking
period, but never exceeded a value of $\mu=0.14$. Over half the data were
taken with $\mu\simle 0.01$.  Thus, effects due to the presence of more than 
one collision in the same bunch crossing (``pile-up'') are minimal.
Primary vertex reconstruction~\cite{ATLAS-CONF-2010-069} is performed using tracks with
$\pt >0.4 \ \GeV$ and $|\eta|<2.5$.  A minimum 
of two tracks is required to form a vertex.
To further reduce the contributions of pile-up, 
events are rejected if more than one primary vertex is reconstructed.

Tracks used in the reconstruction of
jets and UE 
observables are required to have
$\pt > 0.5\ \GeV$, $|\eta|<2.5$,
transverse impact parameter with respect to the primary vertex
$|d_0|<1.5$~mm and longitudinal impact parameter with respect to the primary
vertex $|z_0 \sin \theta | < 1.5$~mm.
Only tracks with at least one pixel hit and 
six SCT hits are considered.  To minimize the contribution of particles
produced by secondary
interactions in the ID, tracks are required to have a hit in the innermost
pixel layer if the extrapolated track passes through an
active portion of that layer.

For each event, jet collections are constructed, corresponding
to the output obtained when the \antikt\ algorithm is
applied to the tracks for five separate values of $R$: 0.2, 0.4,
0.6, 0.8 and 1.0.  
For each jet collection, the leading jet is defined to be
the jet with the largest \ptJet\ satisfying the requirements
$\ptJet >4~\GeV$ and $|\etaJet|<1.5$.  This maximum $|\etaJet|$ cut
ensures that all tracks associated with jets in the fiducial region are 
within the fully-efficient tracking volume. 

\section{Analysis Procedure}
\label{sec:analysis}

The analysis is performed in parallel on each of the five jet collections
using the following procedure.  First, the leading jet is selected 
and events are rejected if there is no jet that satisfies the requirements
described in the previous section.  Next, the $\Delta \phi$ of each track
with respect to the leading jet is calculated and the
tracks in the transverse and away regions are identified.   
In addition to satisfying the selection criteria discussed in 
Section~\ref{sec:reconstruction}, tracks used for the
UE measurements are required to pass the
same pseudorapidity cut as the jets, $|\eta|<1.5$. 
This requirement minimizes the contamination in the UE calculation, 
from constituent tracks of a leading jet with $|\etaJet|>1.5$.

The selected tracks are used to calculate the three event observables \ntrks, \sumpt\ and \ptave,
denoted generically as $ \Obs $,
in the transverse and away regions.
The final results presented here are the distributions and mean values
of these observables for specific ranges (bins) of \ptJet.  To
allow comparisons with MC generators, the data distributions are
corrected for detector acceptance, reconstruction efficiency, and 
for bin migration due to track and jet momentum-resolution effects. 
The corrections used in this unfolding procedure are obtained
by matching the jets reconstructed in simulated MC samples 
with those obtained when the \antikt\ algorithm 
is applied to the primary charged particles produced by the generator.
Primary charged particles are defined as charged particles with a
mean lifetime $\tau > 0.3 \times 10^{- 10}$s, which are produced in
the primary collision or from subsequent decays of particles
with a shorter lifetime. Thus, the charged decay
products of $K^0_S$ and $\Lambda$ particles are not included. 
A reconstructed jet is considered to be matched to a particle-level jet
if their centers are separated by $\sqrt{(\Delta \eta)^2+(\Delta \phi)^2}<R/2$.

Jets below the minimum \ptJet\ cut or outside the maximum $|\etaJet|$ cut are used in the procedure
that corrects for the effects of resolution smearing.
Thus, looser requirements
of $\ptJet\ge 1\ \GeV$  and  $|\etaJet| \le 2.5$
are imposed during the reconstruction.
In addition, since corrections for migration of jets into the
fiducial volume require knowledge of the population outside
that volume, the transverse momentum of the leading jet
outside the fiducial volume (\ptExt) 
is also determined in each reconstructed jet collection for each event 
and, in MC samples, for each collection of
particle-level charged-particle jets.

The events satisfying these requirements are corrected back to the primary charged-particle spectra
satisfying the event-level requirement of at least one 
\antikt\ jet with $ \ptJet \ge 4~\GeV $ and $|\etaJet| < 1.5$
reconstructed from charged primary 
particles with $\pt > 0.5 \ \GeV$ and $|\eta| < 2.5$. 
Data distributions are unfolded using
an iterative method \cite{D'Agostini:1994zf} based on Bayes' theorem,
implemented in the RooUnfold \cite{RooUnfold} software package.
The procedure requires three inputs: a measured input distribution
(stored as a multi-dimensional histogram), a 
response matrix (obtained from simulated data) that provides a mapping
between reconstructed objects and those obtained directly from the
event generator, and an initial choice for the prior probability distribution, or {\it prior} for short.
Each observable \Obs\  
is stored in a three-dimensional histogram (one
histogram for each observable and separate histograms for the transverse
and away regions) where the binning variables are \ptJet, \Obs\
and \ptExt.  To accommodate the decreasing statistics in the data with \ptJet\ and  the variation of
the \ptJet\ and \ptTrack\ resolution
with transverse momentum, these histograms have variable bin width.
The response matrix is stored as a six-dimensional histogram that specifies
the probability that observed values of
\ptJet, \ptExt\ and \Obs\ are measured for given true values of
\ptJet, \ptExt\ and \Obs.  
This response matrix is not unitary because in mapping from generator to
reconstruction some events and jets are lost due to inefficiencies
and some are gained due to misreconstruction or migration of true
objects from outside the fiducial acceptance into the reconstructed 
observables.   

Unfolding the experimental distribution to obtain the corrected distribution is done as follows.
The response matrix, measured data and initial prior are used as inputs to the unfolding algorithm to produce an updated distribution, the {\it posterior}.
This posterior is used as the input prior for another iteration of the algorithm, and this process is repeated. 
The inputs to each iteration of the unfolding algorithm are the baseline response matrix, measured data and the posterior of the previous iteration.
The optimal number of iterations is determined from MC studies.
In this analysis, a total of four iterations are performed for each measured distribution.
The initial prior is taken to be the prediction of the baseline Monte Carlo generator.
Systematic uncertainties associated with this choice 
and with the modeling of the response matrix are discussed in 
Section~\ref{sec:systematics}.

Once the corrected distributions have been obtained in bins of
\ptJet, the mean value of \Obs\ for each \ptJet\ bin is determined
from these distributions.
Some care must be taken to avoid bias when the mean is
calculated since the output of the unfolding procedure is a
histogram and the distribution of the population varies across the bin.
For \sumpt\ and \ntrks,
the cumulative distribution function of the unfolded distribution is calculated
and fit to a cubic spline and the mean is determined from the results
of the spline fit.  This step reduces the bias between the binned and
unbinned calculation of the mean from a few percent to less than $0.5\%$.
The \ptave\ distributions have sufficiently fine binning that the bias
is below $0.5\%$ without this step.

\section{Systematic Uncertainties}
\label{sec:systematics}
A summary of the systematic uncertainties and how they affect the
measurements is presented in Table~\ref{table:uncertainties}.
The following sources of systematic uncertainty have been
considered:
\begin{enumerate}
\item The track reconstruction efficiency and momentum reconstruction
uncertainty.
\item Potential bias arising from the unfolding procedure.
\item Misidentification of the leading jet due to cases where the leading
reconstructed jet does not correspond to the true leading jet.
\item The uncertainty in the response matrix, which is derived using a 
particular MC sample and thus depends on the details of the event
generator.
\item Uncertainty in the calculation of the means from the distributions
of observables due to discretization effects arising from the 
finite bin-width of the unfolded distributions.
\end{enumerate}
\begin{table*}[t]
\caption{\label{table:uncertainties}The systematic uncertainties associated 
with measurement of the mean values of \sumpt, \ntrks\ and \ptave.}
\begin{center}
\begin{tabular}{|c|c|c|c|c|c|c|}
\hline
\multicolumn{7} {|c| } {Relative Systematic Uncertainties } \\
\hline
& \multicolumn{3}{|c|}{{\sc transverse} region} & \multicolumn{3}{|c|}{{\sc away} region} \\
Source & $ \sumpt (\% ) $ & $ \ntrks (\% ) $ & $ \ptave (\% ) $ & $ \sumpt (\% )$ & $ \ntrks (\% )$ & $ \ptave (\% )$ \\
\hline
Tracking Reconstruction & 2.1--2.5 & 2.0--2.3 & 0.2--0.3 & 2.1--2.6 & 2.0--2.3 & 0.1--0.2\\
\hline
Unfolding Procedure&  1.5--6.0 & 1.5--4.0 & 1.0--4.0 & 1.5--5.0 & 1.5--4.0 & 1.0--4.0\\
\hline
Leading Jet Misidentification & $\le 1.0$ & $\le 1.0$ & $\le 0.5$ & $\le 1.0$ & $\le 1.0$ & $\le 0.5$ \\
\hline
Response Matrix & 0.5--2.1 & 0.5--1.6 & 0.5--1.6 & 0.5--2.2 & 0.5--1.6 & 0.5--1.2 \\
\hline
Discretization Effects & $ \le 0.5$ & $ \le 0.5 $ & $ \le 0.5 $ & $ \le 0.5$ & $ \le 0.5 $ & $ \le 0.5 $ \\
\hline
\hline
Total & 2.7--6.6 & 2.6--4.7 & 1.3--4.1 & 2.7--5.7 & 2.6--4.7 & 1.3--4.1 \\
\hline
\end{tabular}
\end{center}
\end{table*}

Uncertainties on the tracking efficiency are $\eta$-dependent and vary
between 2\%\ (for $|\etaTrack | \le 1.3$) and 
7\%\ (for $2.3 < |\etaTrack | \le 2.5$ ) 
\cite{mb2paper}.  The dominant source of this uncertainty comes from possible
inaccuracies in the 
description of the detector material in the simulation.
The effect of this uncertainty on the measured observables is assessed by randomly removing from the data a fraction of the tracks 
consistent with the uncertainty on the tracking efficiency and recalculating the observable.
The resulting uncertainty is then assumed
to be symmetric.
Uncertainties on the track momentum resolution are parameterized
as an additional $\eta$-dependent broadening of the resolution in track curvature
with values that vary from $0.4$~TeV$^{-1}$ to
$0.9$~TeV$^{-1}$~\cite{ATLAS-COM-CONF-2011-046}.  
Systematic uncertainties on the tracking performance lead to 
relative uncertainties 
on the mean values of \sumpt\ and \ntrks\ that vary with $R$ from 
$2.1\%$ ($R=0.2$) to $2.6\%$ ($R=1.0$).
Uncertainties on \ptave\ are
below the percent level for all values of $R$ and \ptJet.

The performance of the unfolding procedure is studied by unfolding the distributions measured in 
simulated MC control samples and comparing them to the known generator-level distributions.
These closure tests have been performed using all
the simulated samples described in Section~\ref{sec:samples}.
In each case, before the test is performed, the input MC
sample is reweighted so that the truth 
distributions of \ptJet\ and of the observable \Obs\ 
reproduce the unfolded distributions in the data.  The resulting deviations 
on the mean value of \Obs\ are
lowest ($1\%$--$1.5 \%$) at low \ptJet\ and increase to $4\%$--$6\% $
at large \ptJet\ for all three observables. 

At least three effects can result in differences between the leading
reconstructed jet and the true leading jet.
One possibility is that a jet with $ \ptJet\ < 4~\GeV $  or with $|\etaJet| > \jetEtaMax $ 
is reconstructed to be inside the acceptance (``feed-in'' jets).
Another effect is that due to differences in the distributions of true
and reconstructed particles the \antikt\ algorithm, when applied to the reconstructed data, 
produces jets that do not match any jets obtained when the algorithm is applied to true particles (``accidental jets'').
A third possibility is that a non-leading jet is identified as a leading jet due to resolution smearing and inefficiencies in the track reconstruction.
The unfolding procedure intrinsically corrects for migration into the fiducial region.
Residual uncertainties on the contribution from misidentification of the leading jet have been assessed 
by reweighting the simulation to reproduce the observed distributions of the
subleading \ptJet\ and the azimuthal angle between the leading and subleading jet,
and applying the unfolding procedure to the reweighted sample.
Changes in the resulting output from the unfolding differ from the default by $<1\%$.
Studies with simulated data indicate that the rate for accidental jets
is below $0.1\%$ for all values of $R$ and \ptJet.  Therefore, the
systematic uncertainty due to accidental jets is judged to be
negligible. 
The systematic uncertainty on the fraction of non-leading jets
that are misidentified as the leading jet
due to uncertainties in the
tracking efficiency is already included
in the tracking systematic uncertainty described above.

Systematic uncertainties due to discretization effects (finite bin size)
have been studied using simulated data
by comparing the mean calculated 
from the binned data to the mean calculated from unbinned data.
The differences are below $0.5\%$ for all observables and all values of $R$.

The sensitivity to the number of iterations used for unfolding is determined by 
comparing the baseline results to those obtained when the number of 
iterations is varied.  Changes in the unfolded results
are below 0.5\%\ for all the observables.

The response matrix is derived using a 
particular MC sample; therefore, it 
depends on the details of the event generator and tune.  The sensitivity
of the result to differences between the baseline MC sample and the
data have been
studied by comparing the baseline results to those obtained when
the data are unfolded using an alternative response matrix
constructed after the MC sample has been reweighted 
to reproduce the raw \ptJet\ and UE observables.
Relative differences between the baseline measurements and those
obtained with the reweighted response matrix
are below $ 0.25\% $ for all three observables.
The effects of statistical uncertainties in the response matrix are 
evaluated using the bootstrap method~\cite{erfon:bootstrap} 
to create 50 statistically independent samples 
of the reweighted MC simulation
and repeating the unfolding procedure on the data for each sample.
The RMS (relative to the baseline) of the resulting ensemble of unfolded 
results is less than $0.1\%$.

When comparing the mean values of UE observables for different values of $R$,
the correlations among the systematic uncertainties must be properly treated.
Uncertainties due to track reconstruction efficiency and 
due to discretization effects are fully correlated among the measurements
and thus do not contribute to the systematic uncertainty on the ratio
of the mean responses measured for different $R$.
Uncertainties due to unfolding are
partially correlated.  The systematic uncertainties on the 
ratios of UE observables for different values of $R$ are
determined from the deviations from the baseline ratios, 
of the ratios obtained from MC samples where
the input spectra are varied concurrently for all jet collections.
The uncertainties on the ratios are typically below $1.5\%$ except
for the highest bin in \ptJet, where the uncertainty on
the ratio of  $R=0.2$ to $R=0.6$ rises
to  $6\%$ for \sumpt\ in the transverse region.

The same events are used to reconstruct all the jet collections; therefore,
the statistical uncertainties are also correlated among the measurements.
The statistical uncertainties on the ratio of the observables measured with 
one value of $R$ to those measured with a different $R$
are obtained by applying a bootstrap
method to the data.
\label{result-figs}
\begin{figure*}[!ptb]
\begin{center}
\subfigure[]{\includegraphics[width=0.42\textwidth]{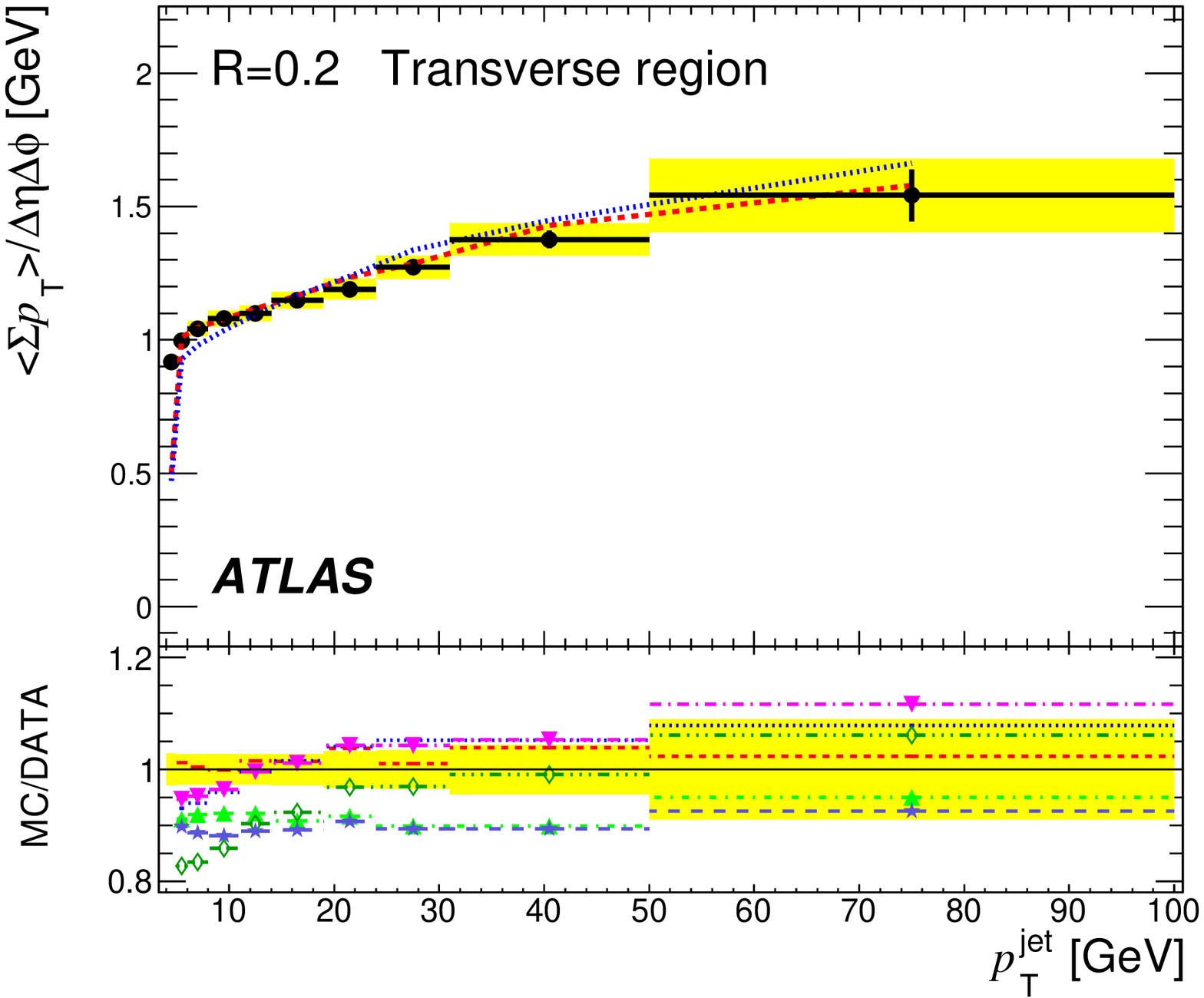}}
\subfigure[]{\includegraphics[width=0.42\textwidth]{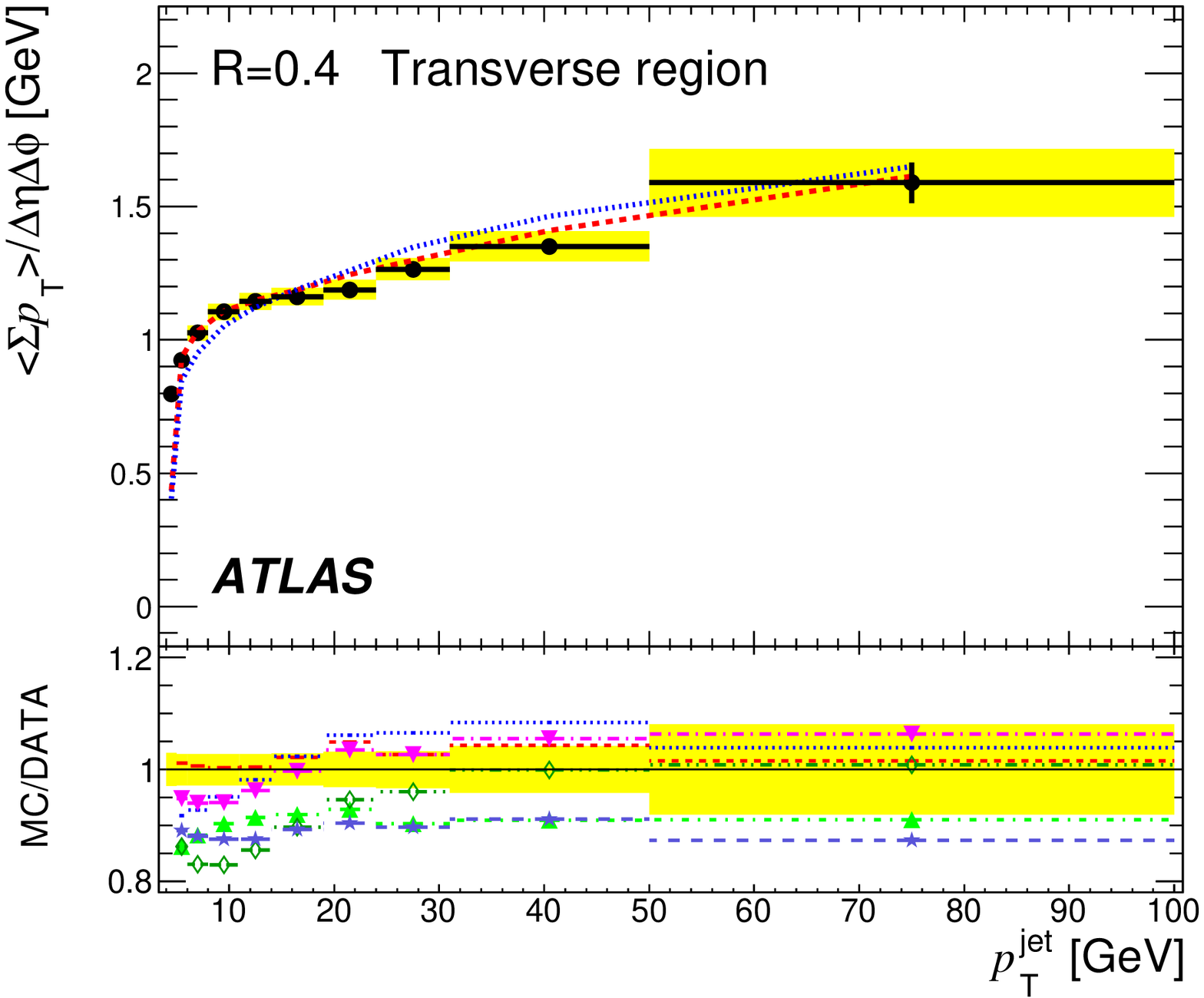}}
\subfigure[]{\includegraphics[width=0.42\textwidth]{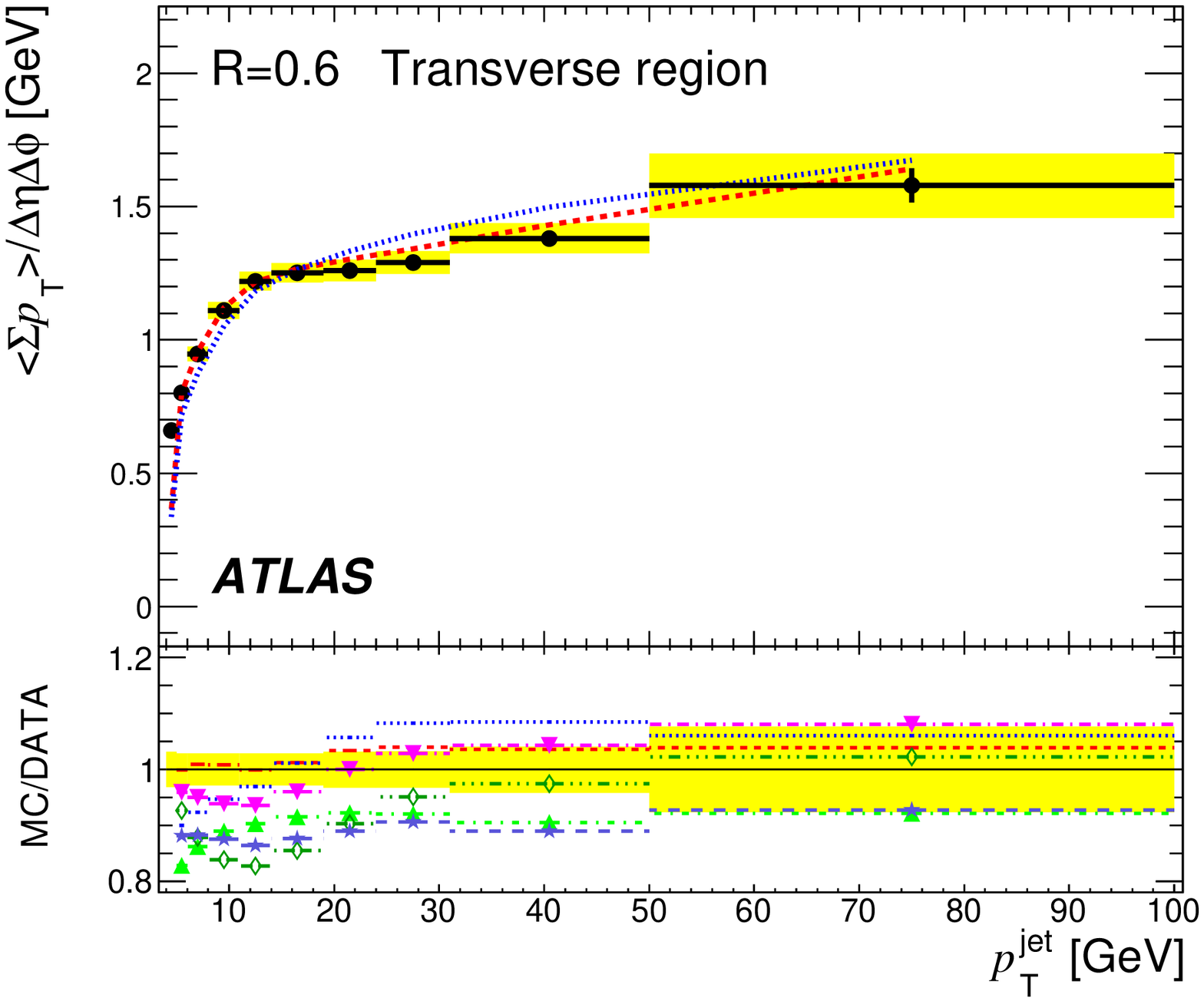}}
\subfigure[]{\includegraphics[width=0.42\textwidth]{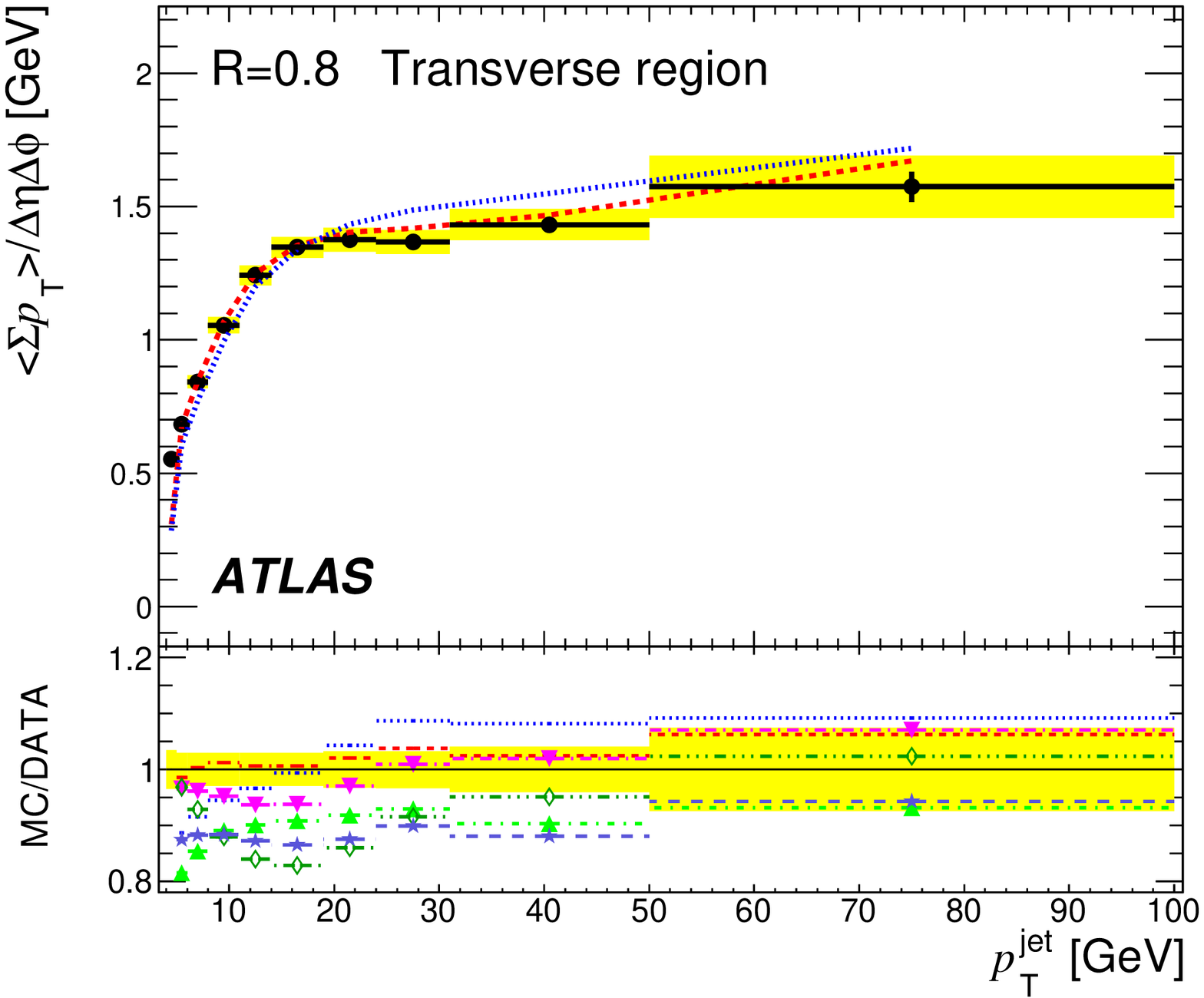}}
\subfigure[]{\includegraphics[width=0.42\textwidth]{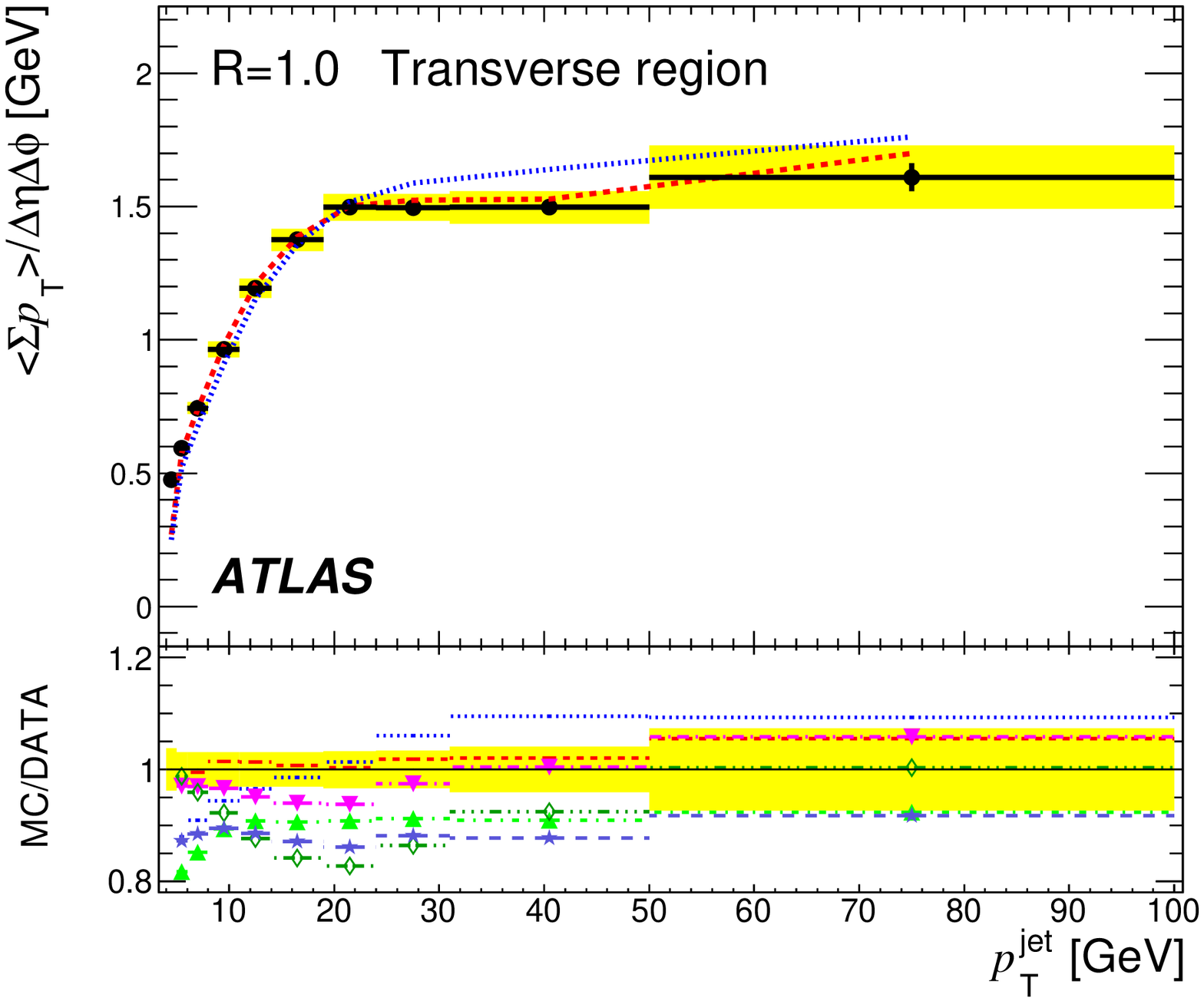}}
{\includegraphics[width=0.42\textwidth]{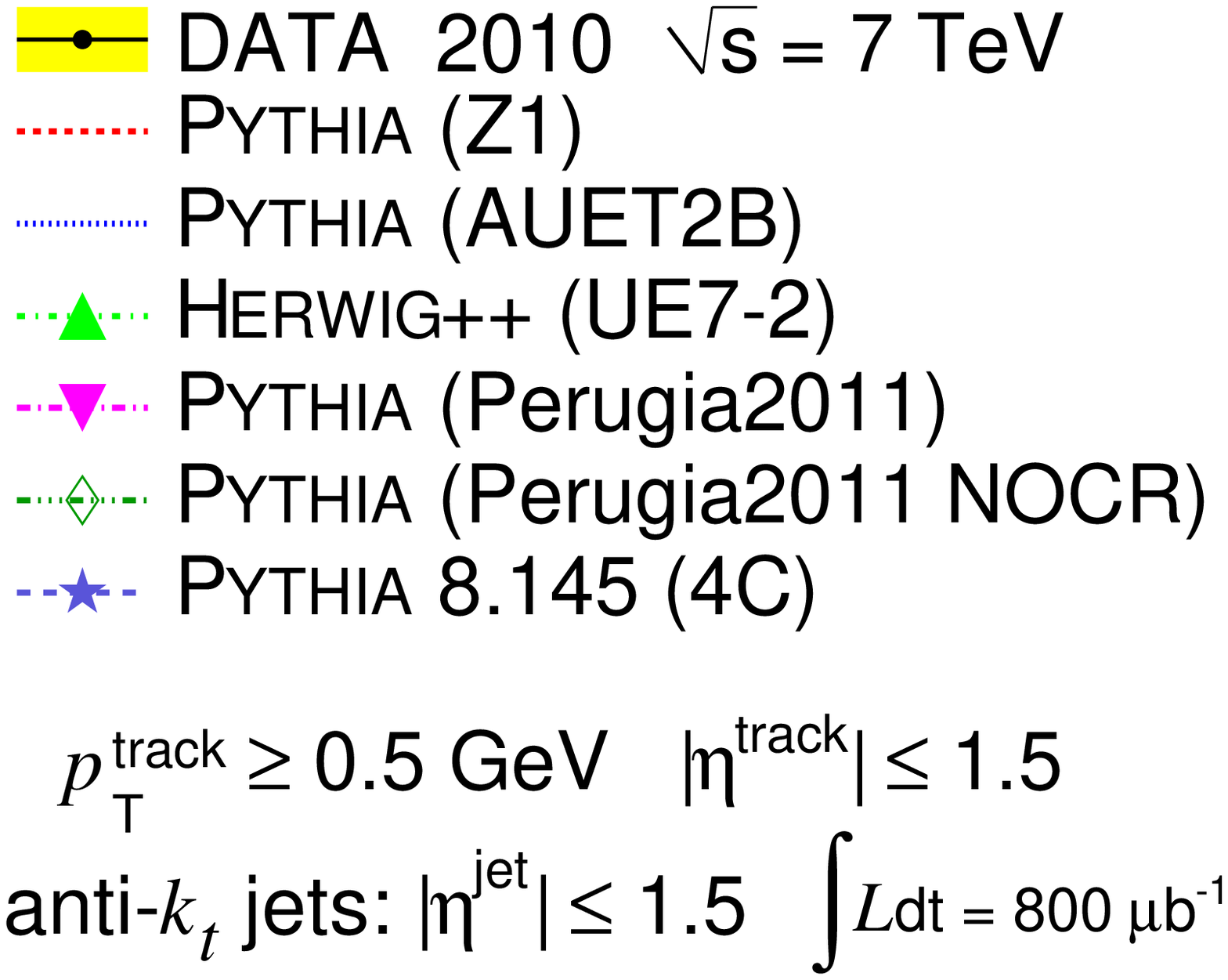}}
\end{center}
\caption{The mean value, per unit $ \eta - \phi $, of \sumpt\ in the transverse region, as a function of \ptJet\ for
(a) $R=0.2$,
(b) $R=0.4$,
(c) $R=0.6$,
(d) $R=0.8$ and
(e) $R=1.0$.
The shaded band shows the combined statistical and systematic
uncertainty on the data.
The data are compared to the predictions 
obtained with \PythiaS\ using the AUET2B and Z1 tunes.  The bottom
insert shows the ratio of MC predictions to data for several recent MC
tunes.
}
\label{fig:sumptbinstrans}
\end{figure*}

\begin{figure*}[!ptb]
\begin{center}
\subfigure[]{\includegraphics[width=0.42\textwidth]{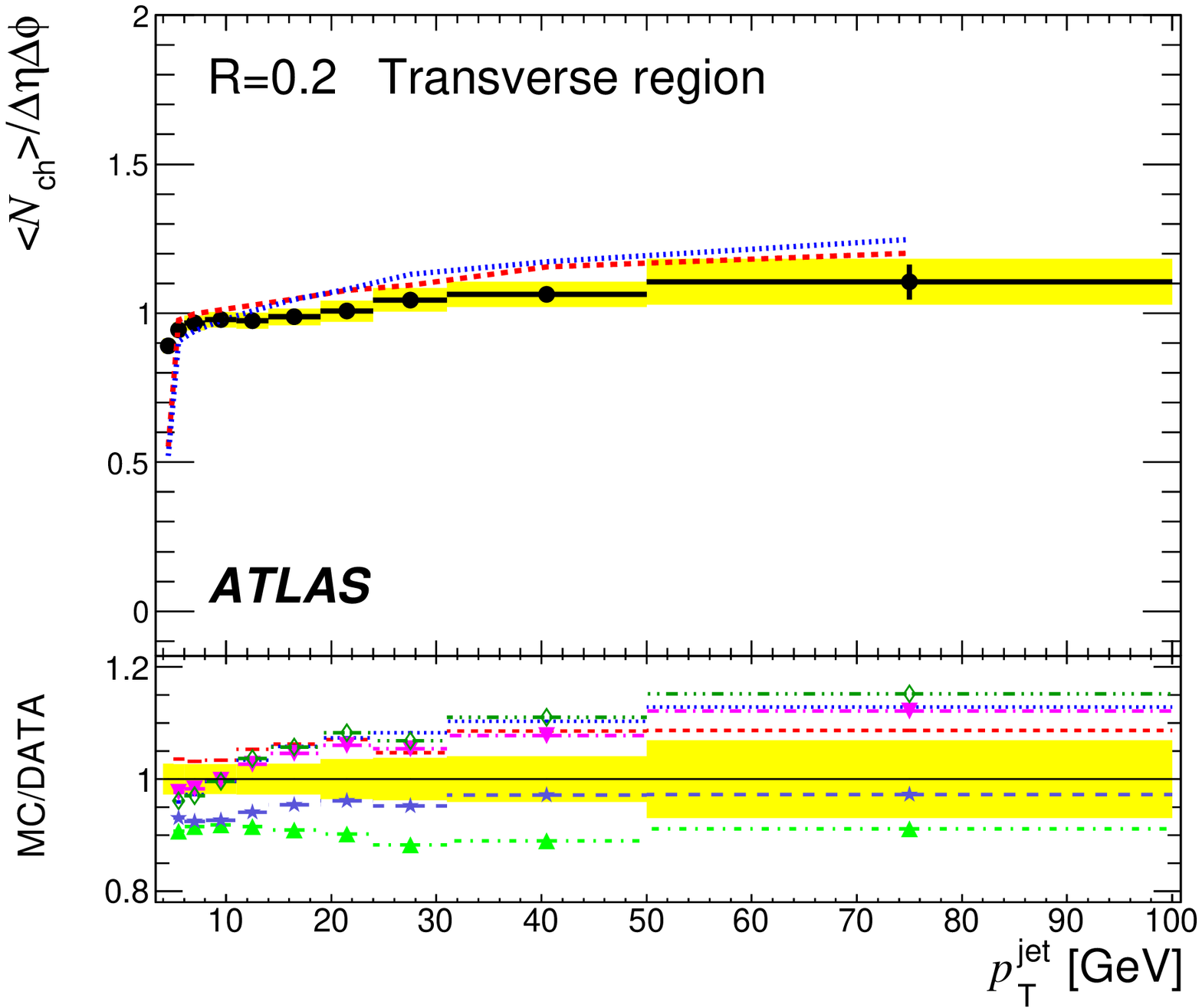}}
\subfigure[]{\includegraphics[width=0.42\textwidth]{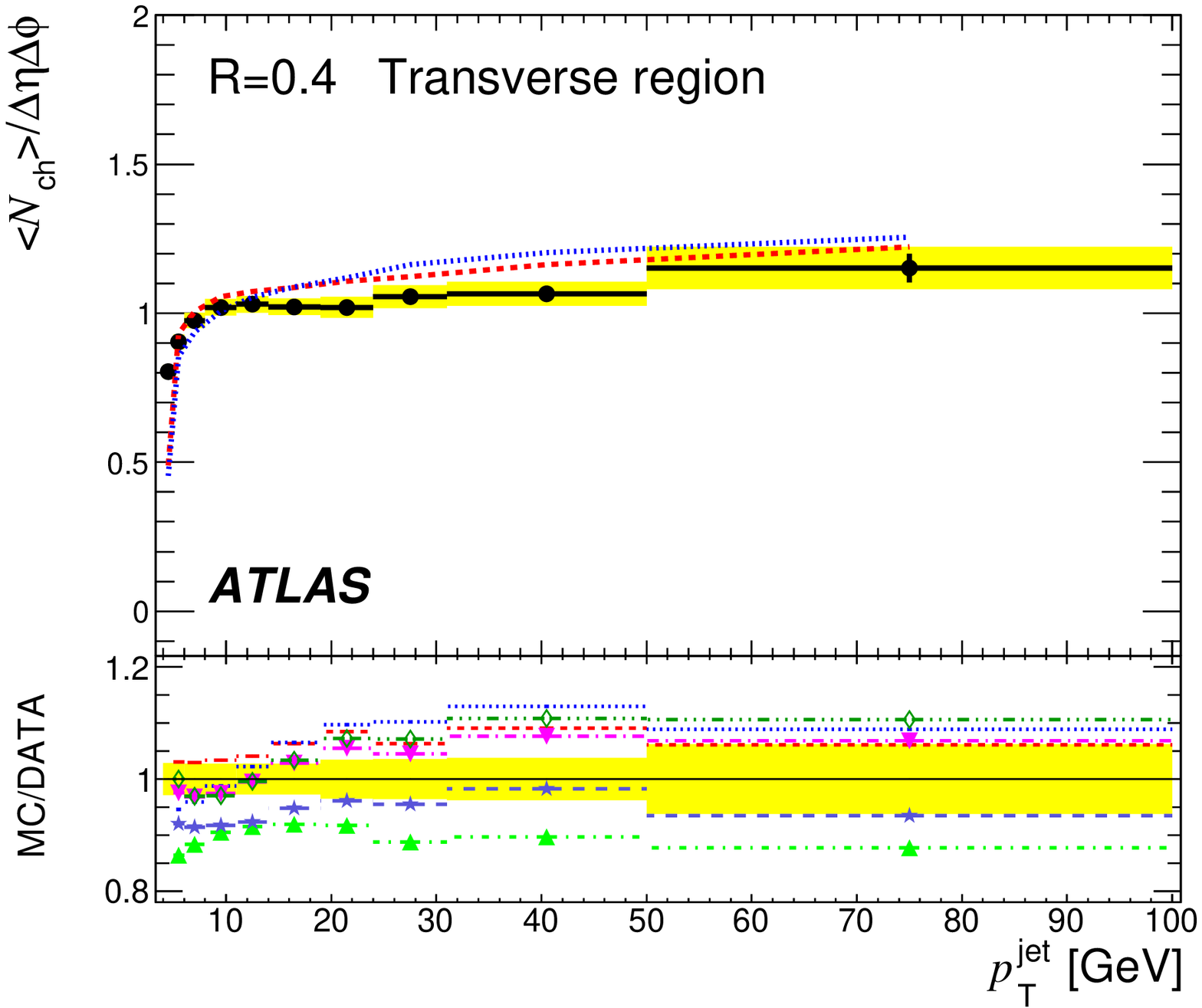}}
\subfigure[]{\includegraphics[width=0.42\textwidth]{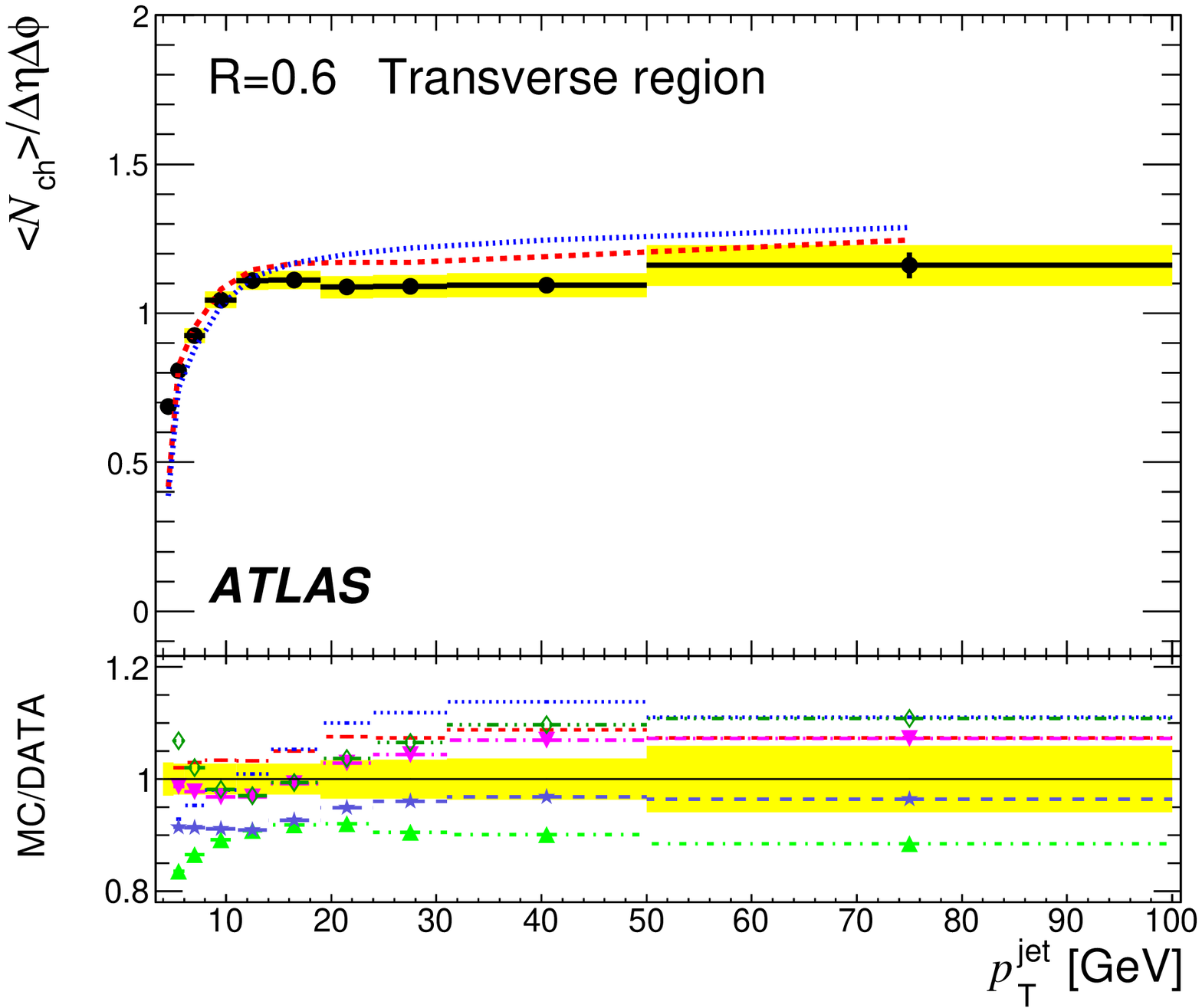}}
\subfigure[]{\includegraphics[width=0.42\textwidth]{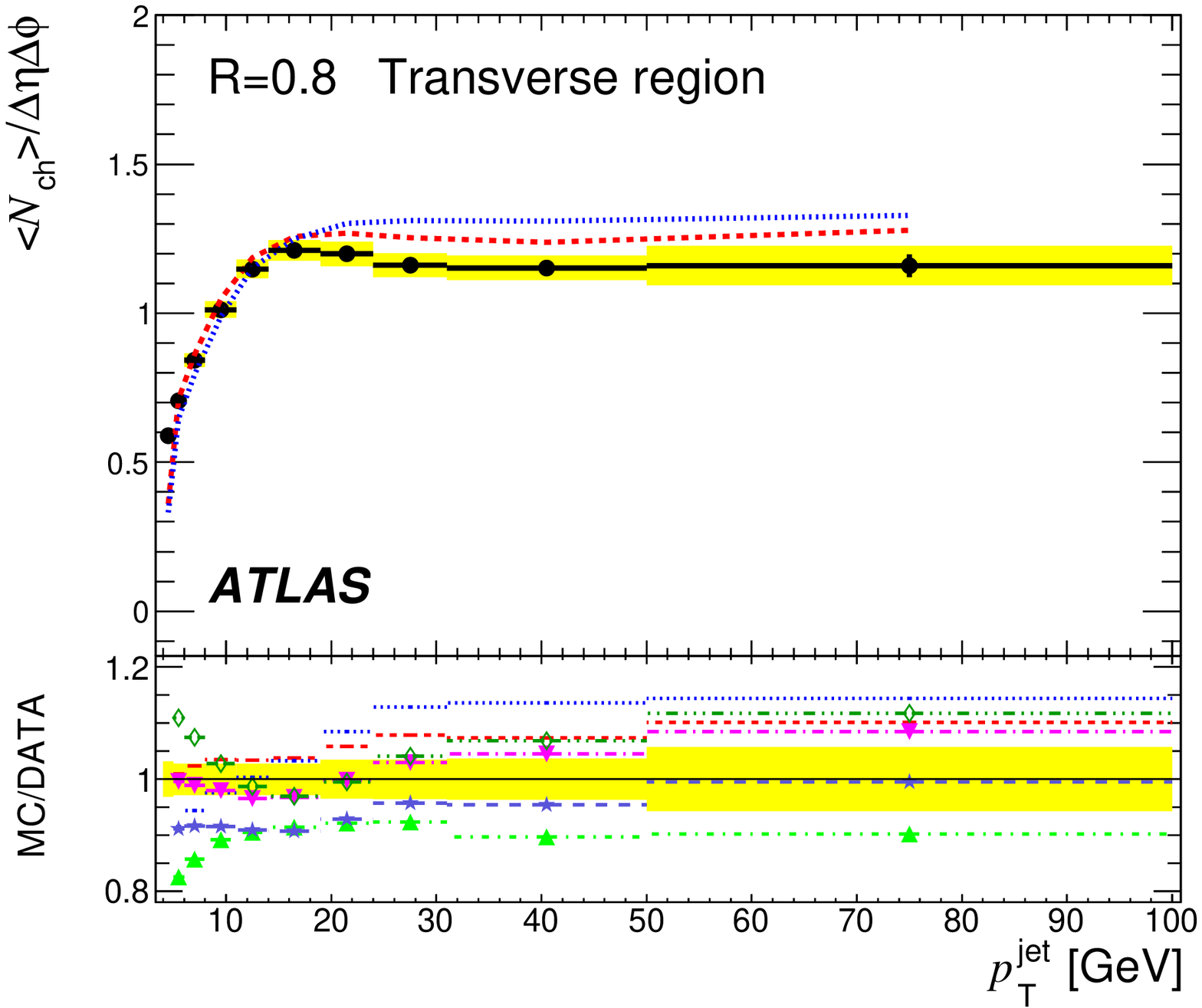}}
\subfigure[]{\includegraphics[width=0.42\textwidth]{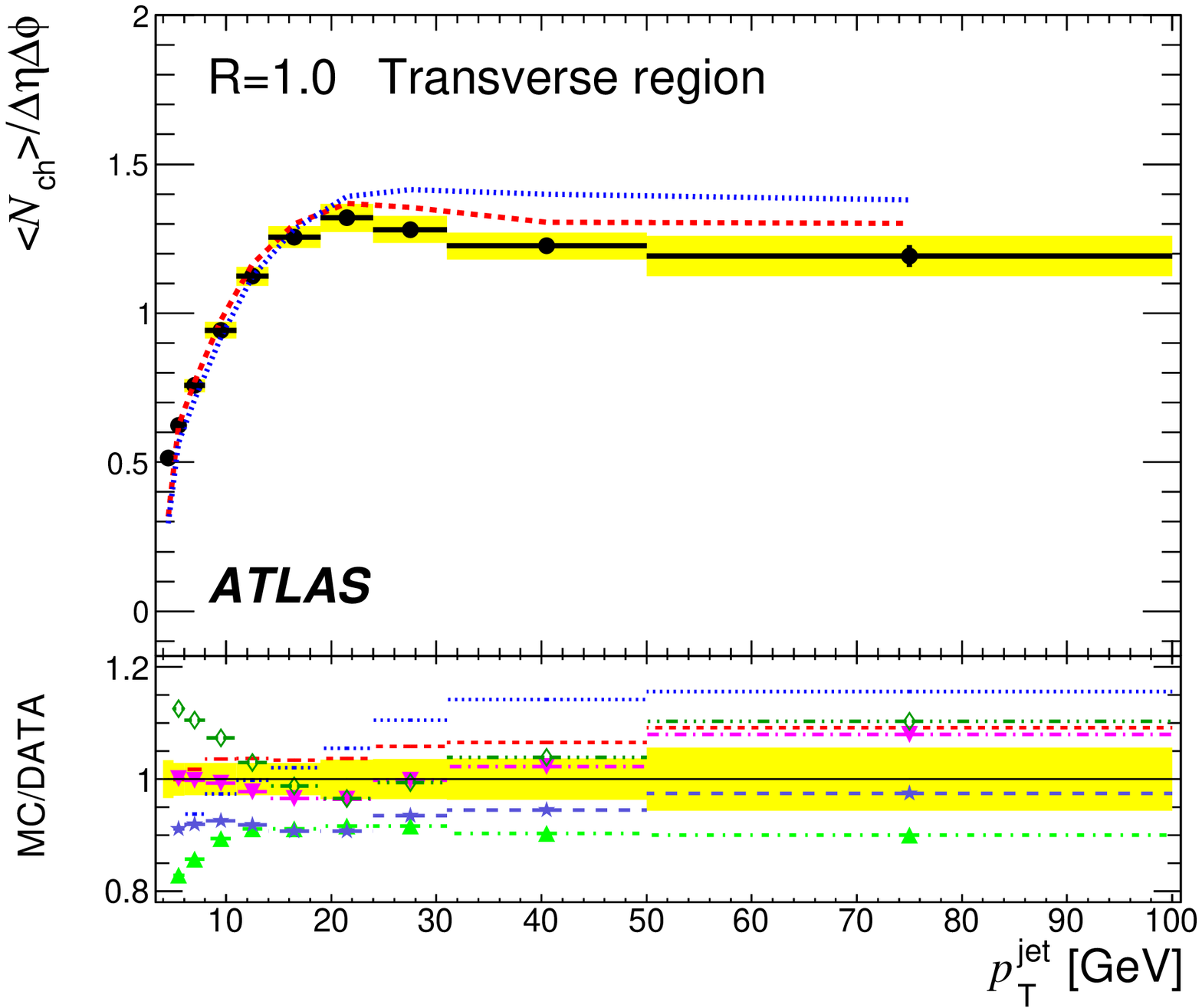}}
{\includegraphics[width=0.42\textwidth]{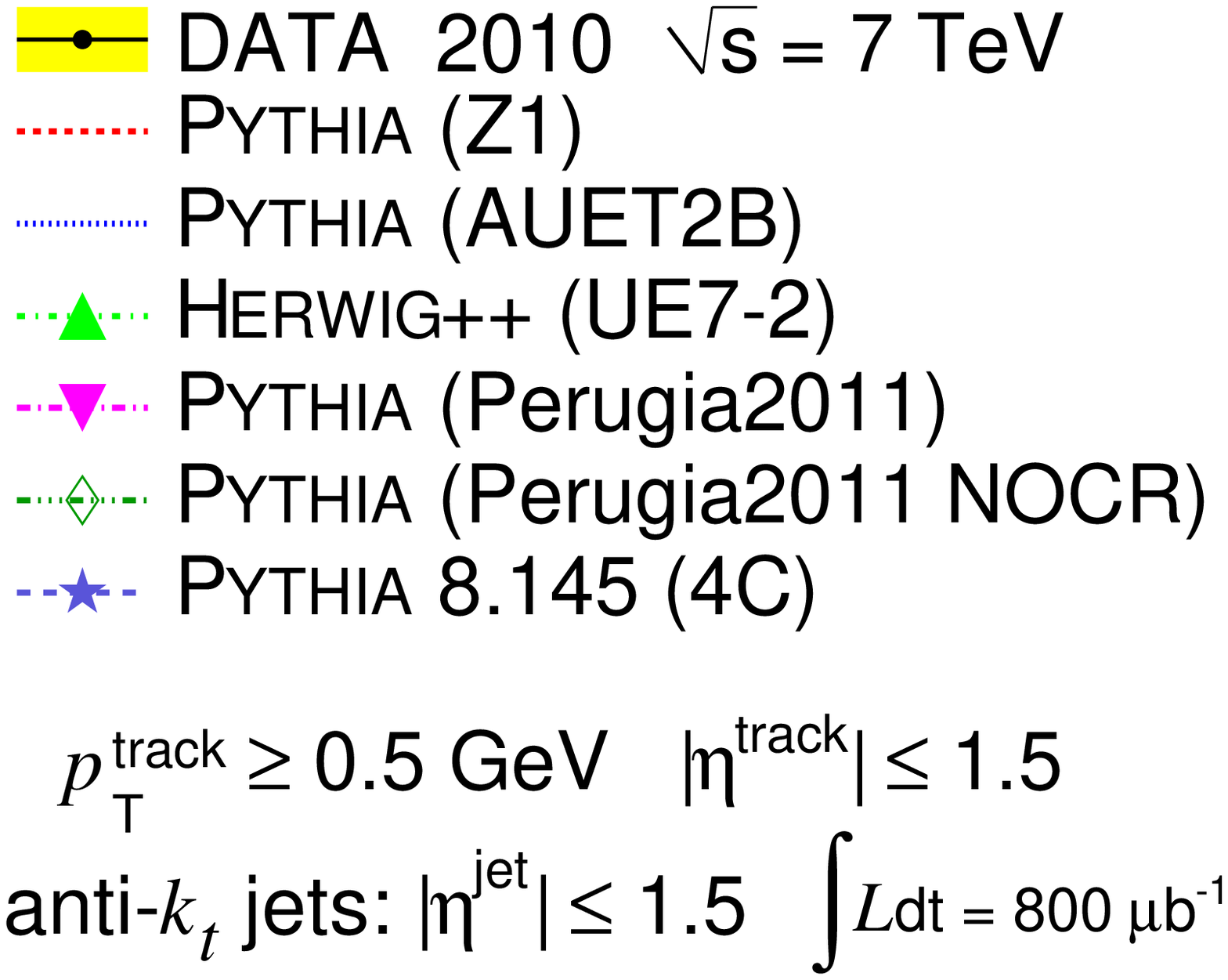}}
\end{center}
\caption{The mean value, per unit $ \eta - \phi $, of \ntrks\ in the transverse region, as a function of
\ptJet\ for
(a) $R=0.2$,
(b) $R=0.4$,
(c) $R=0.6$,
(d) $R=0.8$ and
(e) $R=1.0$.
The shaded band shows the combined statistical and systematic
uncertainty on the data.
The data are compared to the predictions 
obtained with \PythiaS\ using the AUET2B and Z1 tunes.  The bottom
insert shows the ratio of MC predictions to data for several recent MC
tunes.
}
\label{fig:ntrksbinstrans}
\end{figure*}

\begin{figure*}[!ptb]
\begin{center}
\subfigure[]{\includegraphics[width=0.42\textwidth]{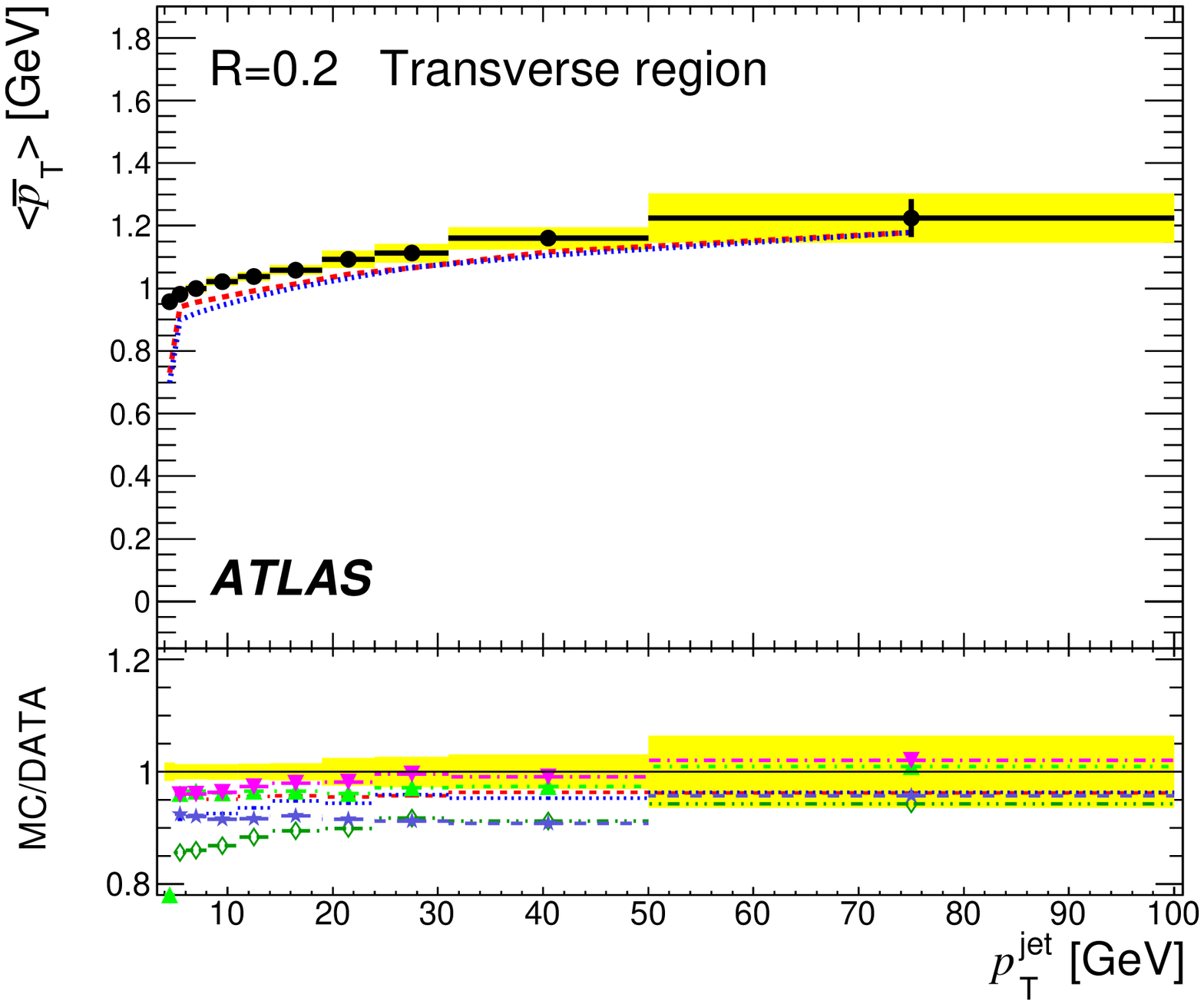}}
\subfigure[]{\includegraphics[width=0.42\textwidth]{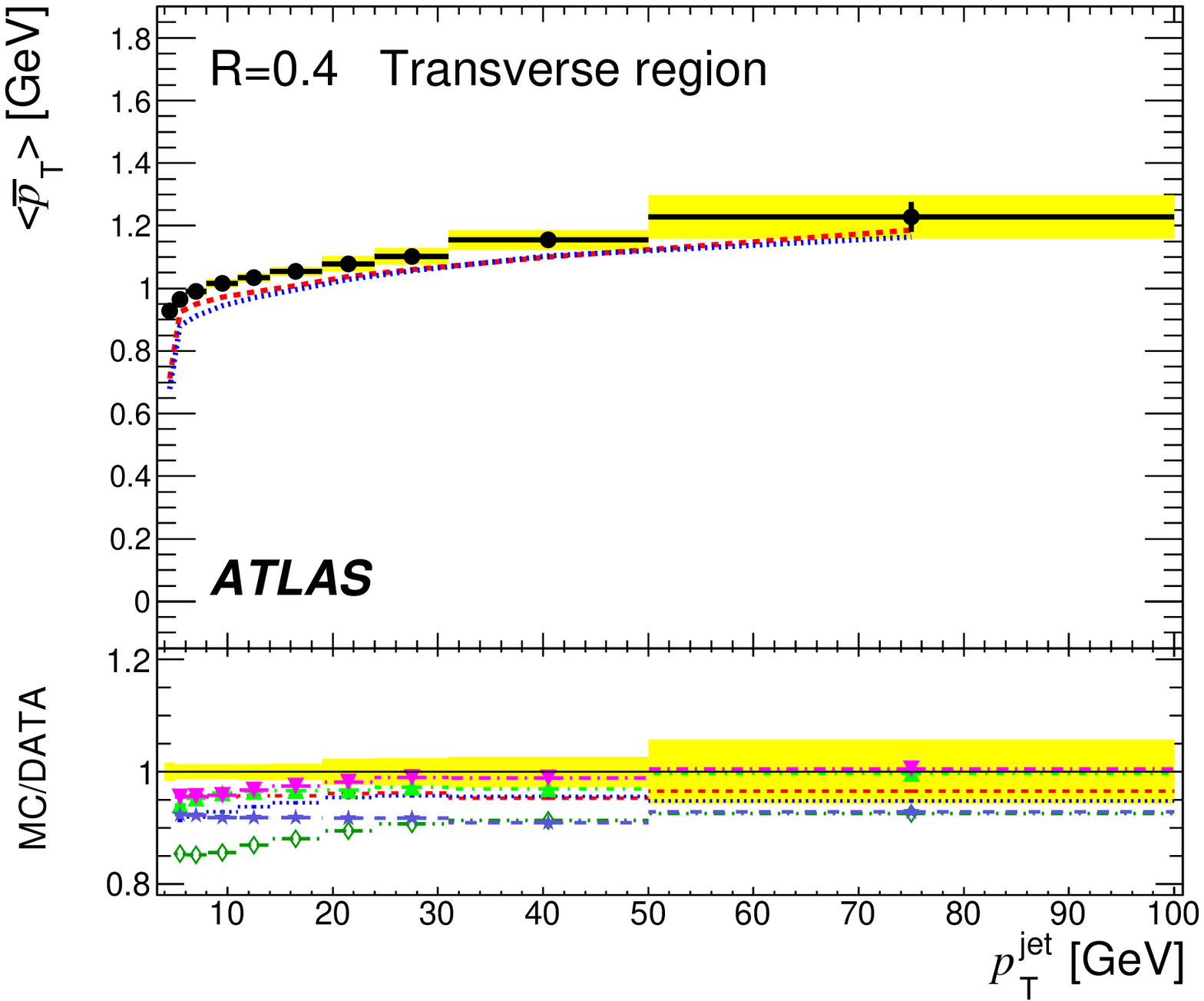}}
\subfigure[]{\includegraphics[width=0.42\textwidth]{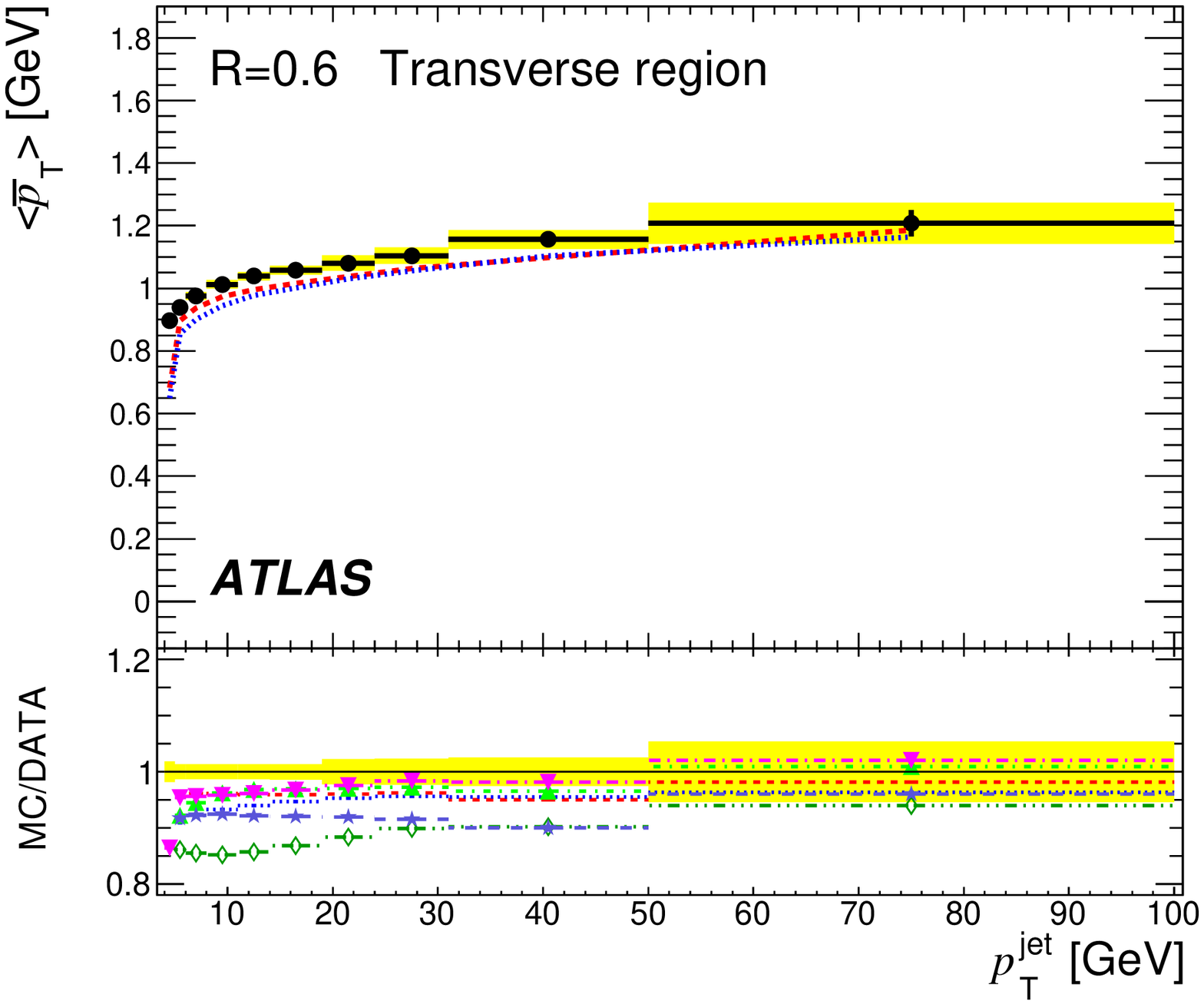}}
\subfigure[]{\includegraphics[width=0.42\textwidth]{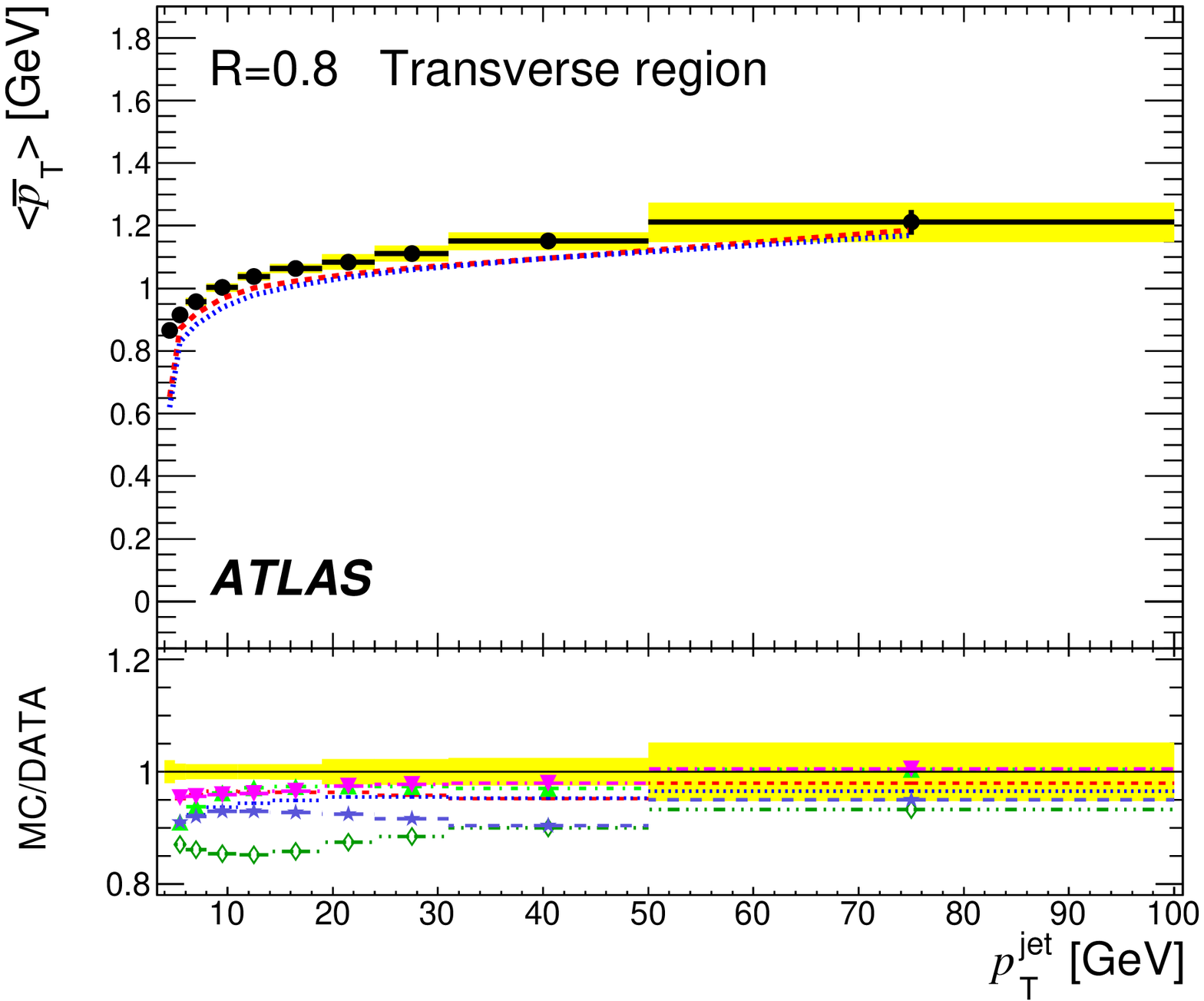}}
\subfigure[]{\includegraphics[width=0.42\textwidth]{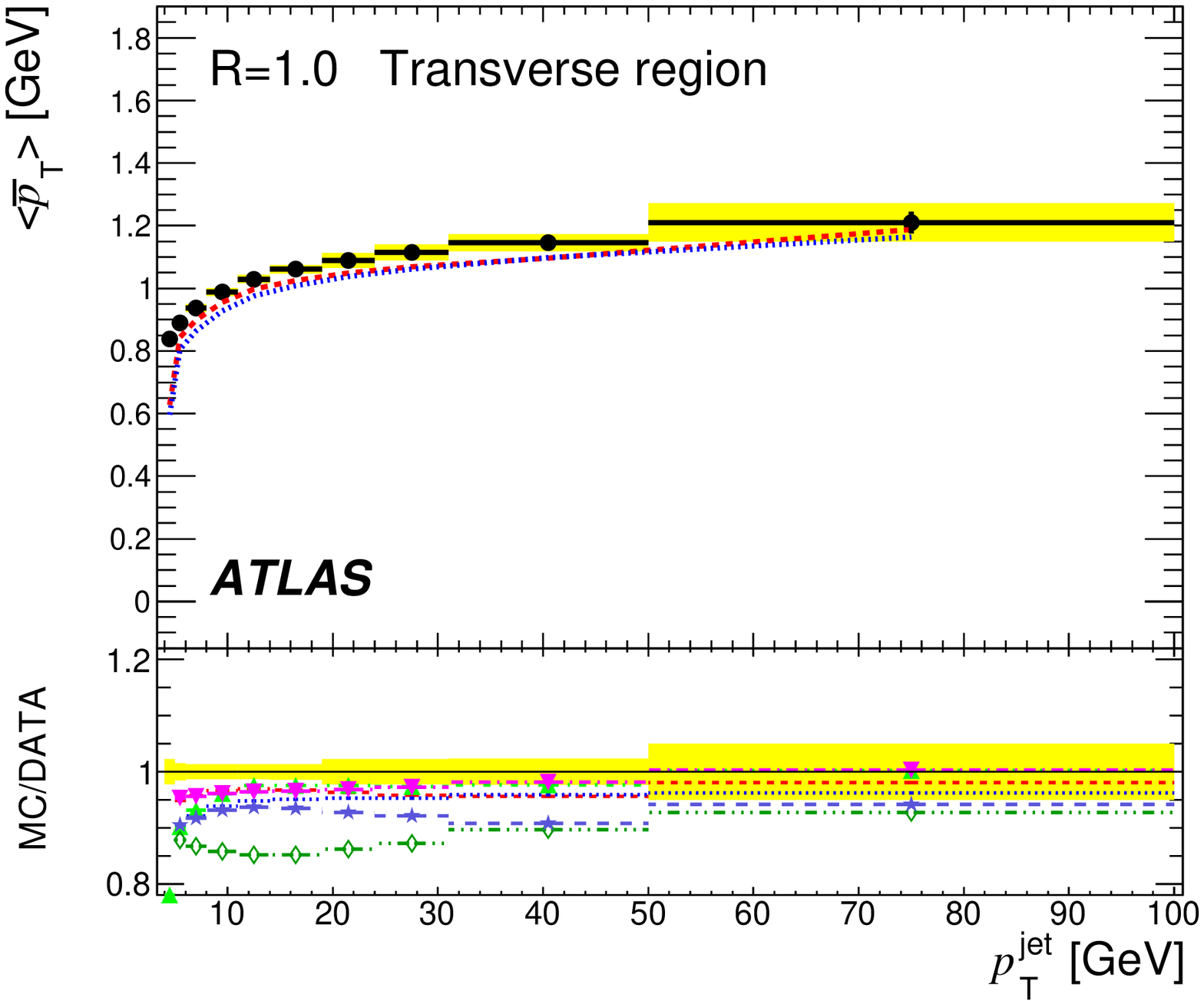}}
{\includegraphics[width=0.42\textwidth]{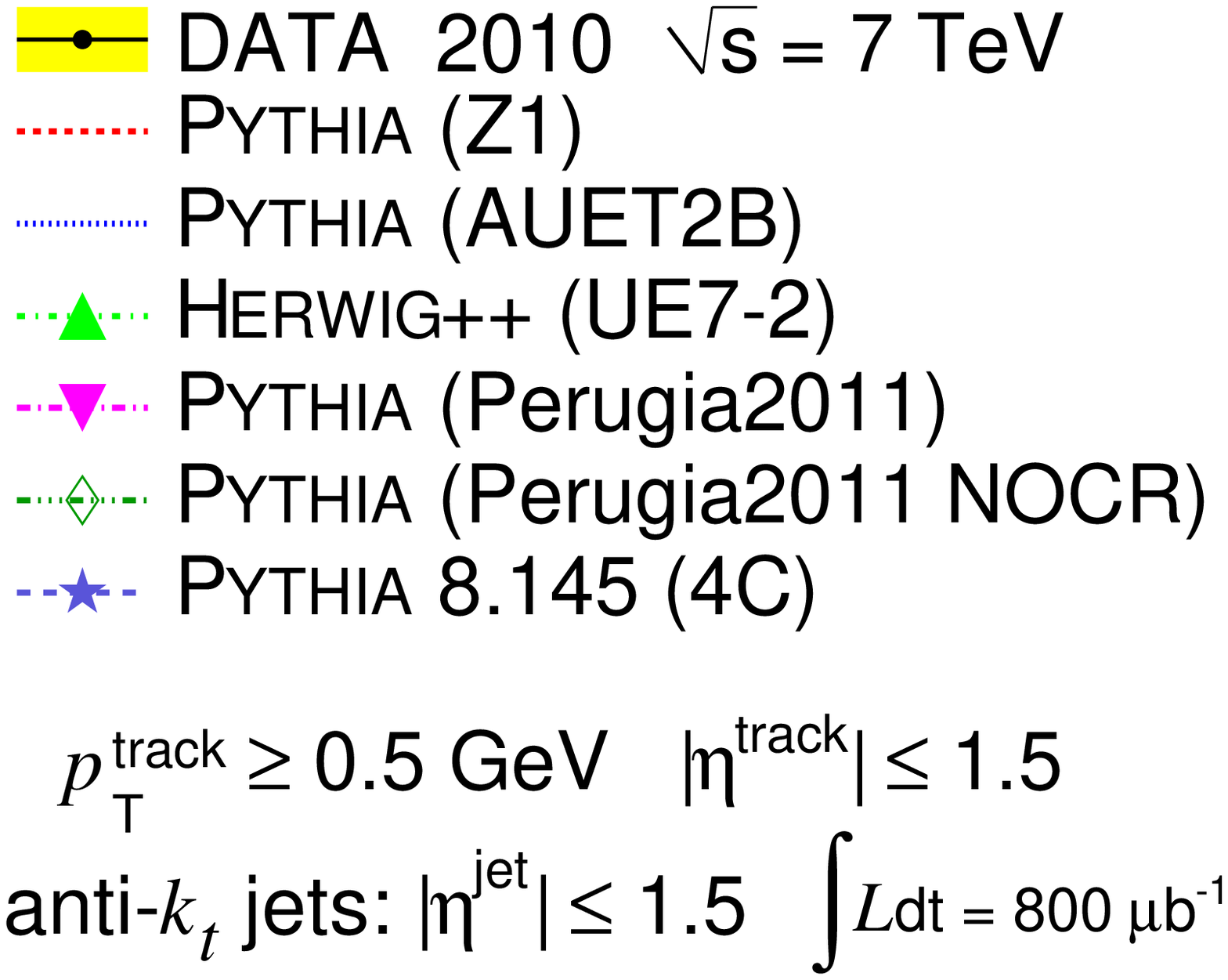}}
\end{center}
\caption{The mean value of \ptave\ in the transverse region, as a function of
\ptJet\ for
(a) $R=0.2$,
(b) $R=0.4$,
(c) $R=0.6$,
(d) $R=0.8$ and
(e) $R=1.0$.
The shaded band shows the combined statistical and systematic
uncertainty on the data.
The data are compared to the predictions 
obtained with \PythiaS\ using the AUET2B and Z1 tunes.  The bottom
insert shows the ratio of MC predictions to data for several recent MC
tunes.
}
\label{fig:ptavebinstrans}
\end{figure*}
\begin{figure*}[!ptb]
\begin{center}
\subfigure[]{\includegraphics[width=0.42\textwidth]{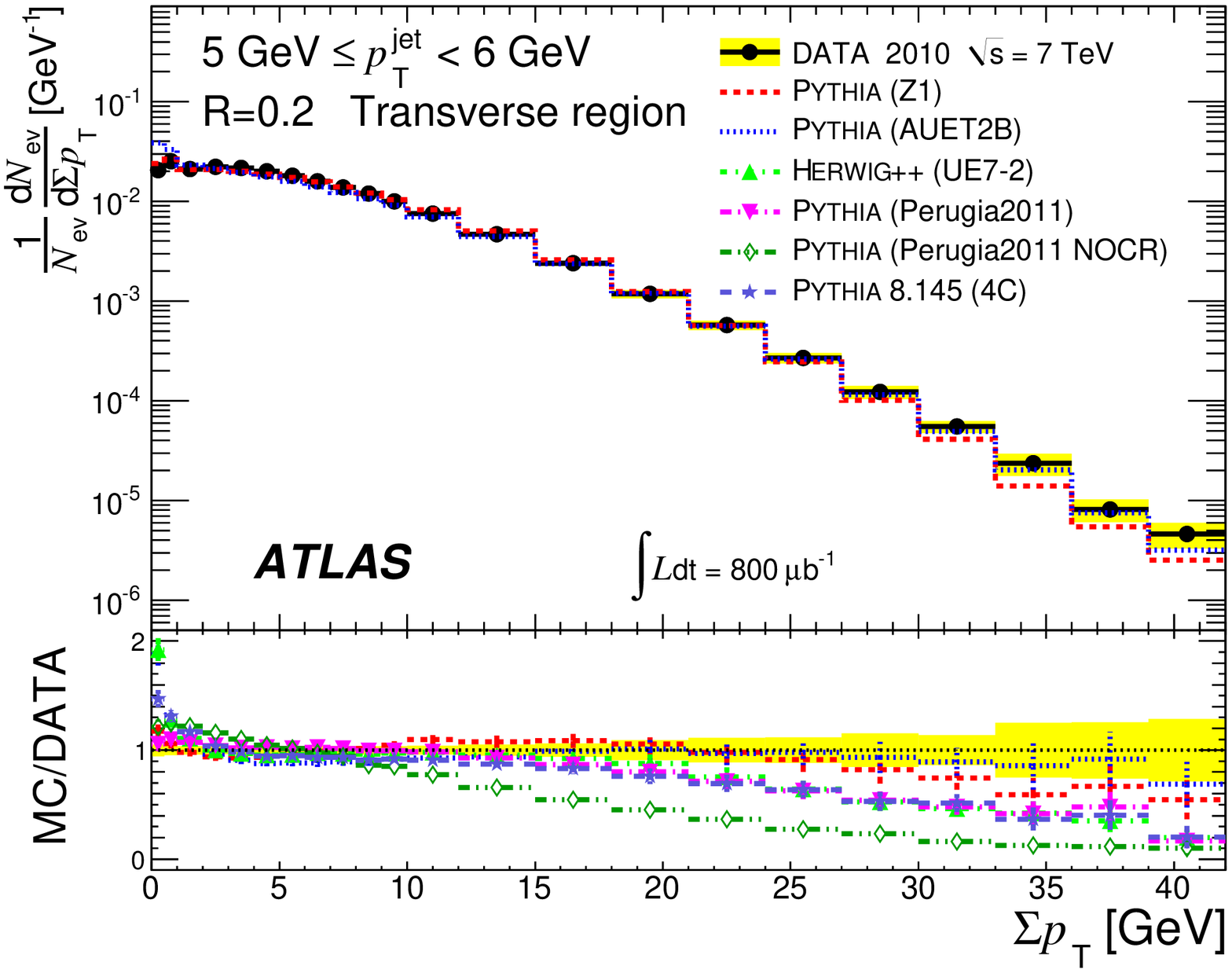}\label{fig:sumptbinsa}}
\subfigure[]{\includegraphics[width=0.42\textwidth]{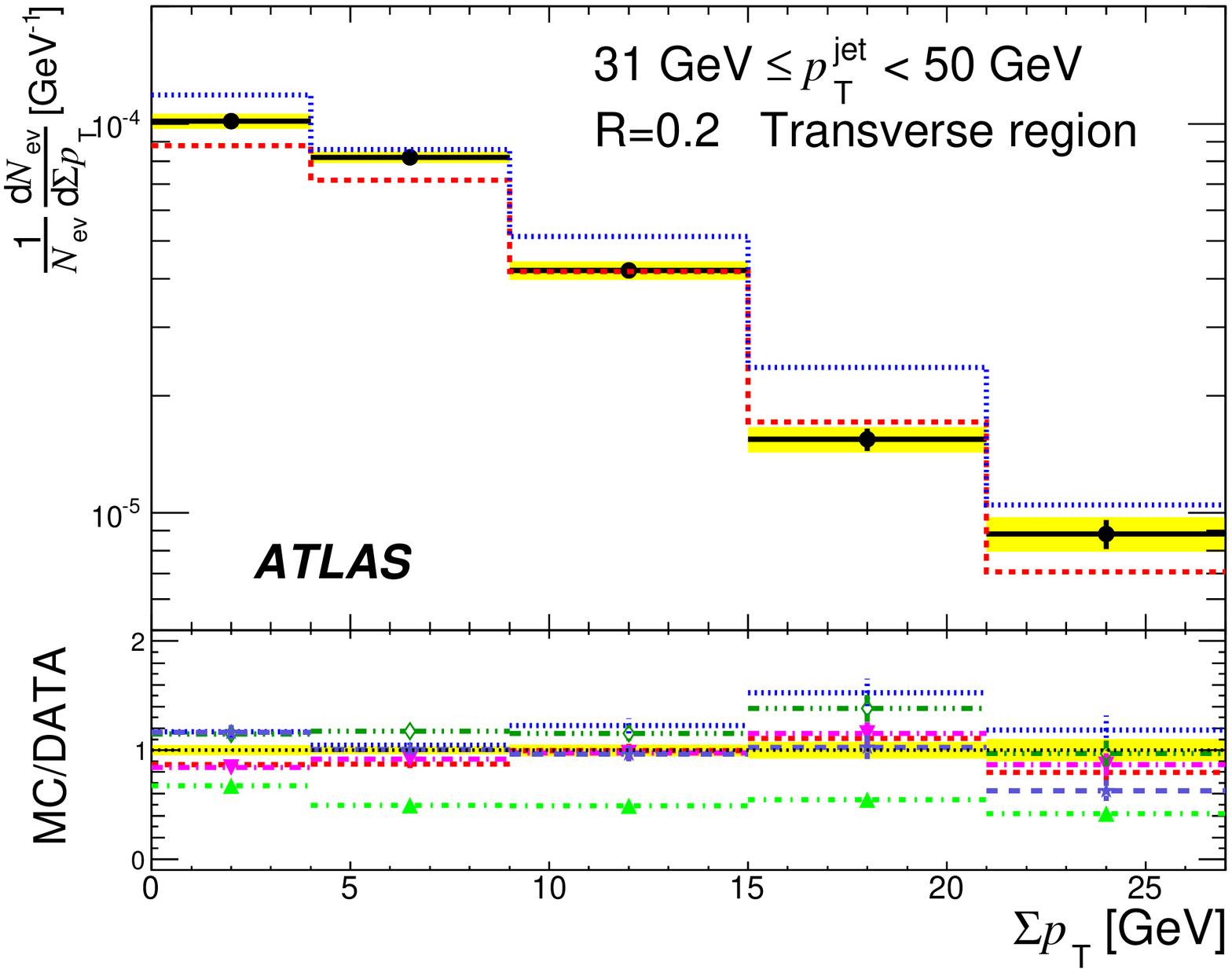}}
\subfigure[]{\includegraphics[width=0.42\textwidth]{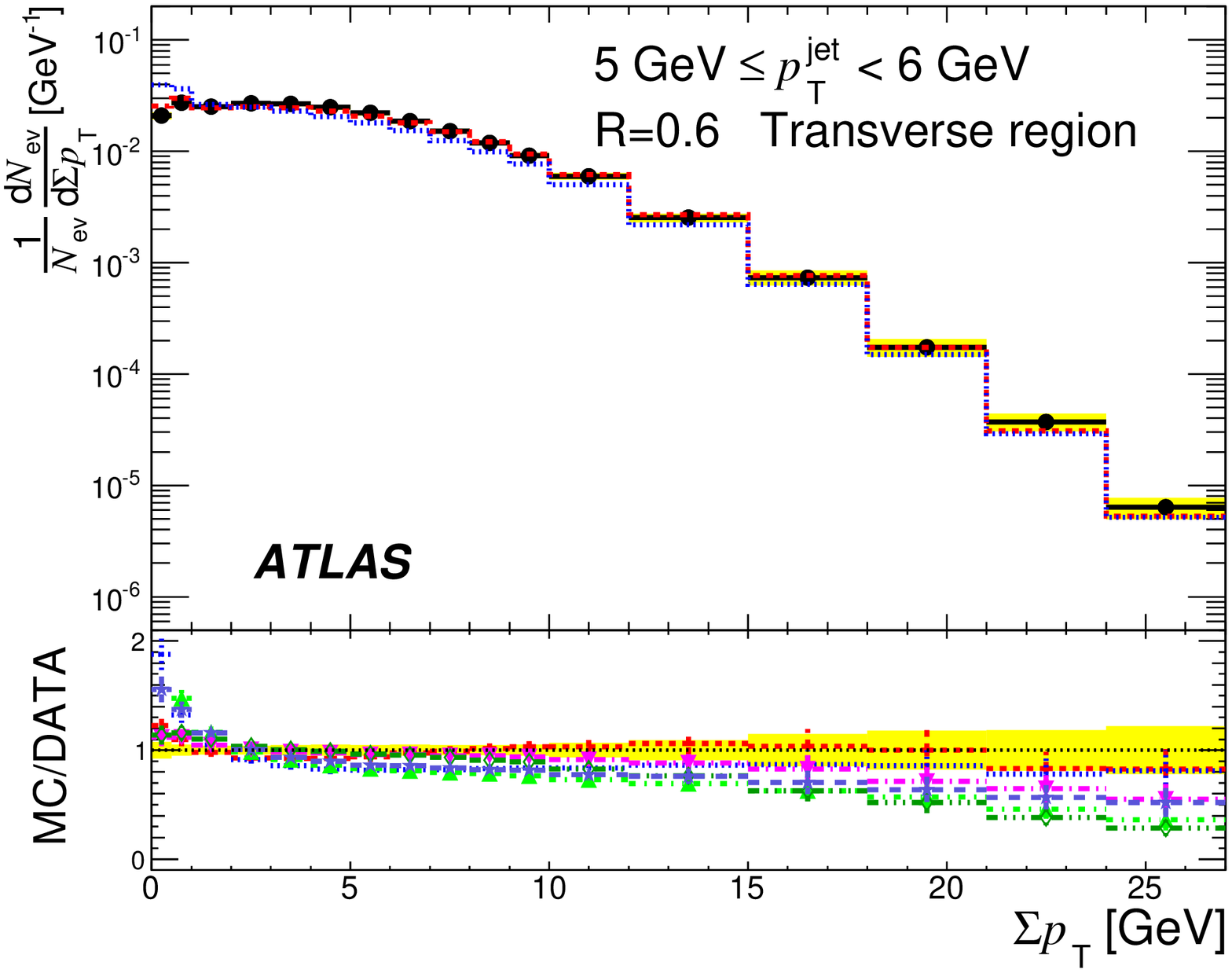}}
\subfigure[]{\includegraphics[width=0.42\textwidth]{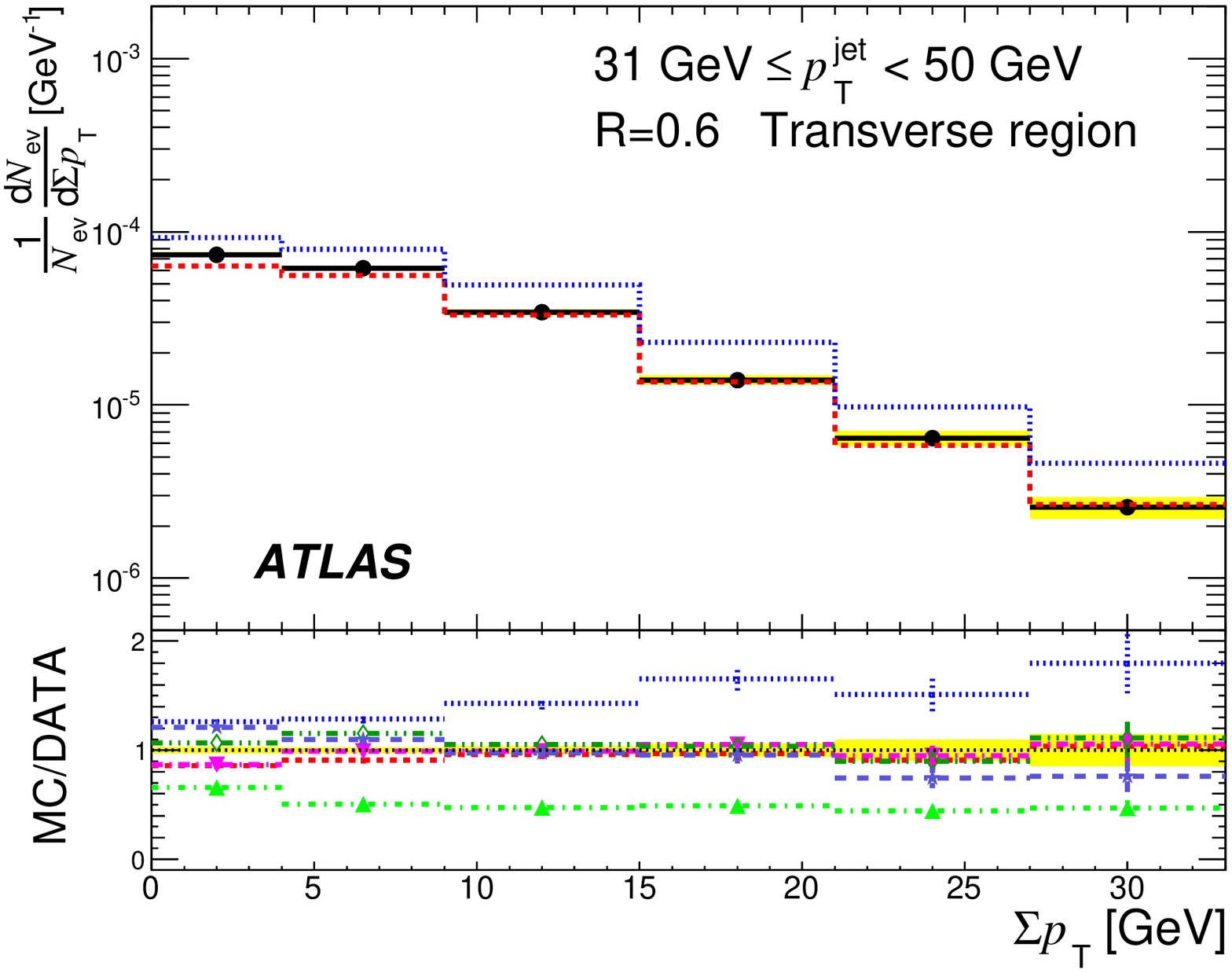}}
\subfigure[]{\includegraphics[width=0.42\textwidth]{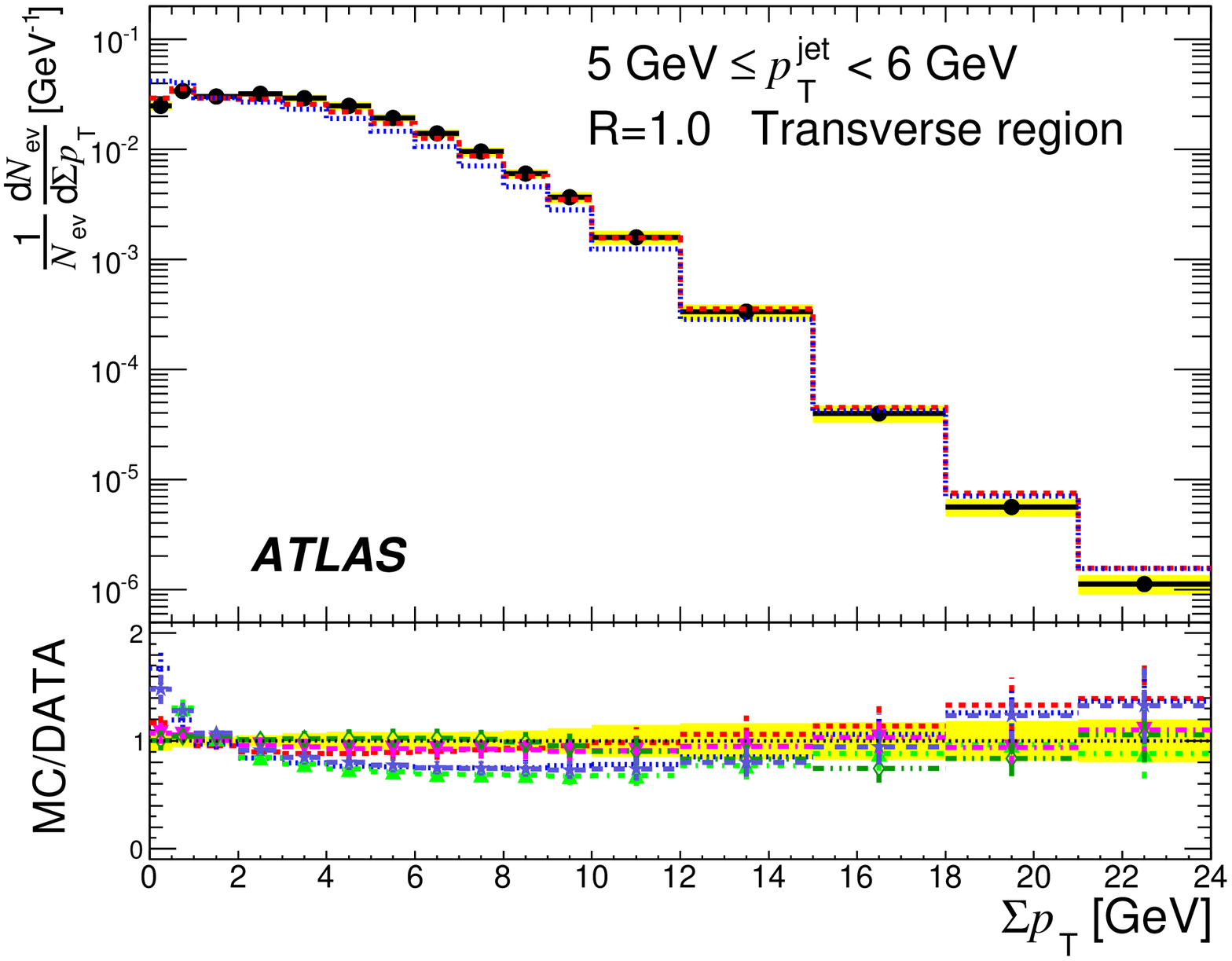}}
\subfigure[]{\includegraphics[width=0.42\textwidth]{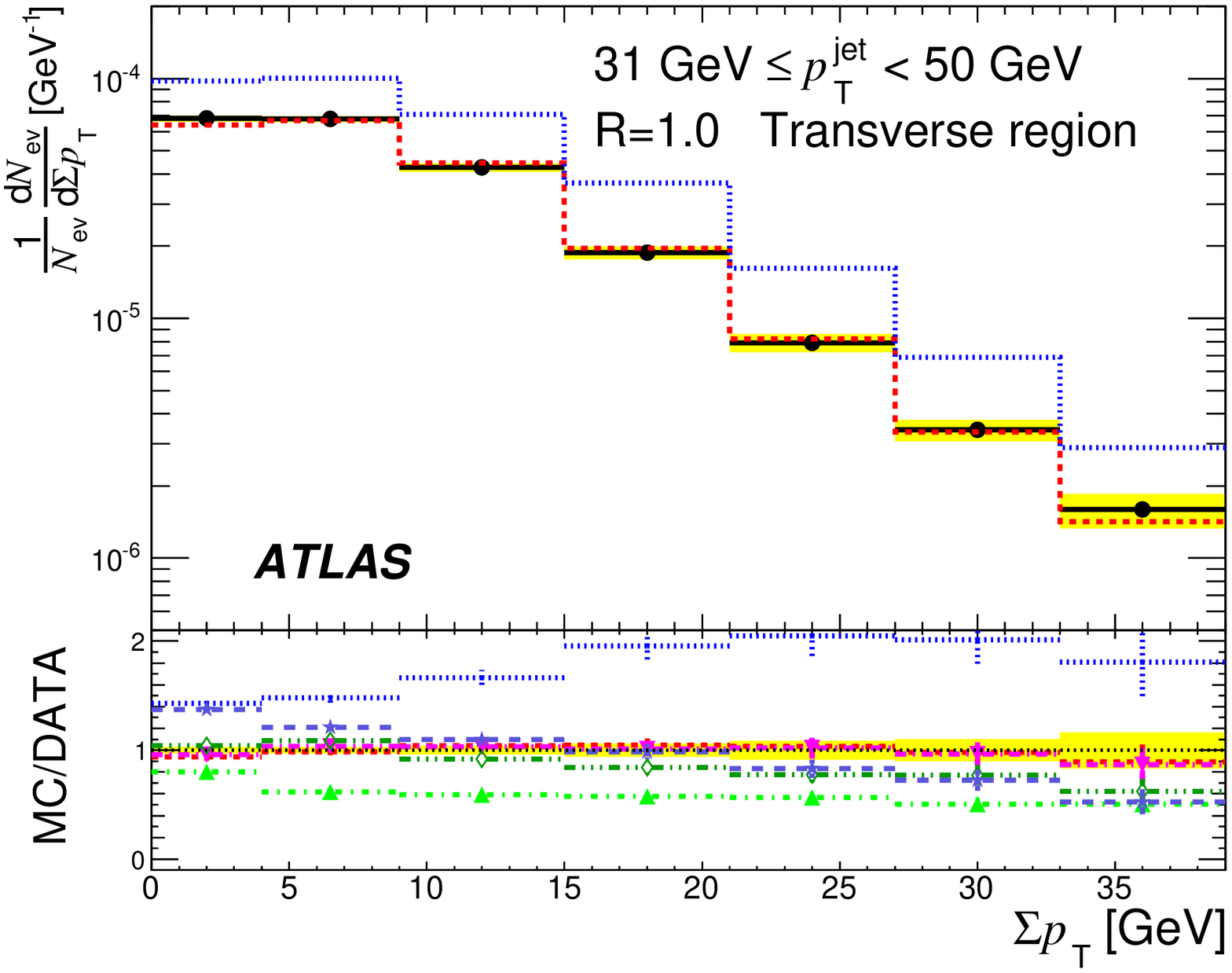}}
\end{center}
\caption{Distributions of \sumpt\ in the 
transverse region for
(a) $R=0.2$ and $5 \; \GeV \le \ptJet < 6 \;\GeV$,
(b) $R=0.2$ and $31 \; \GeV \le \ptJet < 50 \;\GeV$,
(c) $R=0.6$ and $5 \; \GeV \le \ptJet < 6 \;\GeV$,
(d) $R=0.6$ and $31 \; \GeV \le \ptJet < 50 \;\GeV$ 
(e) $R=1.0$ and $5 \; \GeV \le \ptJet < 6 \;\GeV$ and 
(f) $R=1.0$ and $31 \; \GeV \le \ptJet < 50 \;\GeV$ .
The shaded band shows the combined statistical and systematic 
uncertainty on the data.
The histograms show the predictions of several MC models with the same legend as in Fig. \subref{fig:sumptbinsa} and also Figs. [\ref{fig:sumptbinstrans}, \ref{fig:ntrksbinstrans}, \ref{fig:ptavebinstrans}].
}
\label{fig:sumptbins}
\end{figure*}

\begin{figure*}[!ptb]
\subfigure[]{\includegraphics[width=0.42\textwidth]{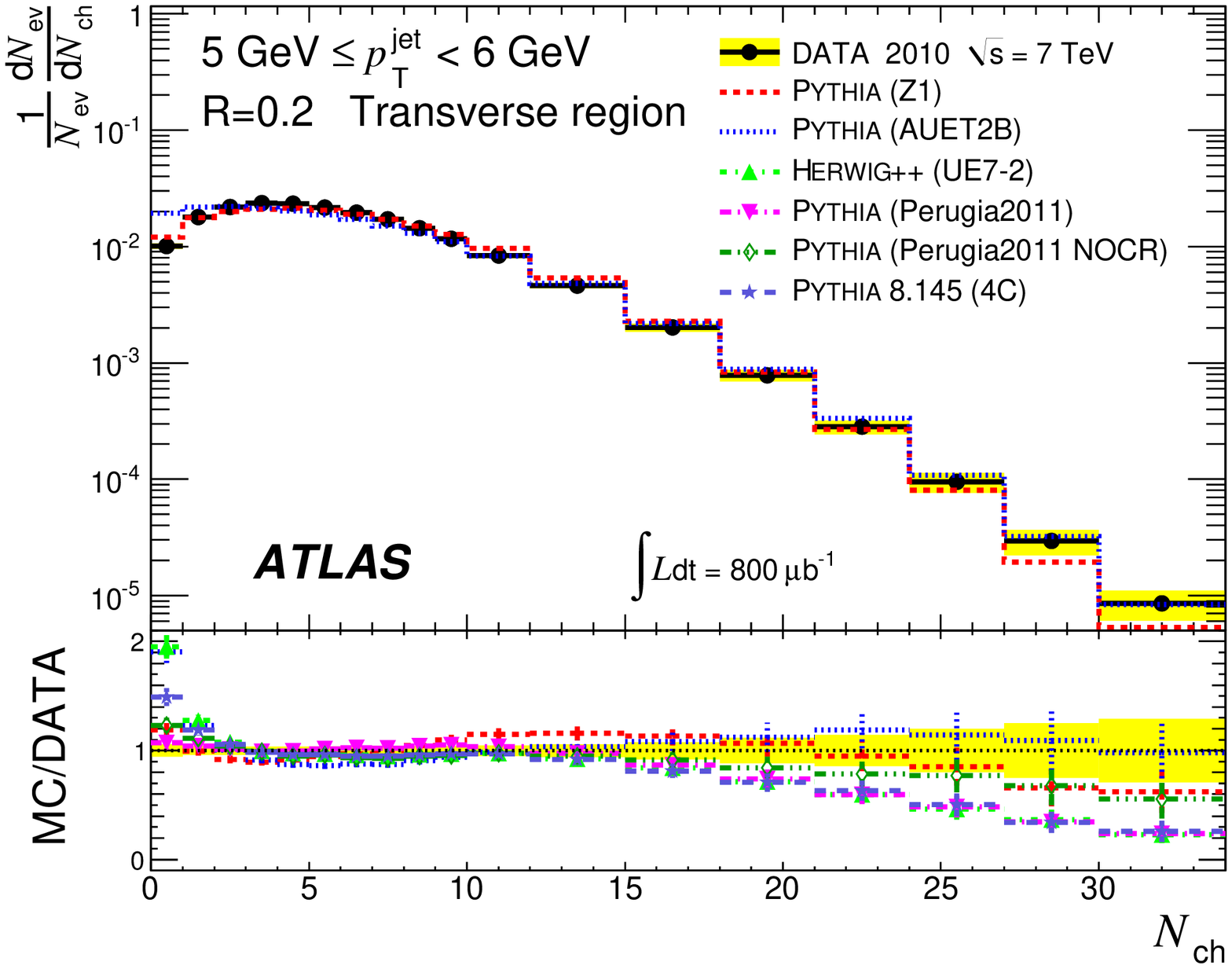}\label{fig:ntrksbinsa}}
\subfigure[]{\includegraphics[width=0.42\textwidth]{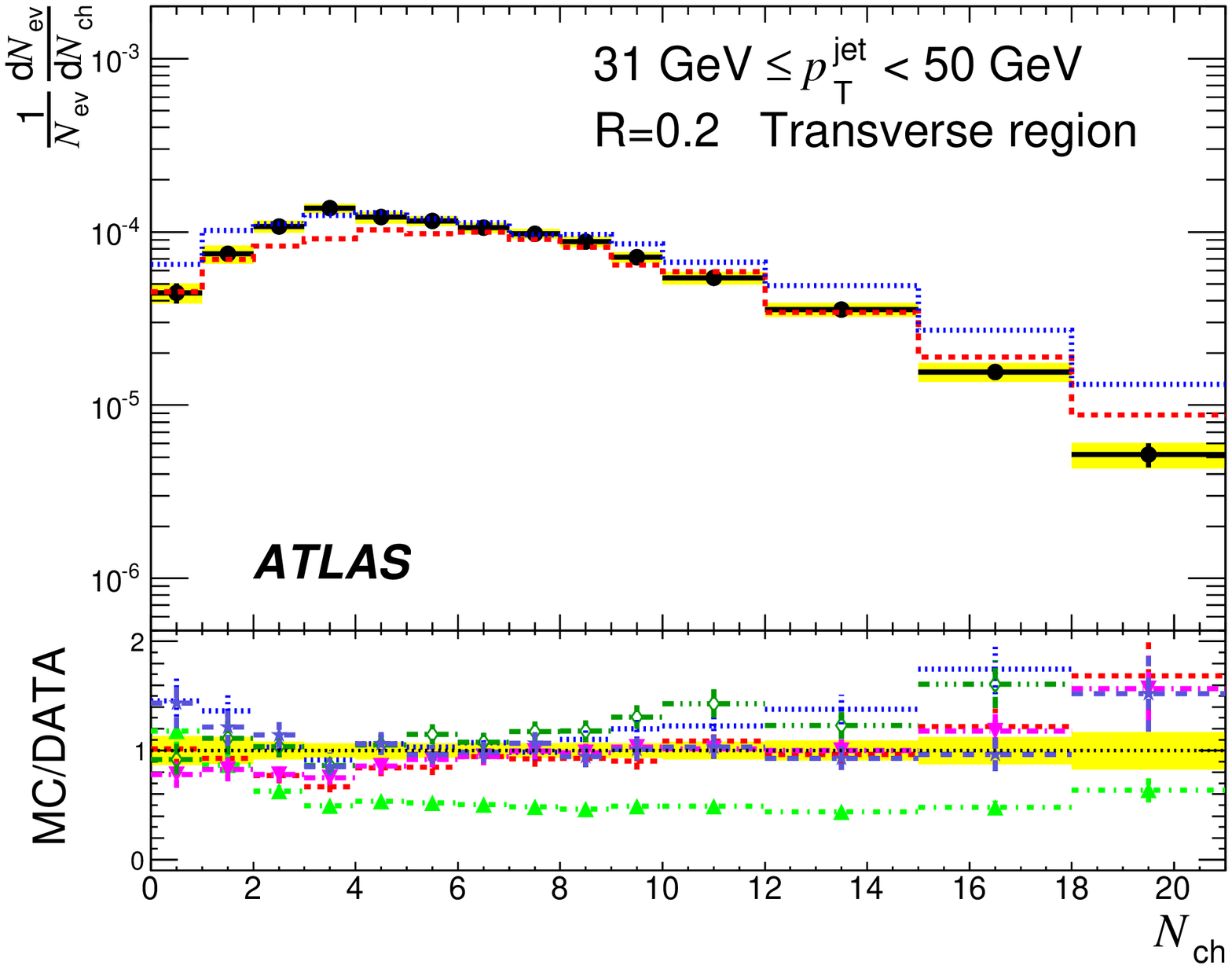}}
\subfigure[]{\includegraphics[width=0.42\textwidth]{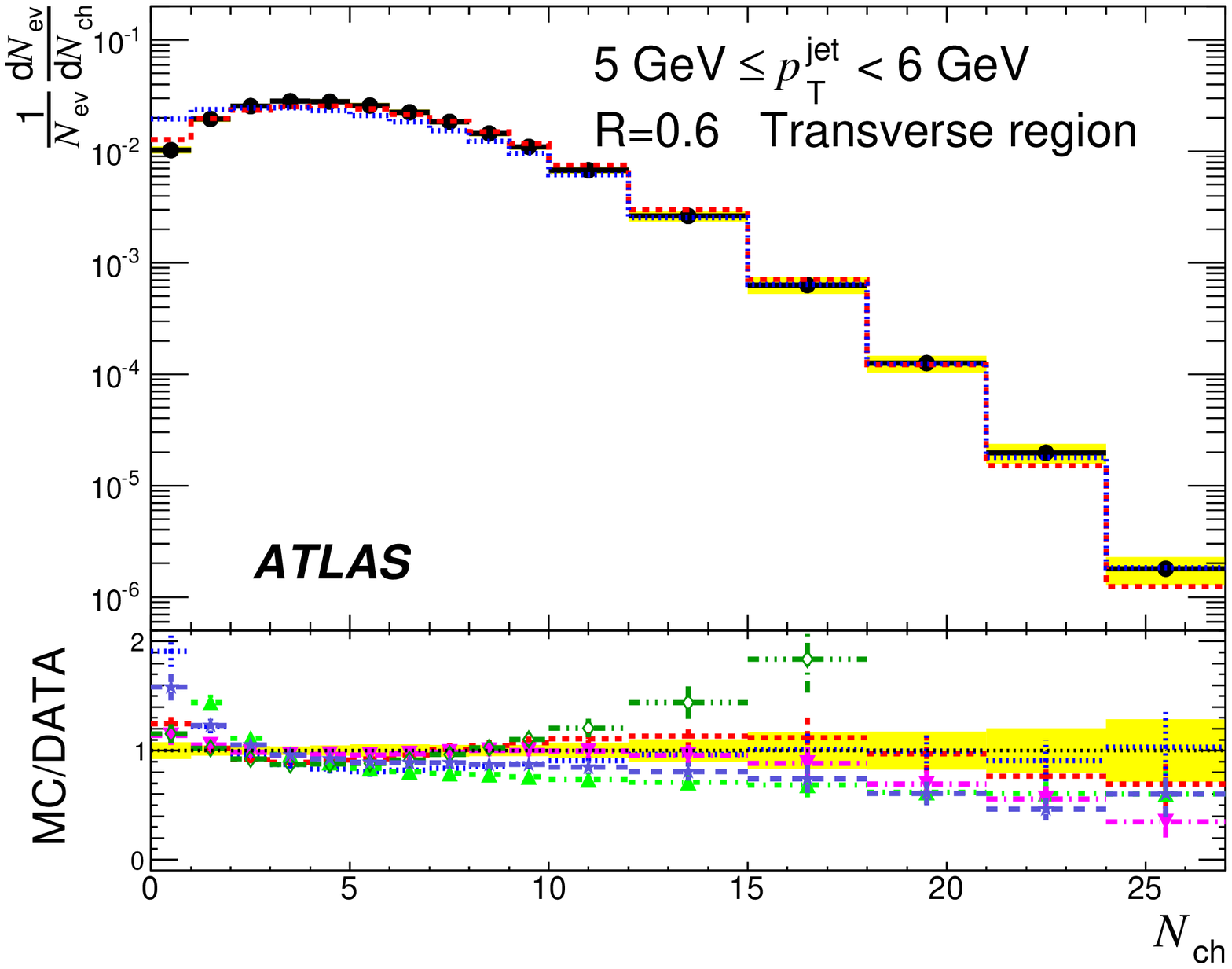}}
\subfigure[]{\includegraphics[width=0.42\textwidth]{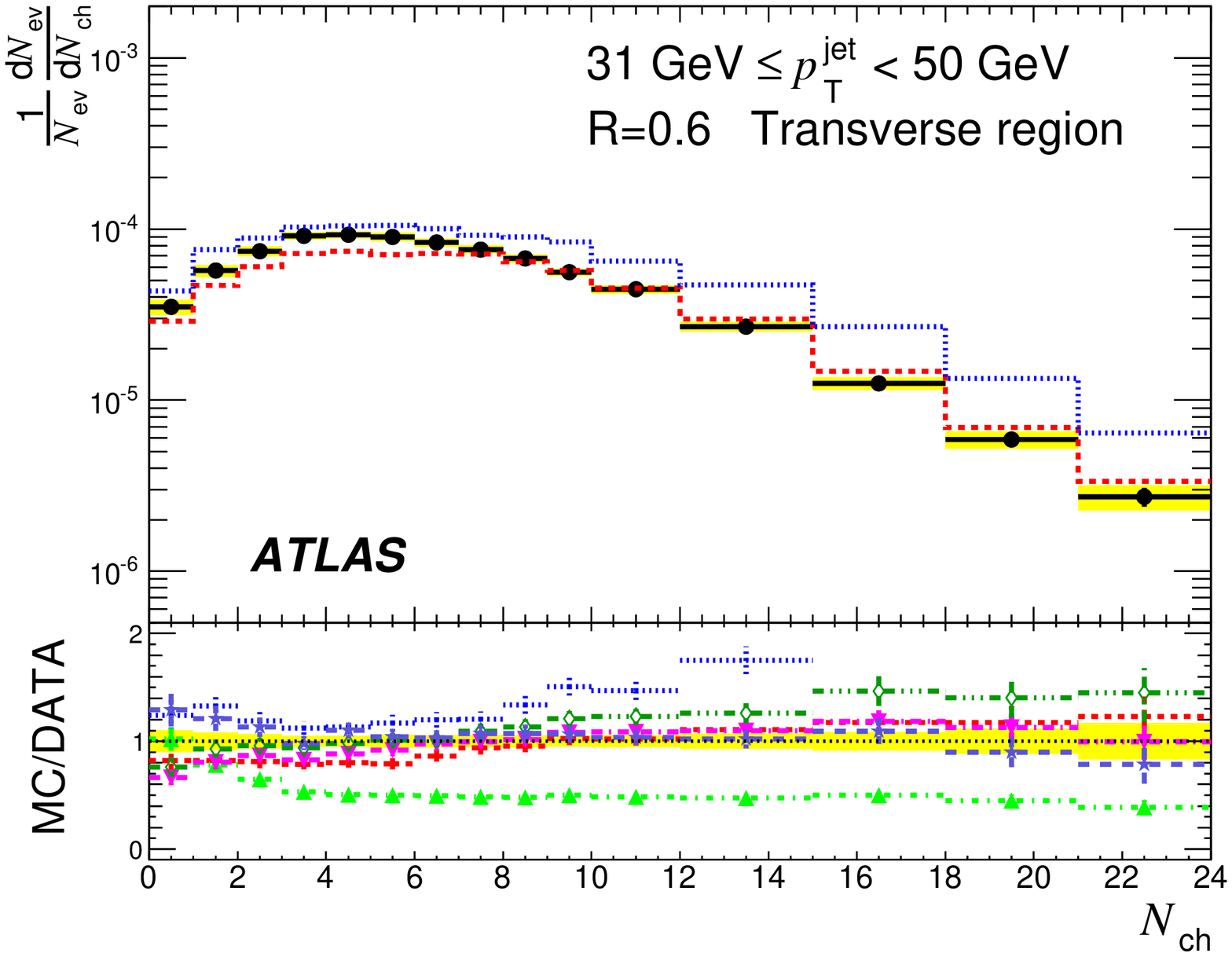}}
\subfigure[]{\includegraphics[width=0.42\textwidth]{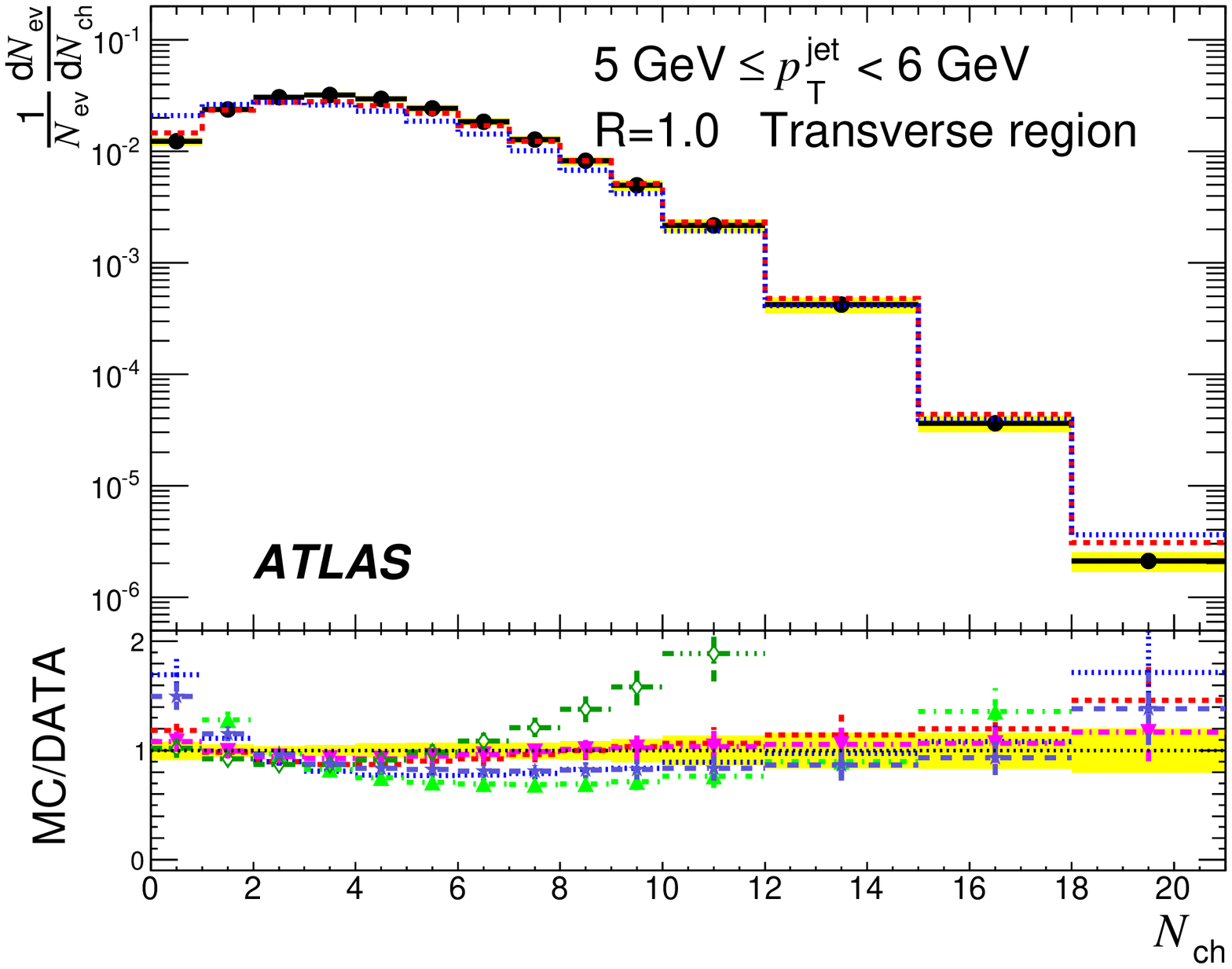}}
\subfigure[]{\includegraphics[width=0.42\textwidth]{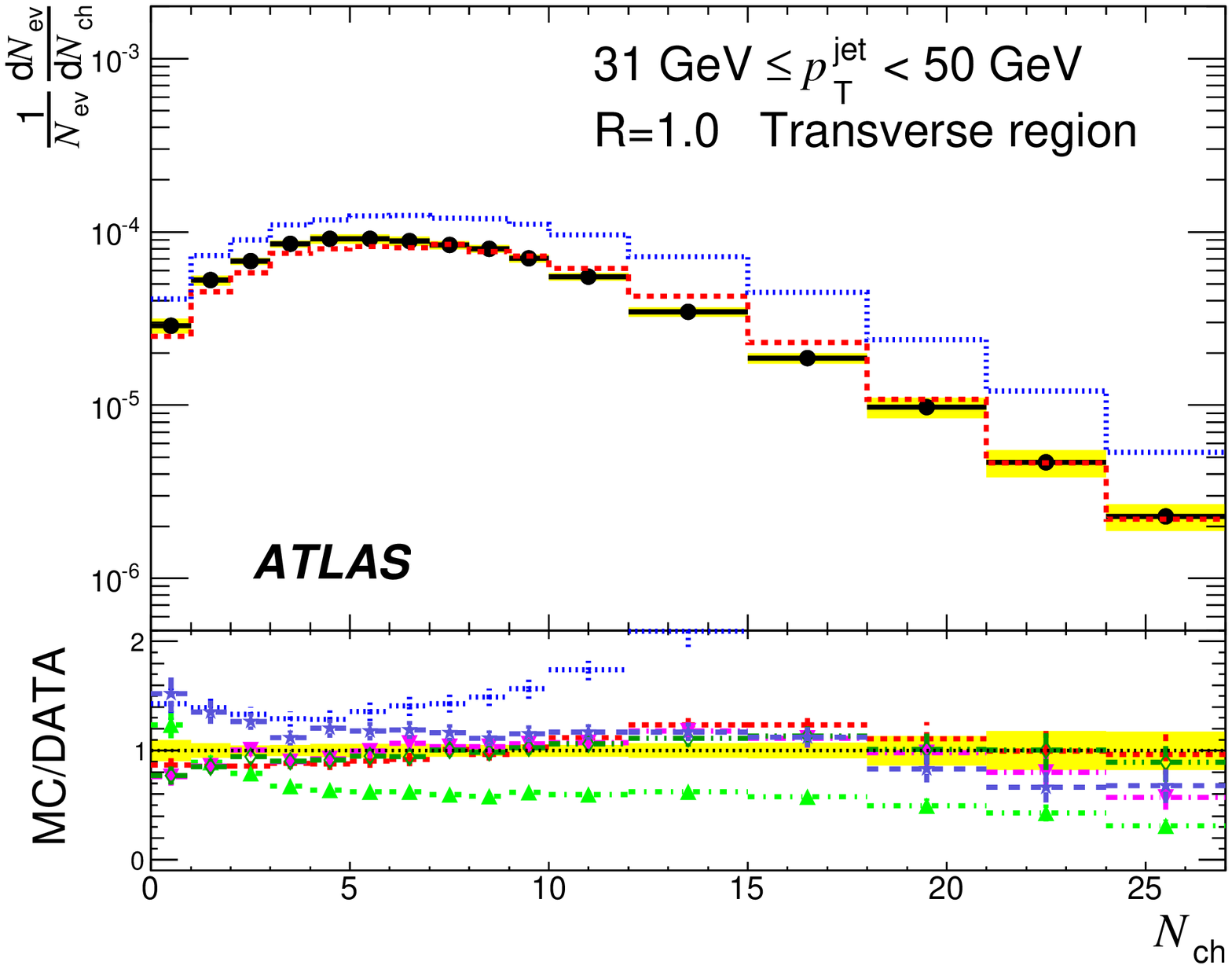}}
\caption{Distributions of \ntrks\ in the 
transverse region for
(a) $R=0.2$ and $5 \; \GeV \le \ptJet < 6 \;\GeV$,
(b) $R=0.2$ and $31 \; \GeV \le \ptJet < 50 \;\GeV$,
(c) $R=0.6$ and $5 \; \GeV \le \ptJet < 6 \;\GeV$, 
(d) $R=0.6$ and $31 \; \GeV \le \ptJet < 50 \;\GeV$ 
(e) $R=1.0$ and $5 \; \GeV \le \ptJet < 6 \;\GeV$ and 
(f) $R=1.0$ and $31 \; \GeV \le \ptJet < 50 \;\GeV$.
The shaded band shows the combined statistical and systematic 
uncertainty on the data.  
The histograms show the predictions of several MC models with the same legend as in Fig. \subref{fig:ntrksbinsa} and also Figs. [\ref{fig:sumptbinstrans}, \ref{fig:ntrksbinstrans}, \ref{fig:ptavebinstrans}].
}
\label{fig:ntrksbins}
\end{figure*}

\begin{figure*}[!ptb]
\subfigure[]{\includegraphics[width=0.42\textwidth]{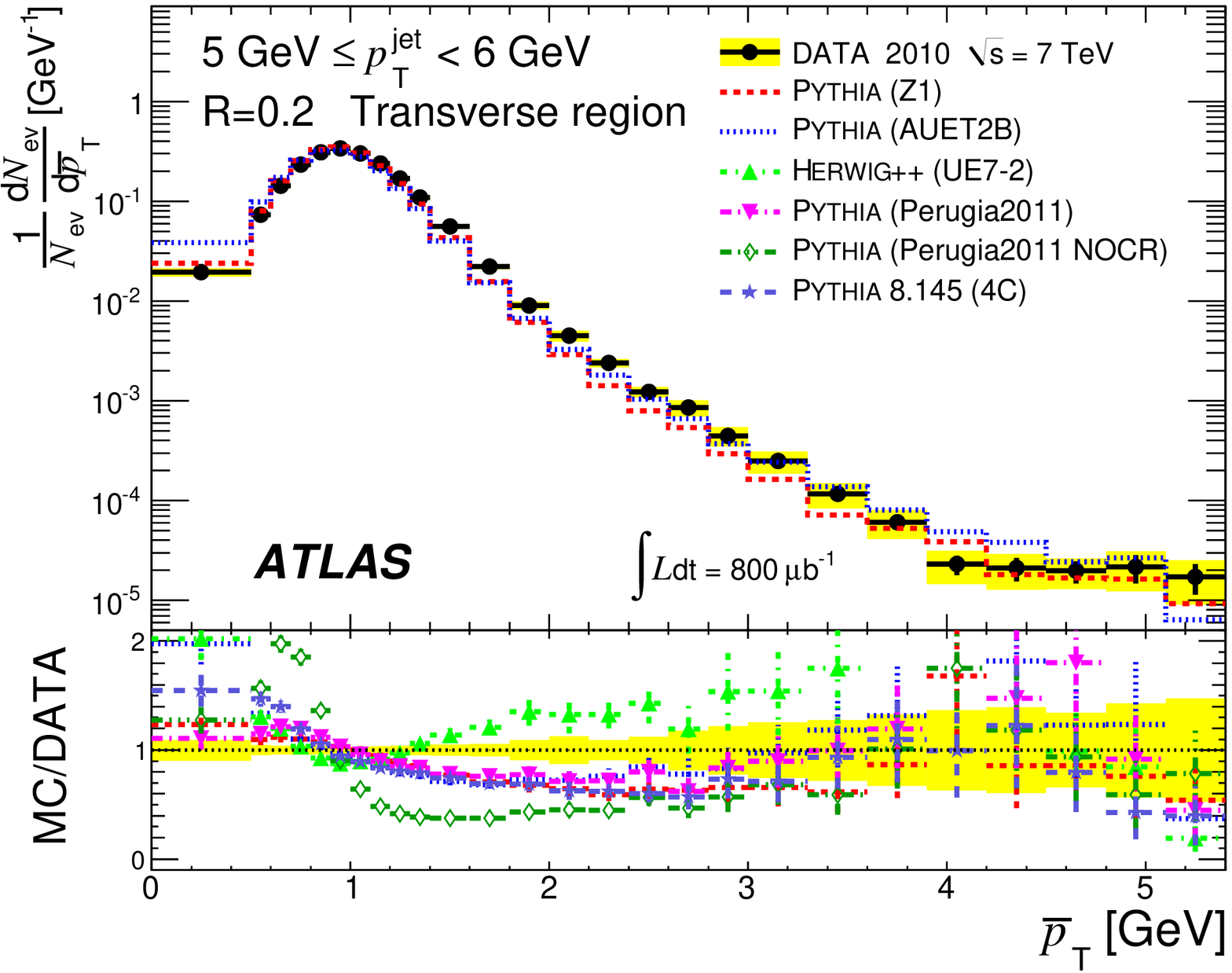}\label{fig:ptavebinsa}}
\subfigure[]{\includegraphics[width=0.42\textwidth]{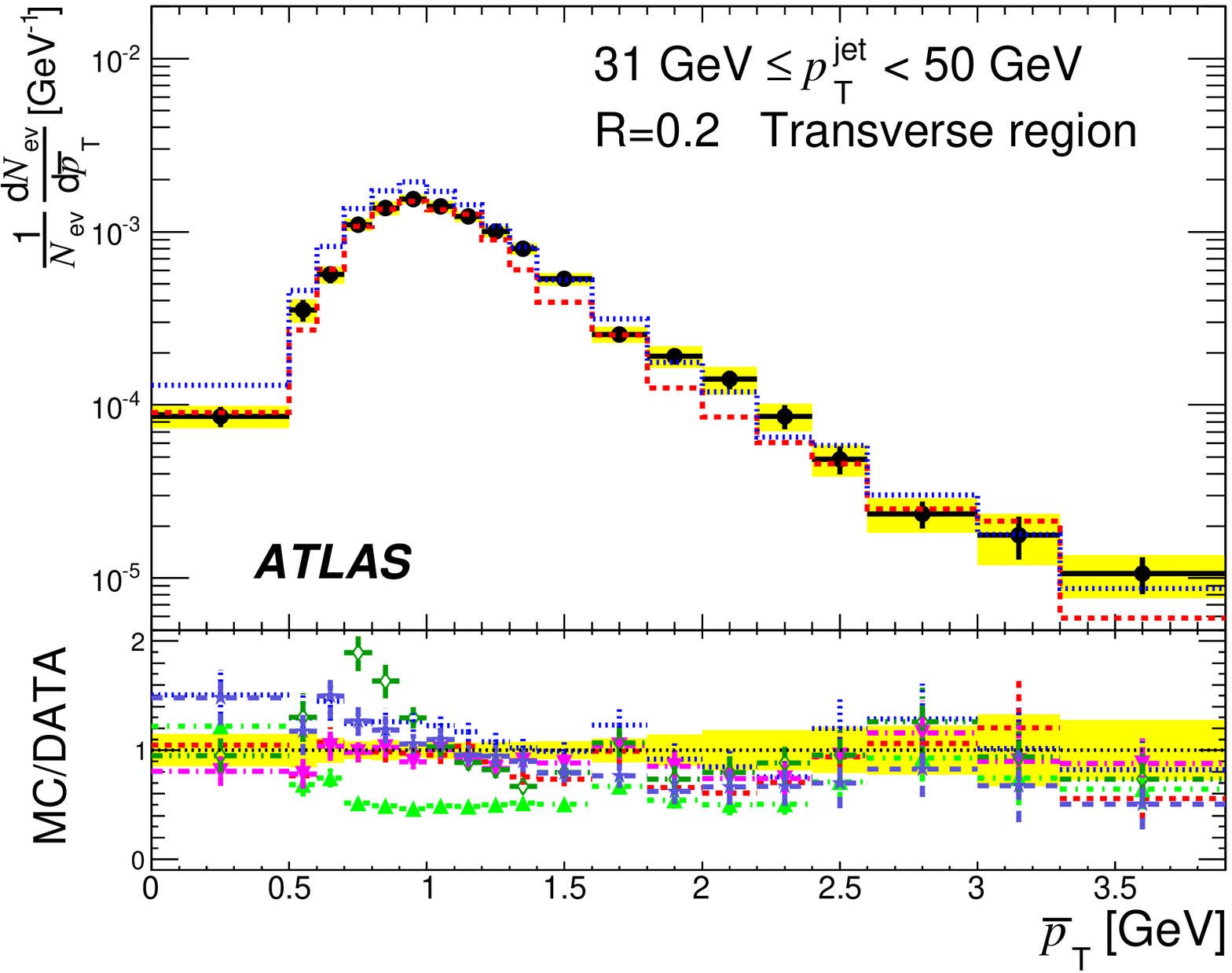}}
\subfigure[]{\includegraphics[width=0.42\textwidth]{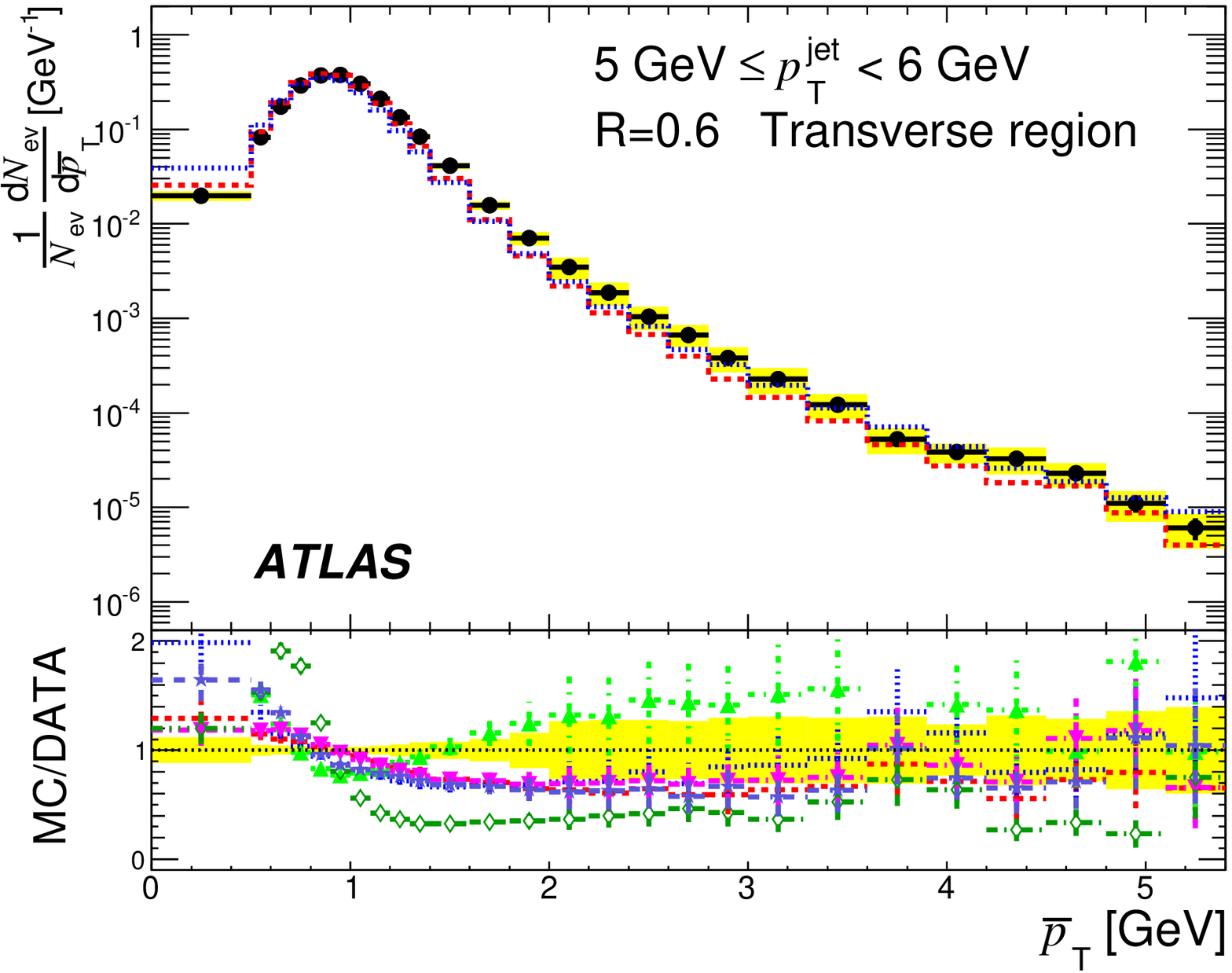}}
\subfigure[]{\includegraphics[width=0.42\textwidth]{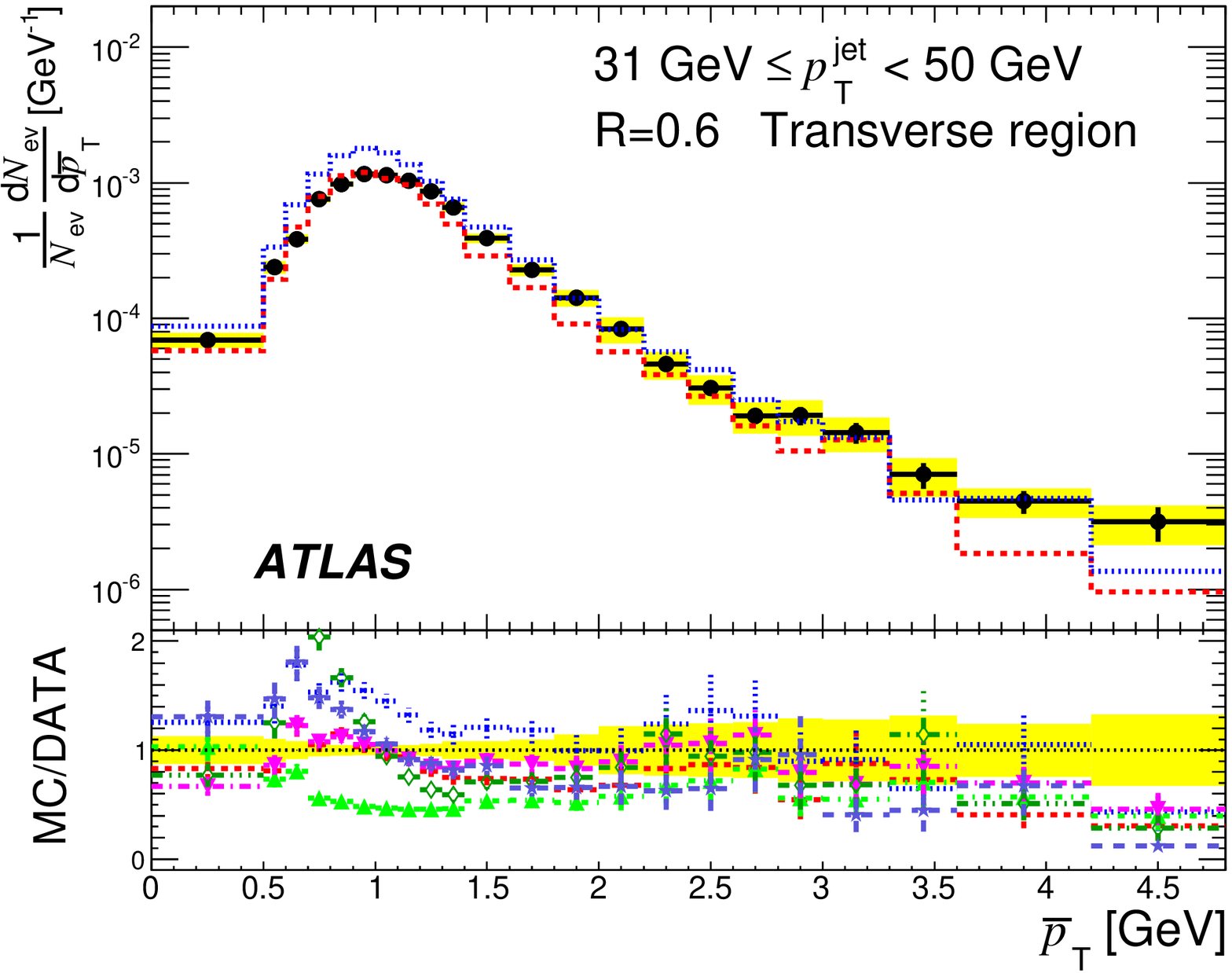}}
\subfigure[]{\includegraphics[width=0.42\textwidth]{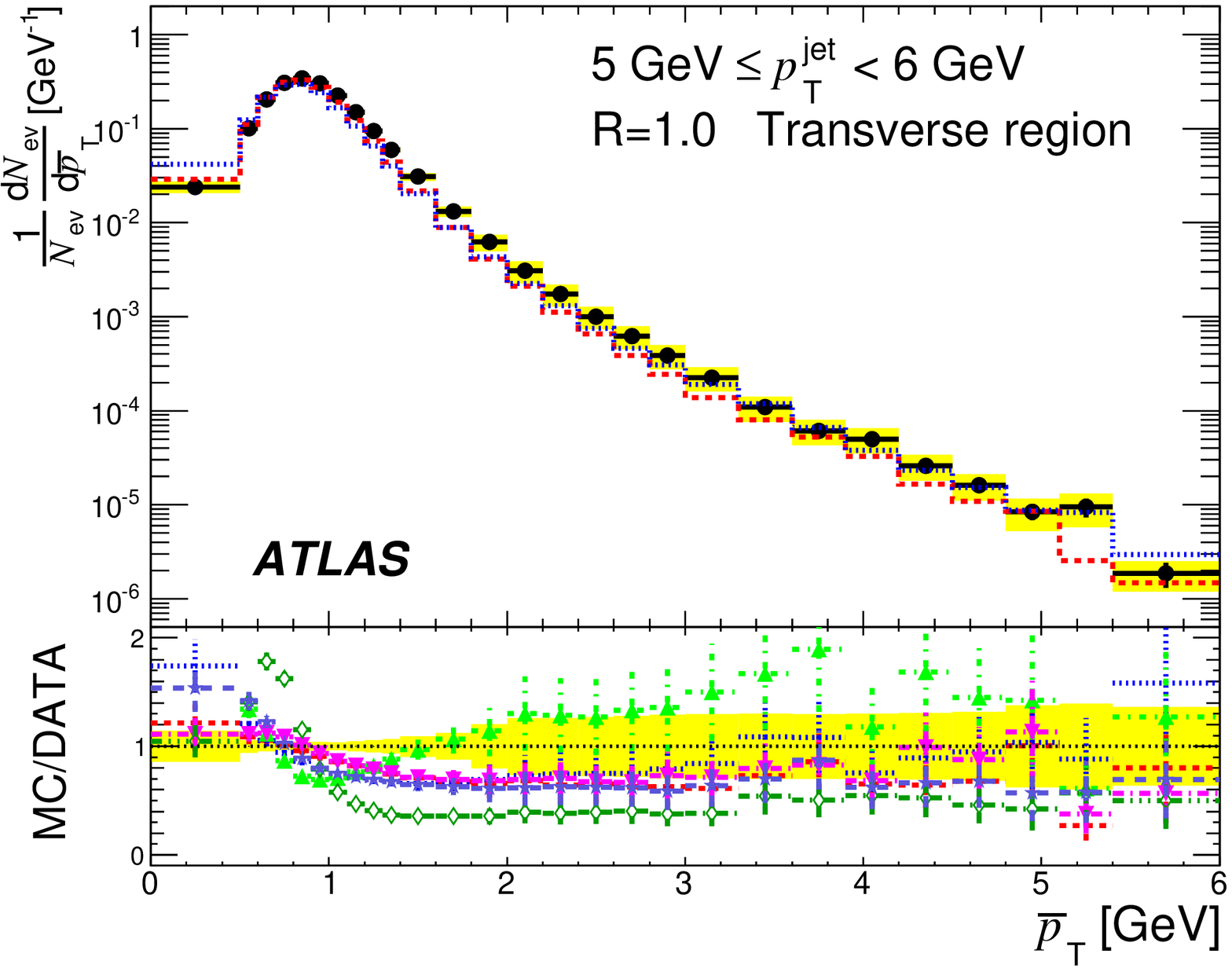}}
\subfigure[]{\includegraphics[width=0.42\textwidth]{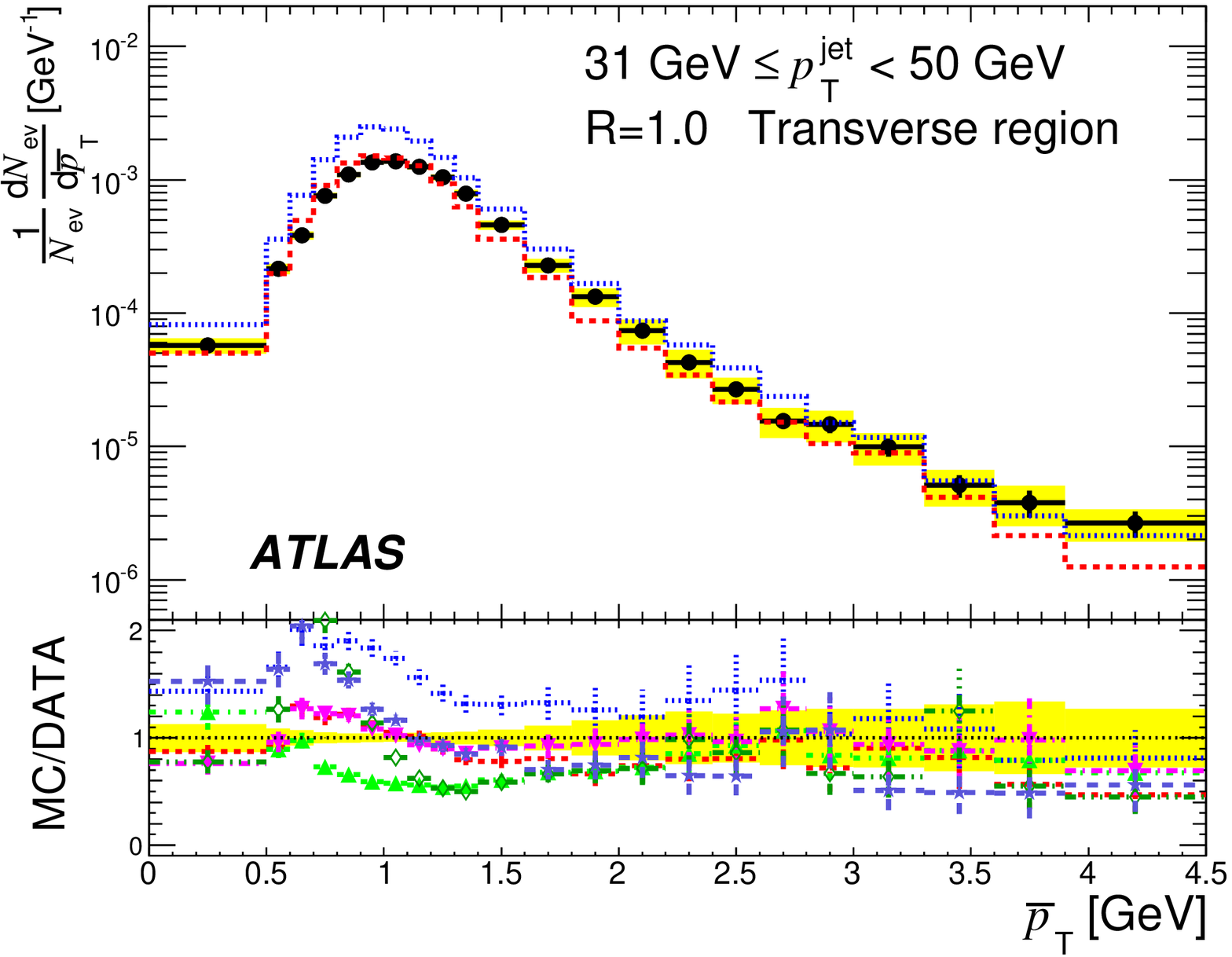}}
\caption{Distributions of \ptave\ in the 
transverse region for
(a) $R=0.2$ and $5 \; \GeV \le \ptJet < 6 \;\GeV$,
(b) $R=0.2$ and $31 \; \GeV \le \ptJet < 50 \;\GeV$,
(c) $R=0.6$ and $5 \; \GeV \le \ptJet < 6 \;\GeV$, 
(d) $R=0.6$ and $31 \; \GeV \le \ptJet < 50 \;\GeV$ 
(e) $R=1.0$ and $5 \; \GeV \le \ptJet < 6 \;\GeV$ and 
(f) $R=1.0$ and $31 \; \GeV \le \ptJet < 50 \;\GeV$.
The shaded band shows the combined statistical and systematic 
uncertainty on the data.
The histograms show the predictions of several MC models with the same legend as in Fig. \subref{fig:ptavebinsa} and also Figs. [\ref{fig:sumptbinstrans}, \ref{fig:ntrksbinstrans}, \ref{fig:ptavebinstrans}].
}
\label{fig:ptavebins}
\end{figure*}

\section{Measurements of UE Distributions}
\label{sec:results}
The dependence on \ptJet\ of the mean values of the unfolded \sumpt, \ntrks\ and \ptave\ distributions
is shown in Figures~\ref{fig:sumptbinstrans}, \ref{fig:ntrksbinstrans} and \ref{fig:ptavebinstrans}, respectively.
The dependence is shown for all five values of $R$ for the transverse region.
To facilitate comparisons with previous measurements, these mean
values are reported as densities per unit $\eta$-$\phi$.
Therefore, the mean values of \sumpt\ and \ntrks\ are divided by $ 2 \pi = \Delta \phi \times \Delta \eta$,
where $ \Delta \phi = 2 \pi / 3 $ ($\pi / 3 \le |\Delta \phi| \le 2 \pi / 3 $)
and $ \Delta \eta = 3 $ ($-\jetEtaMax \le \eta \le \jetEtaMax$ contributes $ \Delta \eta = 3 $).
The qualitative behavior of the distributions is the same for all five $R$ values.  
The mean values of \sumpt\ rise rapidly with \ptJet\ for low \ptJet\ and continue to rise slowly for high \ptJet.
The mean values of \ptave\ exhibit qualitatively similar behavior as those of $ \sumpt $.
The mean value of \ntrks\ rises rapidly with \ptJet\
for low \ptJet\ and approaches a plateau for high \ptJet.  
The systematic differences in the measurements as $R$ is varied are discussed in
Section~\ref{sec:rdependence}.

The unfolded data are compared to several MC generators and tunes.  In general,
the level of agreement between the data and MC samples is reasonable,
with differences below 20\%\ for all observables and all \ptJet\ bins.
The \PythiaS~Z1 sample reproduces the mean values of \sumpt\ within uncertainites for all \ptJet\ bins.
This MC sample tends to slightly overestimate 
$\ntrks$ for $\ptJet \simge 15~\gev$,
and this manifests itself as a slight underestimation of 
$\ptave$ in the same \ptJet\ range.
The \PythiaS~AUET2B sample tends to slightly underestimate 
\sumpt\ for $\ptJet \simle 20~\gev$,
and overestimate 
\sumpt\ at higher $\ptJet$.
\PythiaS~AUET2B reproduces 
\ntrks\ for $\ptJet \simle 15~\gev$,
and overestimates 
\ntrks for $\ptJet \simge 15~\gev$.
This MC sample slightly underestimates the mean values of \ptave\ in all \ptJet bins.
\PythiaS~Perugia2011 exhibits reasonable agreement with the data \sumpt\ distributions,
having tendencies to underestimate the data at low $ \ptJet $ and overestimate it for $ \ptJet \simge 15~\gev $.
This MC sample tends to overestimate \ntrks\ and slightly underestimate $ \ptave $.
The other tunes show somewhat worse agreement for all distributions.
\PythiaE\ tune 4C underestimates all three observables, although the agreement is better for $ \ptave $.
\Herwigpp\ tune UE7-2 underestimates 
the \sumpt\ and \ntrks. 
The \PythiaS\ Perugia 2011 NOCR tune is disfavored by the \ptave\ data.

Figures~\ref{fig:sumptbins} through~\ref{fig:ptavebins} show the 
unfolded distributions 
$(1/\Nev) \mathrm{d}\Nev/\mathrm{d}\sumpt$, $(1/\Nev) \mathrm{d}\Nev/\mathrm{d}\ntrks$ and $(1/\Nev) \mathrm{d}\Nev/\mathrm{d}\ptave$ 
in the transverse region
for three representative values of $R$ and for low (5--6~GeV) and high
(31--50~GeV) bins of \ptJet.    Here \Nev\ is the 
number of events in the sample
with $\ptJet > 4$~GeV and $|\etaJet |<1.5$. 
The \PythiaS\ Z1 tune shows differences with respect
to the data of less than $20\%$--$30\%$ for most
distributions and \ptJet\ bins.  
The \PythiaS\ AUET2B shows discrepancies of up to a factor of two for
the \sumpt\ and \ntrks\ distributions at large values of $ \ptJet $. 
For $\ptJet \approx 5~\gev $, the Perugia 2011 tune undershoots the data for $R=0.2$ by $30\% - 75\%$ for $\ntrks~\gtap~20$ and by $20\% - 80\%$ for $\sumpt~\gtap~20~\gev$.
The differences are smaller for $\ptJet \approx 30~\gev$ and for large $R$ values
\Herwigpp\ shows an excess of events at low \sumpt\ and low \ntrks\ for small \ptJet;  at
high \ptJet\ it does a reasonable job of predicting the shape of the distributions but
understimates the normalization of \sumpt\ and \ntrks\ distributions by about 50\% .

The measured distributions in the away region, together with MC comparisons, are shown
in Appendix~\ref{sec:appendix}. 
The \sumpt, \ntrks\ and \ptave\ all continue to rise with $ \ptJet $, as expected.  
This behavior indicates the presence of a second jet recoiling against the leading jet.

\section{Dependence of the UE on $R$}
\label{sec:rdependence}
\begin{figure*}[p]
\begin{center}
\subfigure[]{\includegraphics[width=0.46\textwidth]{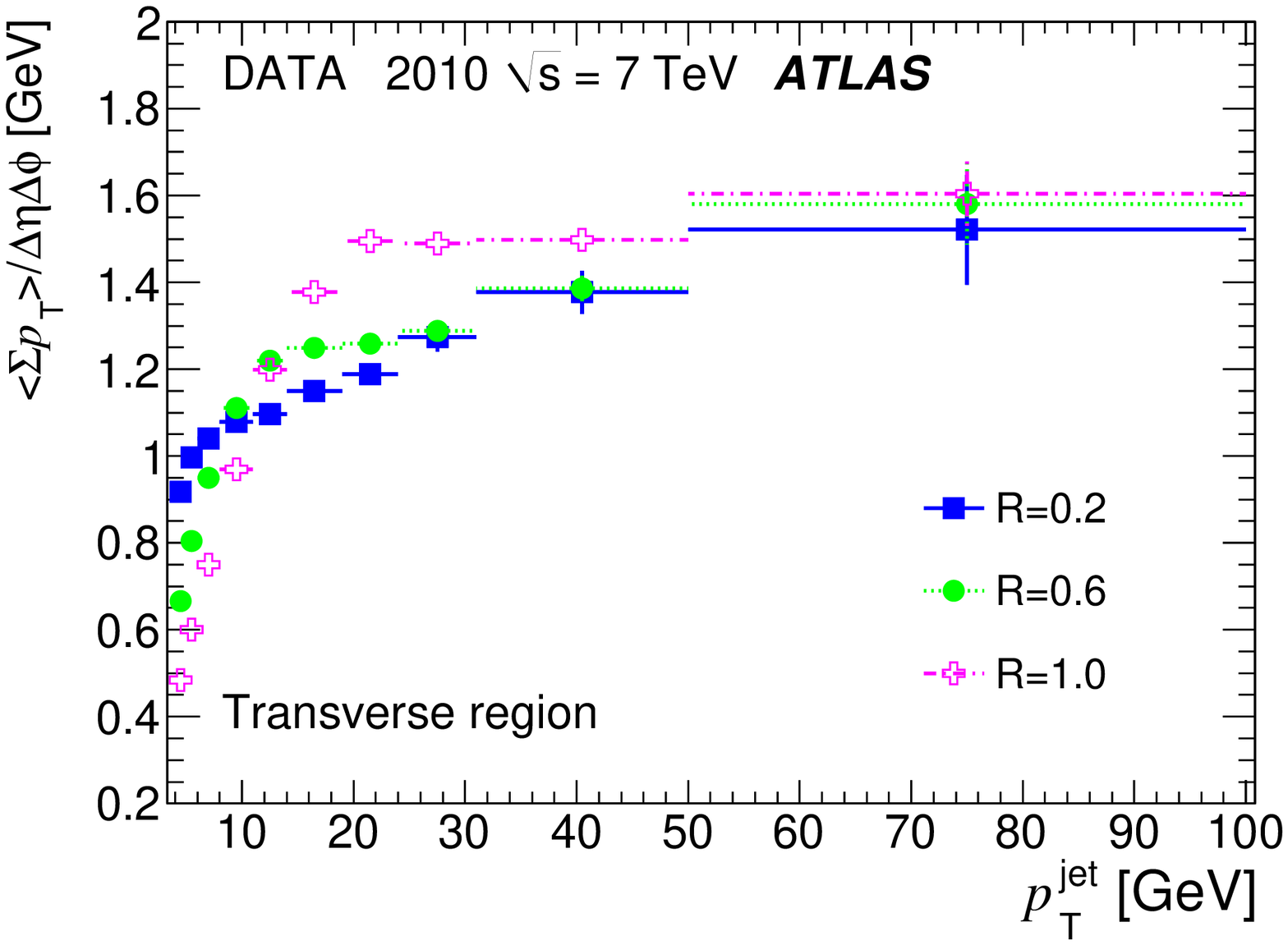}}
\subfigure[]{\includegraphics[width=0.46\textwidth]{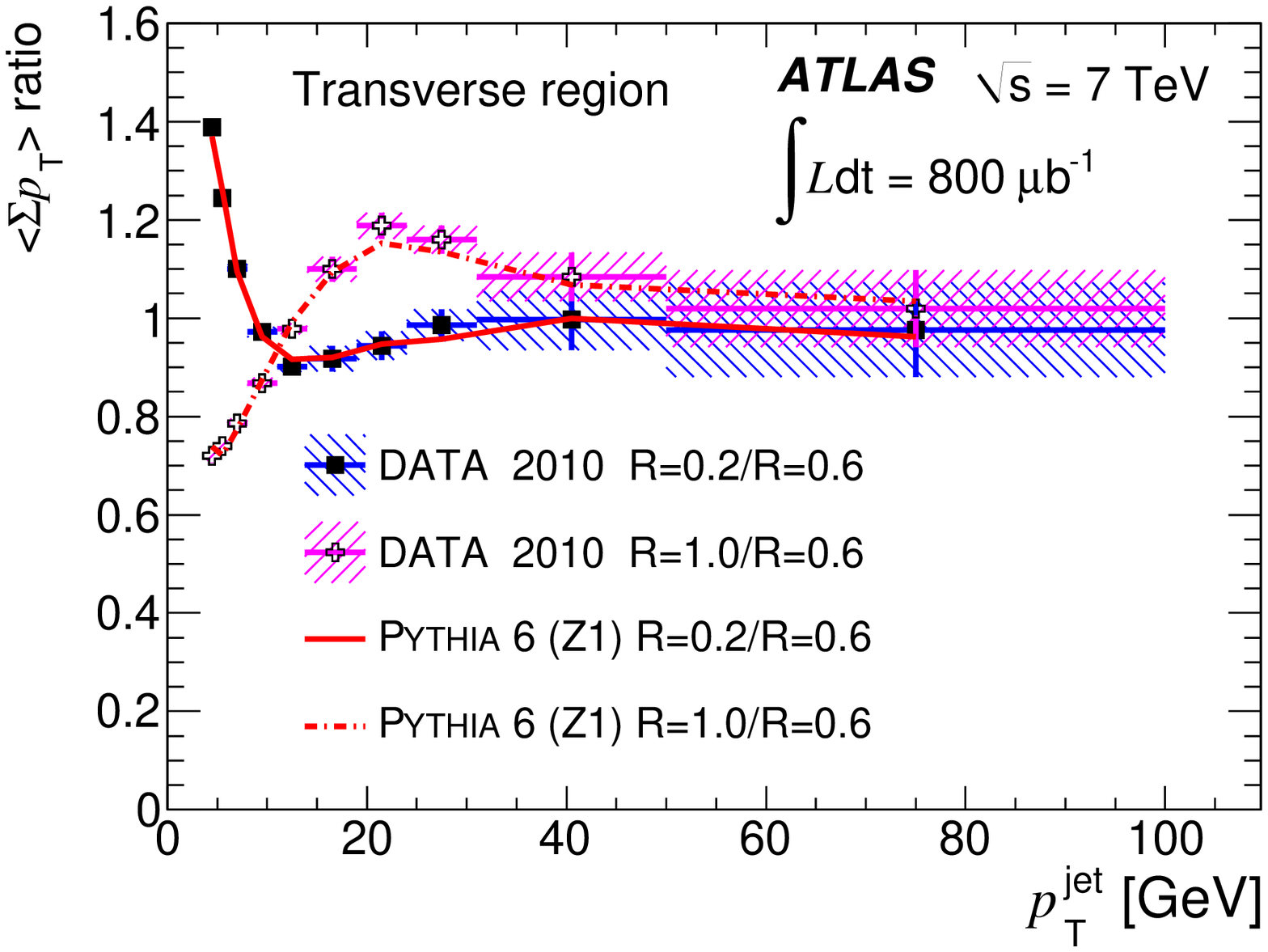}}
\subfigure[]{\includegraphics[width=0.46\textwidth]{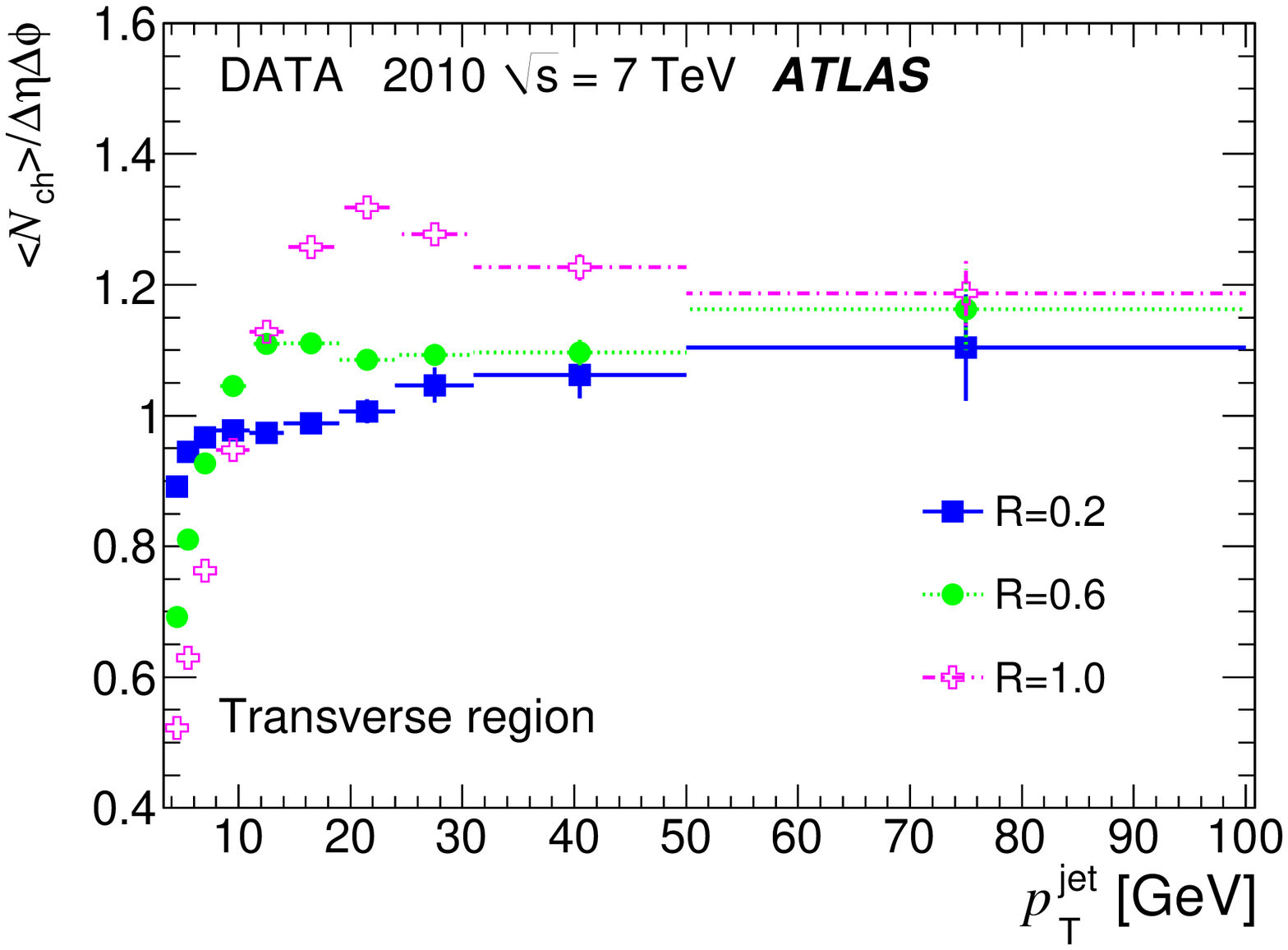}}
\subfigure[]{\includegraphics[width=0.46\textwidth]{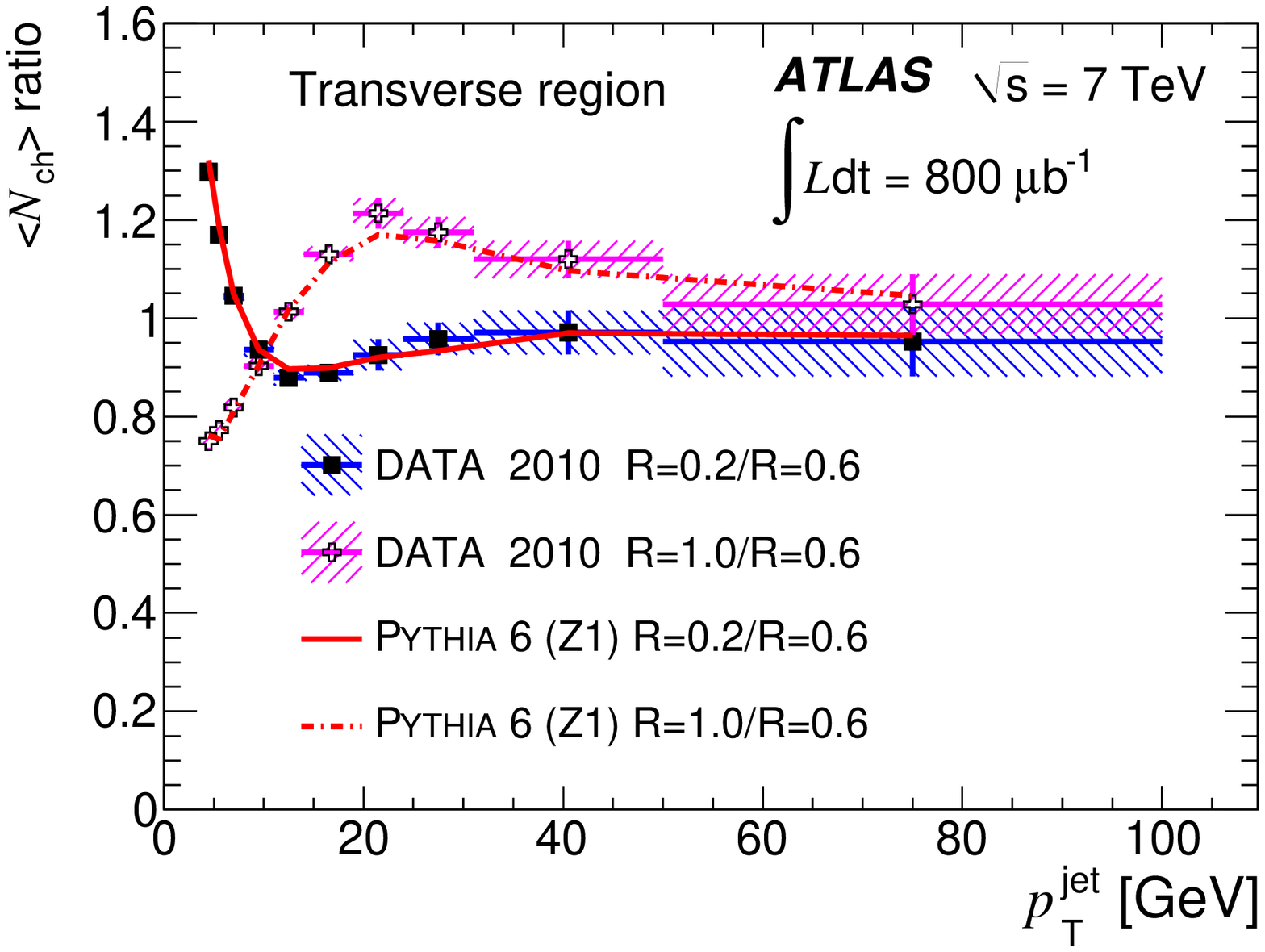}}
\subfigure[]{\includegraphics[width=0.46\textwidth]{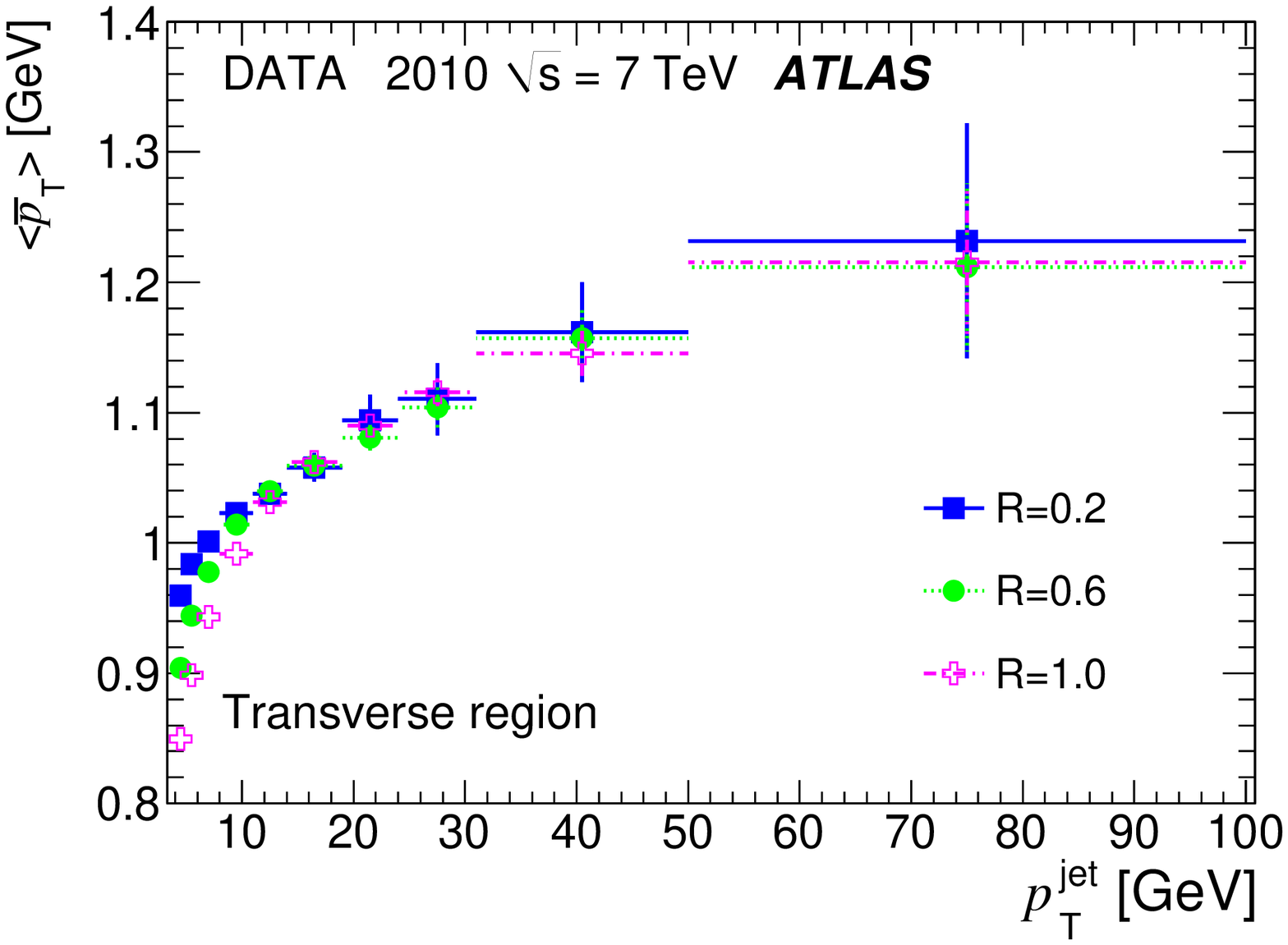}}
\subfigure[]{\includegraphics[width=0.46\textwidth]{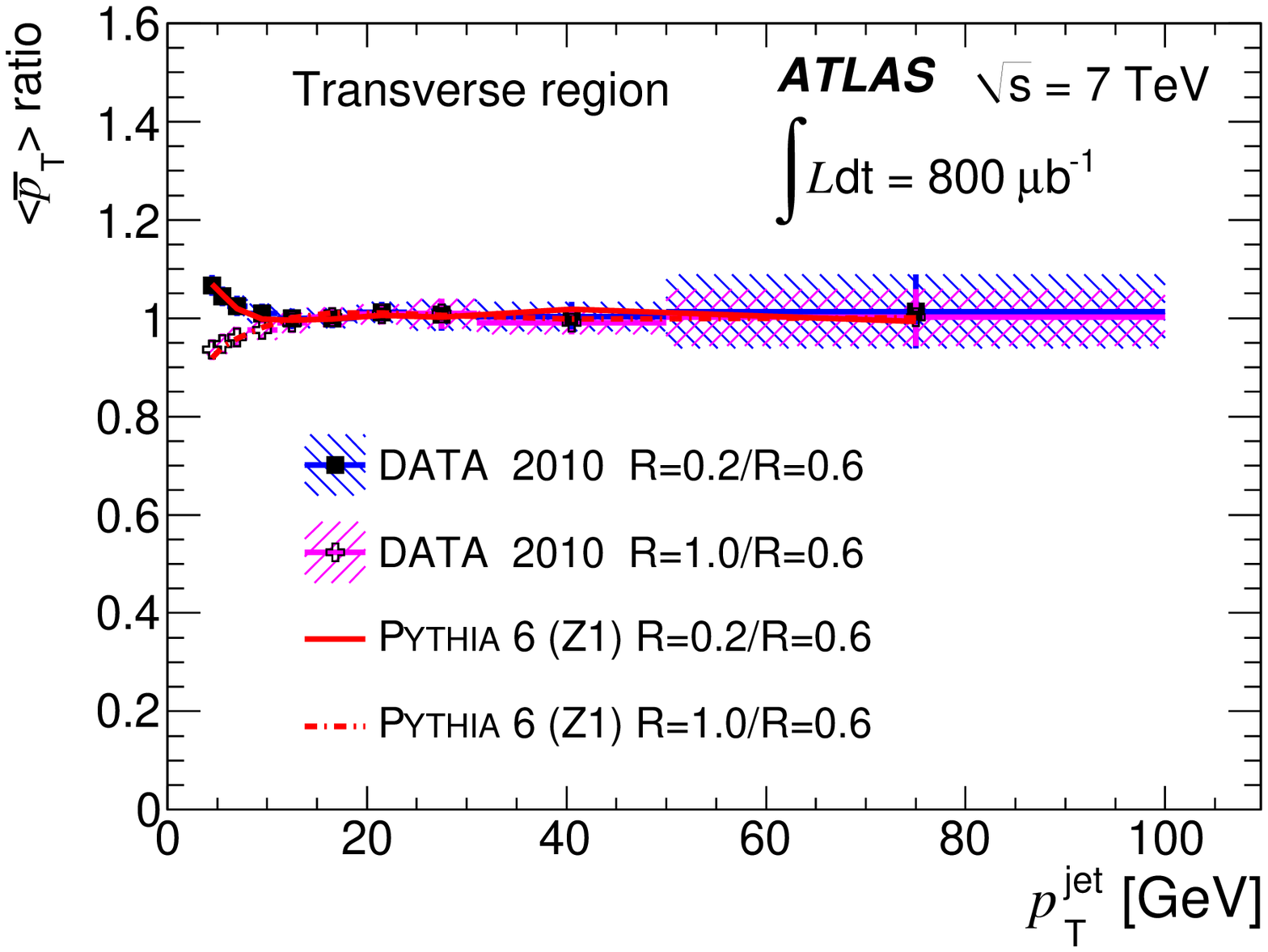}}
\end{center}
\vfil
\caption{The mean value of (a)~\sumpt (per unit $ \eta - \phi$), (c)~\ntrks\ (per unit $ \eta - \phi $) and (e)~\ptave\  
in the transverse region as a function of \ptJet\ for 
$R=0.2$, $R=0.6$ and $R=1.0$; because the systematic uncertainties are
correlated among the different $R$ values, only statistical
uncertainties are shown.  And
the ratio of the mean value of (b)~\sumpt, (d)~\ntrks\ and (f)~\ptave\
in the transverse region
measured for $R=0.2$ and $R=1.0$ to that measured for $R=0.6$.  The
shaded bands show the total uncertainty.
The lines show the predictions of \PythiaS\ with the Z1 tune; predictions
for AUET2B show comparable agreement.
}
\label{fig:ratios}
\end{figure*}

The distributions of the UE observables
change with the value of $R$ used in the jet reconstruction.  The
variations are summarized in Figure~\ref{fig:ratios}, which
compares the results obtained for the observables in the transverse
region for
$R=0.2$, 0.6 and 1.0.
For low \ptJet, the mean values of the \sumpt\ and \ntrks\ densities
are largest for the smallest value of $R$, while  
they are largest for the
highest value of $R$ at high \ptJet.
In contrast, the mean value of \ptave\ in the transverse region
shows little variation with $R$.  

As noted in Section~\ref{sec:introduction}, the phenomenological 
description of jet production and the UE is complex.  
This is especially true in the low \ptJet\ region where 
the distinction between the hard scattering process and the soft physics 
associated with the beam remnants is an artifact of the model used
to parameterize this physics.  Nevertheless, 
several general features of jet production are useful for 
interpreting the observed $R$ dependence of the UE observables. 
For example, the inclusive jet cross section depends on $ R $.
Also, leading jets obtained with different $ R $ parameters 
will not in general have the same reconstructed centroid position,
resulting in differences in the definitions of the transverse and away regions.
Furthermore, the amount of transverse momentum collected in a jet increases with the value of $R$ used for the reconstruction.

Measurements of the inclusive cross section for jets reconstructed
from tracks show that this cross section increases significantly as
the radius parameter $R$ is increased~\cite{PhysRevD.84.054001}.
Both the Z1 and AUET2B tunes of \Pythia~6 do a remarkably
good job of reproducing the measurements therein.
The number of jets reconstructed in the data depends strongly
on the value of $R$ used in the \antikt\ algorithm. For the
events passing the trigger and primary vertex
requirements described in Section~\ref{sec:reconstruction}, 
the fraction of events with at least one 
reconstructed jet with $\ptJet > 4\ \GeV $ and
$|\eta|<1.5$ varies from $4.4\%$ to $19.8\%$
as $R$ is increased from 0.2 to 1.0.  
These additional jets reconstructed for large $R$
largely populate the region of low \ptJet.  
The results shown in Figure~\ref{fig:ratios} indicate that including
these jets reduces the average hadronic activity for $\ptJet \ltap 10~\gev$, 
i.e. the additional $R$=1.0 jets are found in events with less hadronic activity.

The difference in azimuthal angle between the leading jet reconstructed
with $R=0.2$ and the leading jet reconstructed with $R=1.0$ has been studied as a function of $\ptJet$.
These studies demonstrate that event reorientation effects 
due to changes in the reconstructed position of the leading jet 
for different values of $R$ are small.
For $ \ptJet \simeq 4~\gev $, 7\% of events are reoriented by $ \pi / 3 \le | \Delta \phi | \le 2 \pi / 3 $.
For $ \ptJet \simge 15~\gev $ the effect is much smaller; less than 1\% of events are reoriented.

For cases where jets reconstructed with different $R$ parameters
are matched, the \pt\ of the jet reconstructed with larger
$R$ will exceed that of the jet reconstructed with smaller $R$,
leading to migration of events to bins with higher \ptJet.  
Many physical processes influence
the amount of migration as the radius parameter increases.  These effects include
the collection of additional fragmentation particles, the inclusion of
additional hadrons produced via soft gluon radiation from the final-state 
parton, and the sweeping of particles from the UE into the jet cone.
Attempts to compensate for the observed $R$-dependence by correcting \ptJet\
using the average UE energy deposited within the jet cone were
unsuccessful. This might be due to the fact that there is a
correlation between the amount of \pt\ migration and the level of
UE activity in the event because UE activity exhibits long-range
correlations in $\eta$-$\phi$.
Events with higher UE activity will exhibit
larger \pt\ migration as $R$ is increased.  Thus, the ability of the
MC generators to model the variation of the UE with $R$ depends not only on
how well the generator reproduces the mean UE response, but also on how
well it models the fluctuations in the UE and how correlated these fluctuations are spatially.

\section{Conclusion}
\label{sec:conclusion}
Observables sensitive to UE activity in events containing one or more
charged-particle jets produced in \pp\ collisions at $\sqrt{s}=7$~TeV have
been measured. The properties of the UE activity have been studied both
in the transverse and away regions. The jets are reconstructed with the
\antikt\ algorithm, with a radius parameter $R$ varying between 0.2 and
1.0. Measurements of the evolution of the UE activity with $R$ are also
presented. Predictions from several MC tunes have been compared to
the data. The predictions from \PythiaS\ based on tunes that have been
determined using LHC data, namely Z1, AUET2B and Perugia 2011, show 
reasonable agreement with the data,
not only for the mean event activity but also for fluctuations
in UE activity within events. Other tunes, such as
Perugia 2011 NOCR, are disfavored by the data. The comparison of the
predictions from \PythiaE.145 tune 4C and the \Herwigpp\ 2.5.1 UE7-2 tune
to the data indicates that additional tuning of UE parameters is
necessary in these cases.
The measurements presented here provide a testing ground for
further development of the Monte Carlo models.

\label{ratio-figs}
\begin{acknowledgments}
We thank CERN for the very successful operation of the LHC, as well as the
support staff from our institutions without whom ATLAS could not be
operated efficiently.

We acknowledge the support of ANPCyT, Argentina; YerPhI, Armenia; ARC,
Australia; BMWF, Austria; ANAS, Azerbaijan; SSTC, Belarus; CNPq and FAPESP,
Brazil; NSERC, NRC and CFI, Canada; CERN; CONICYT, Chile; CAS, MOST and NSFC,
China; COLCIENCIAS, Colombia; MSMT CR, MPO CR and VSC CR, Czech Republic;
DNRF, DNSRC and Lundbeck Foundation, Denmark; EPLANET and ERC, European Union;
IN2P3-CNRS, CEA-DSM/IRFU, France; GNAS, Georgia; BMBF, DFG, HGF, MPG and AvH
Foundation, Germany; GSRT, Greece; ISF, MINERVA, GIF, DIP and Benoziyo Center,
Israel; INFN, Italy; MEXT and JSPS, Japan; CNRST, Morocco; FOM and NWO,
Netherlands; RCN, Norway; MNiSW, Poland; GRICES and FCT, Portugal; MERYS
(MECTS), Romania; MES of Russia and ROSATOM, Russian Federation; JINR; MSTD,
Serbia; MSSR, Slovakia; ARRS and MVZT, Slovenia; DST/NRF, South Africa;
MICINN, Spain; SRC and Wallenberg Foundation, Sweden; SER, SNSF and Cantons of
Bern and Geneva, Switzerland; NSC, Taiwan; TAEK, Turkey; STFC, the Royal
Society and Leverhulme Trust, United Kingdom; DOE and NSF, United States of
America.

The crucial computing support from all WLCG partners is acknowledged
gratefully, in particular from CERN and the ATLAS Tier-1 facilities at
TRIUMF (Canada), NDGF (Denmark, Norway, Sweden), CC-IN2P3 (France),
KIT/GridKA (Germany), INFN-CNAF (Italy), NL-T1 (Netherlands), PIC (Spain),
ASGC (Taiwan), RAL (UK) and BNL (USA) and in the Tier-2 facilities
worldwide.
\end{acknowledgments}
\bibliographystyle{apsrev4-1}
\bibliography{TrackJetUEPaper}
\appendix
\section{UE Distributions in the Away Region}
\label{sec:appendix}
\begin{figure*}[p]
\begin{center}
\subfigure[]{\includegraphics[width=0.44\textwidth]{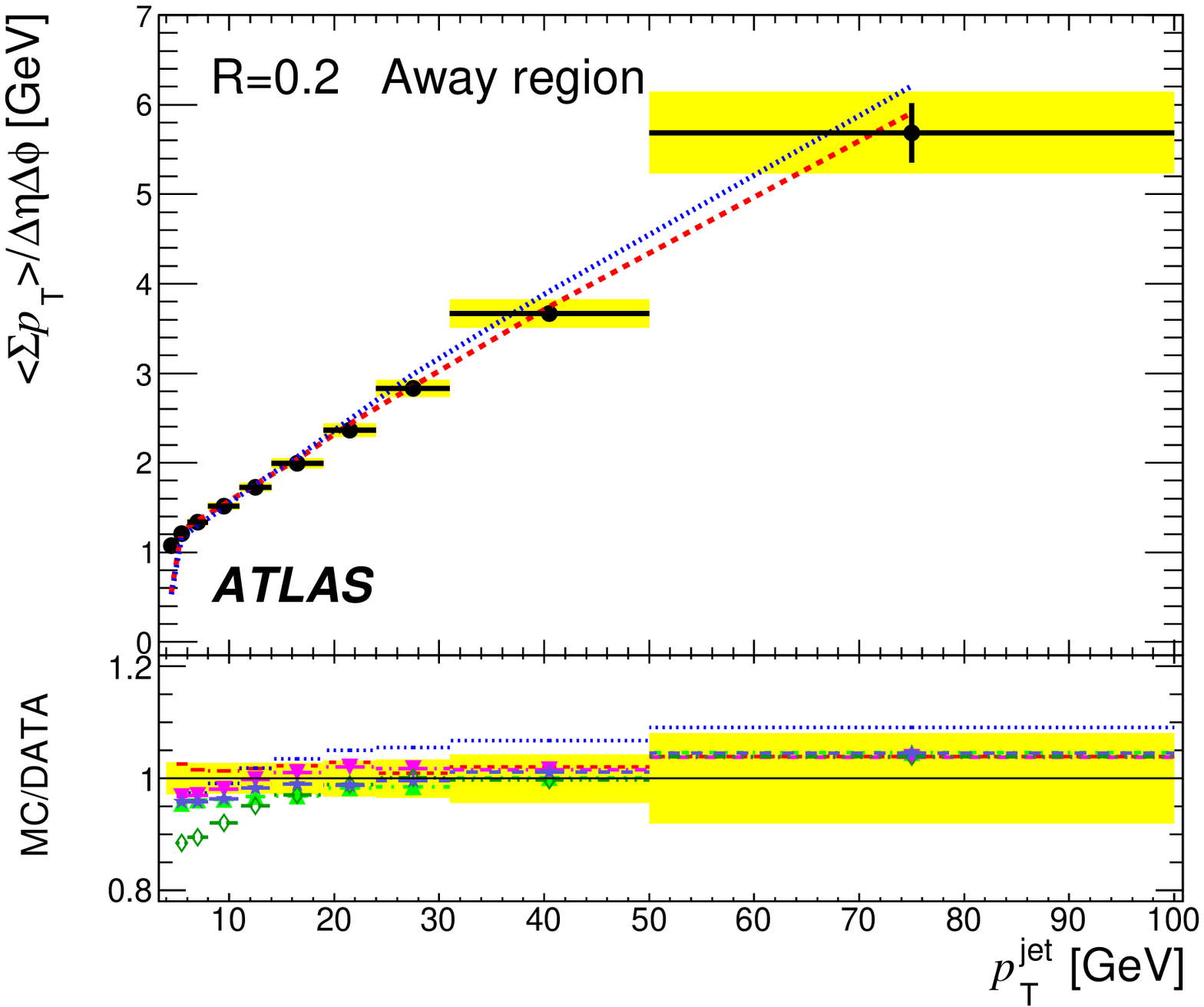}}
\subfigure[]{\includegraphics[width=0.44\textwidth]{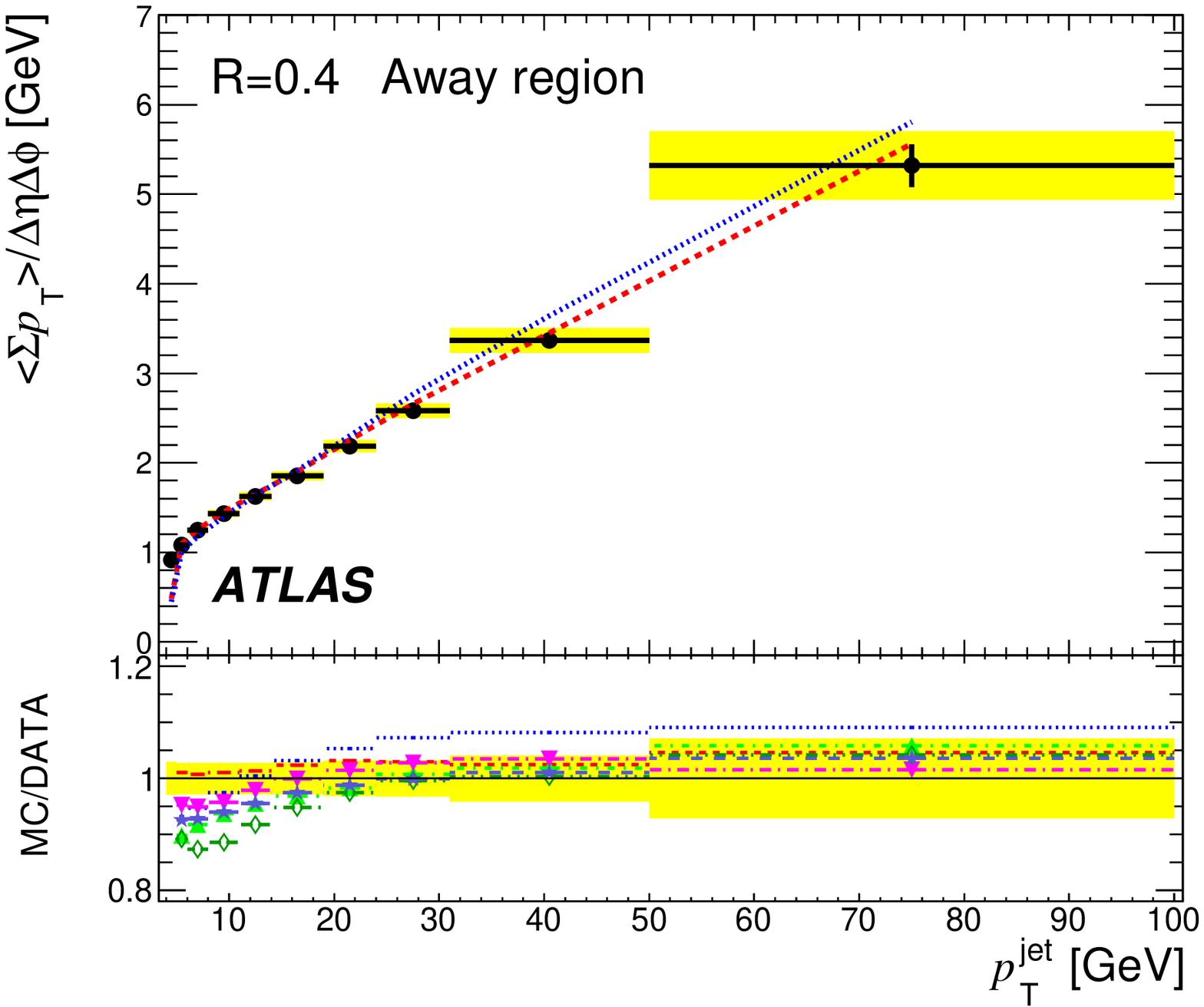}}
\subfigure[]{\includegraphics[width=0.44\textwidth]{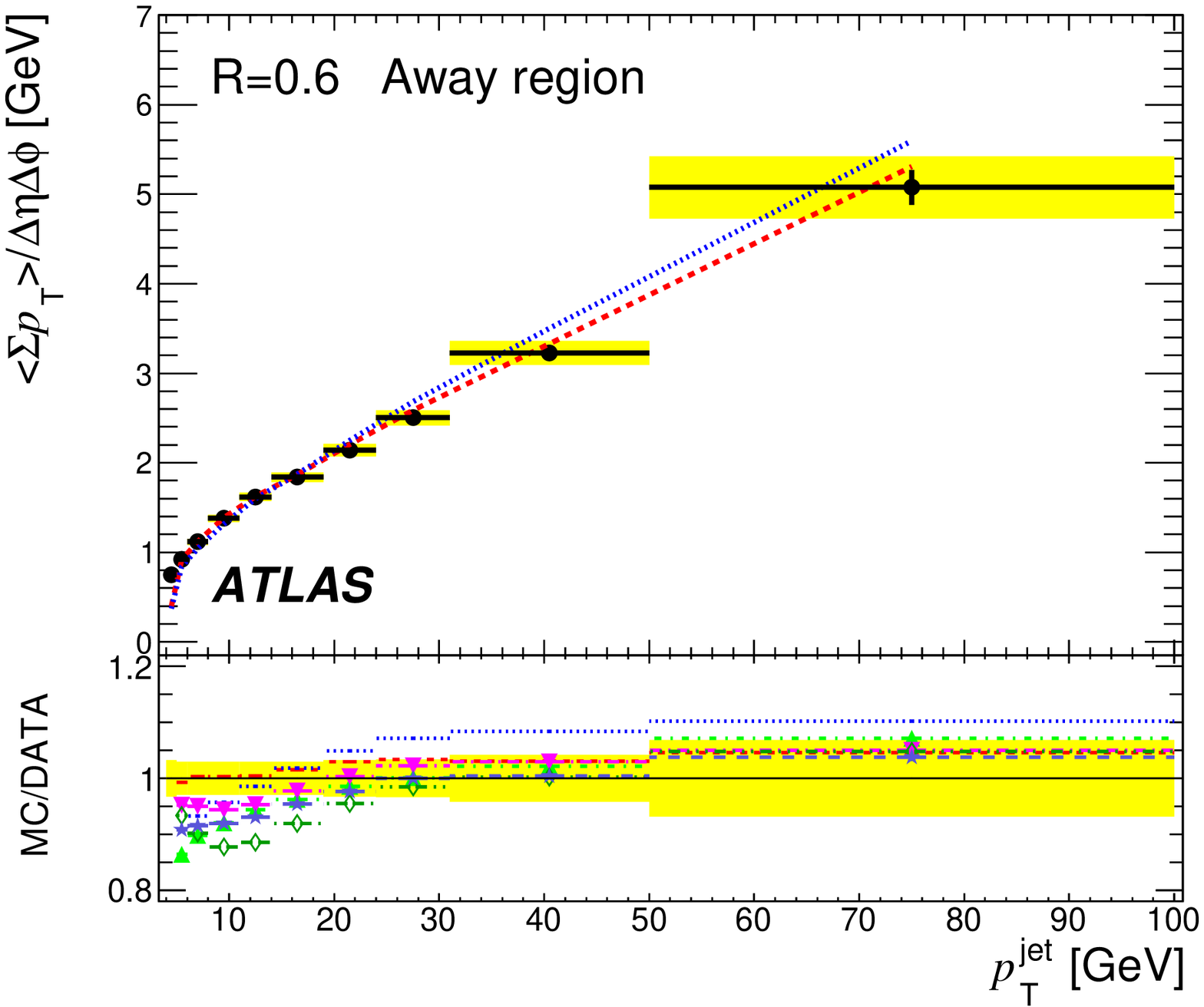}}
\subfigure[]{\includegraphics[width=0.44\textwidth]{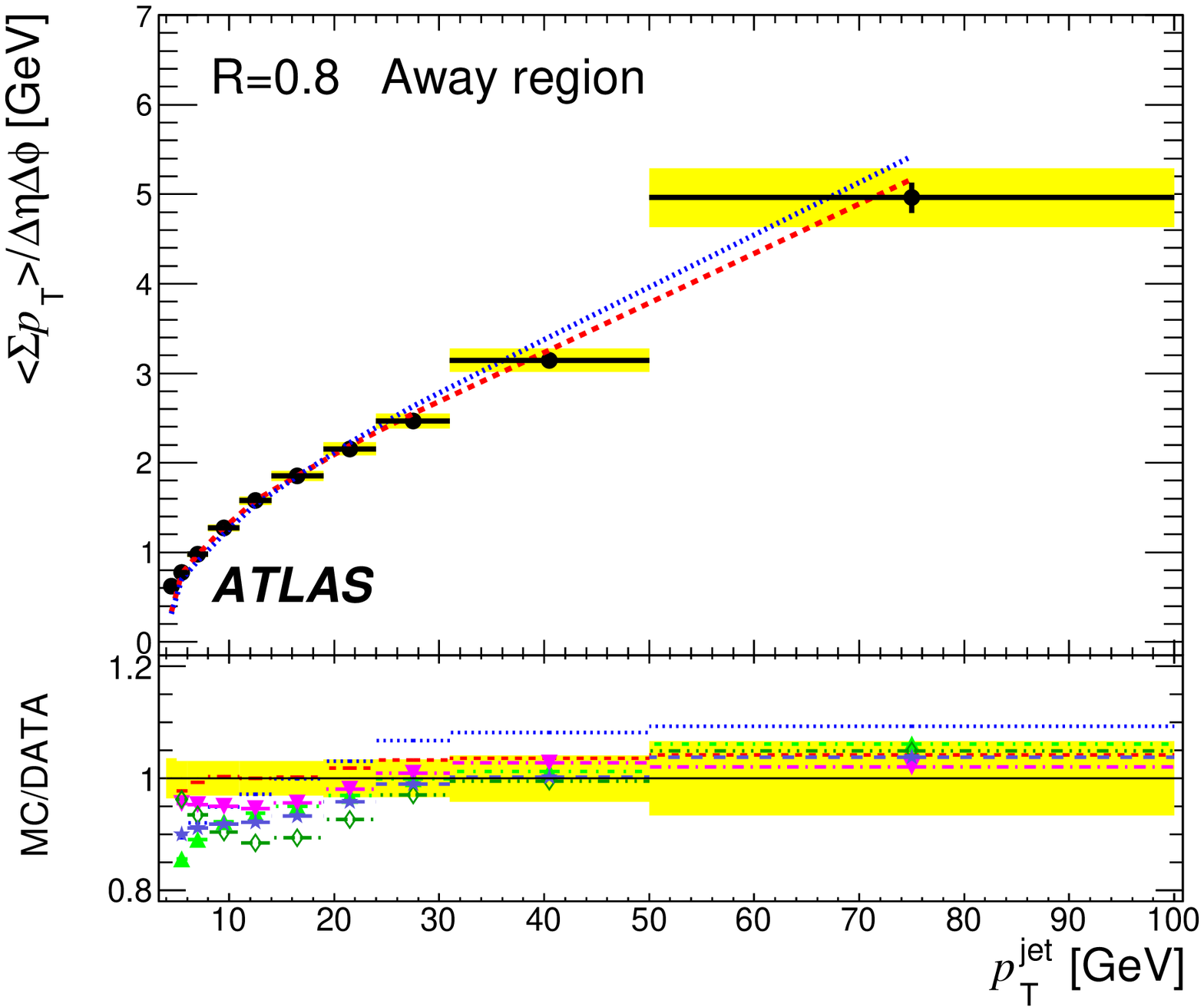}}
\subfigure[]{\includegraphics[width=0.44\textwidth]{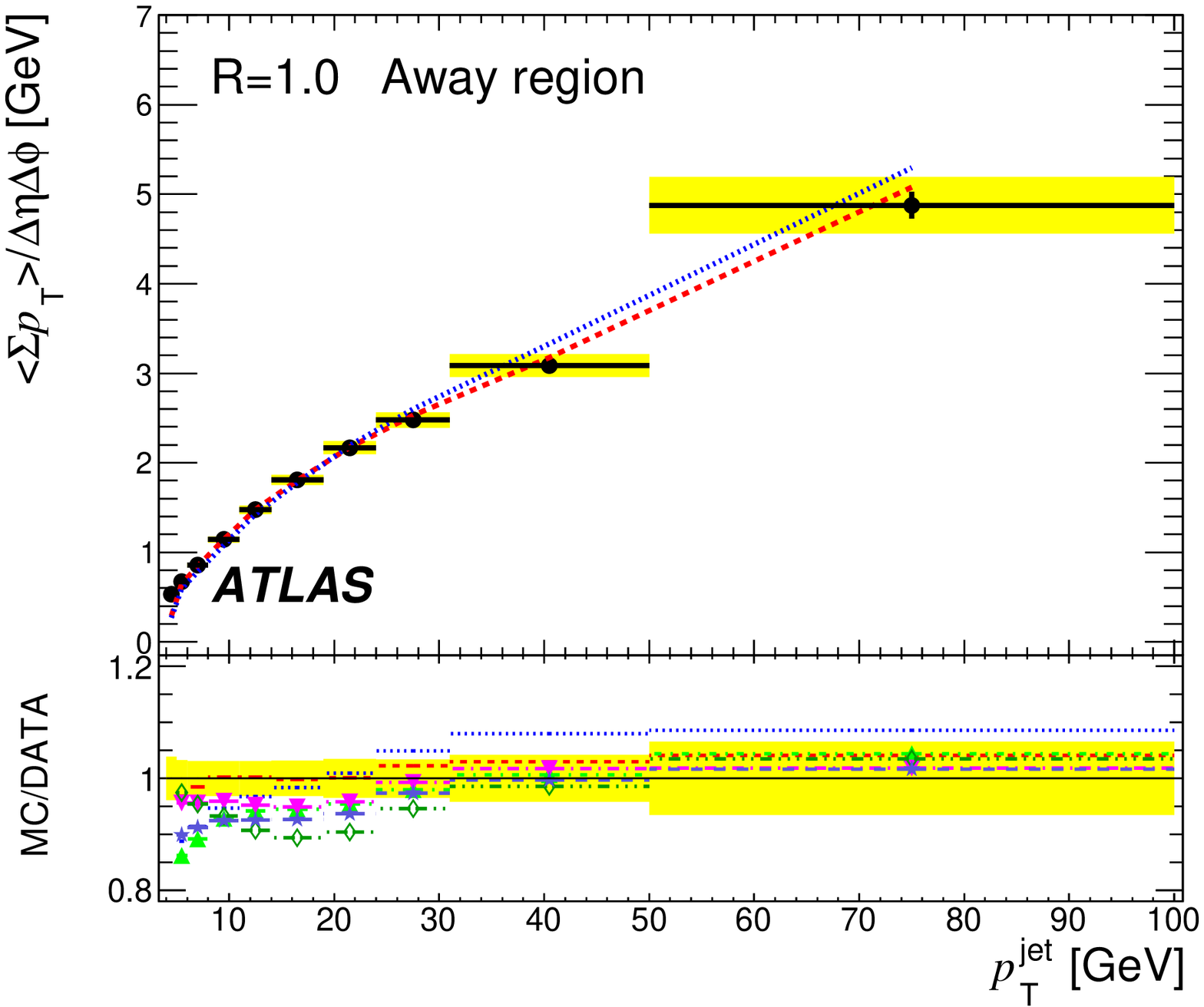}}
{\includegraphics[width=0.44\textwidth]{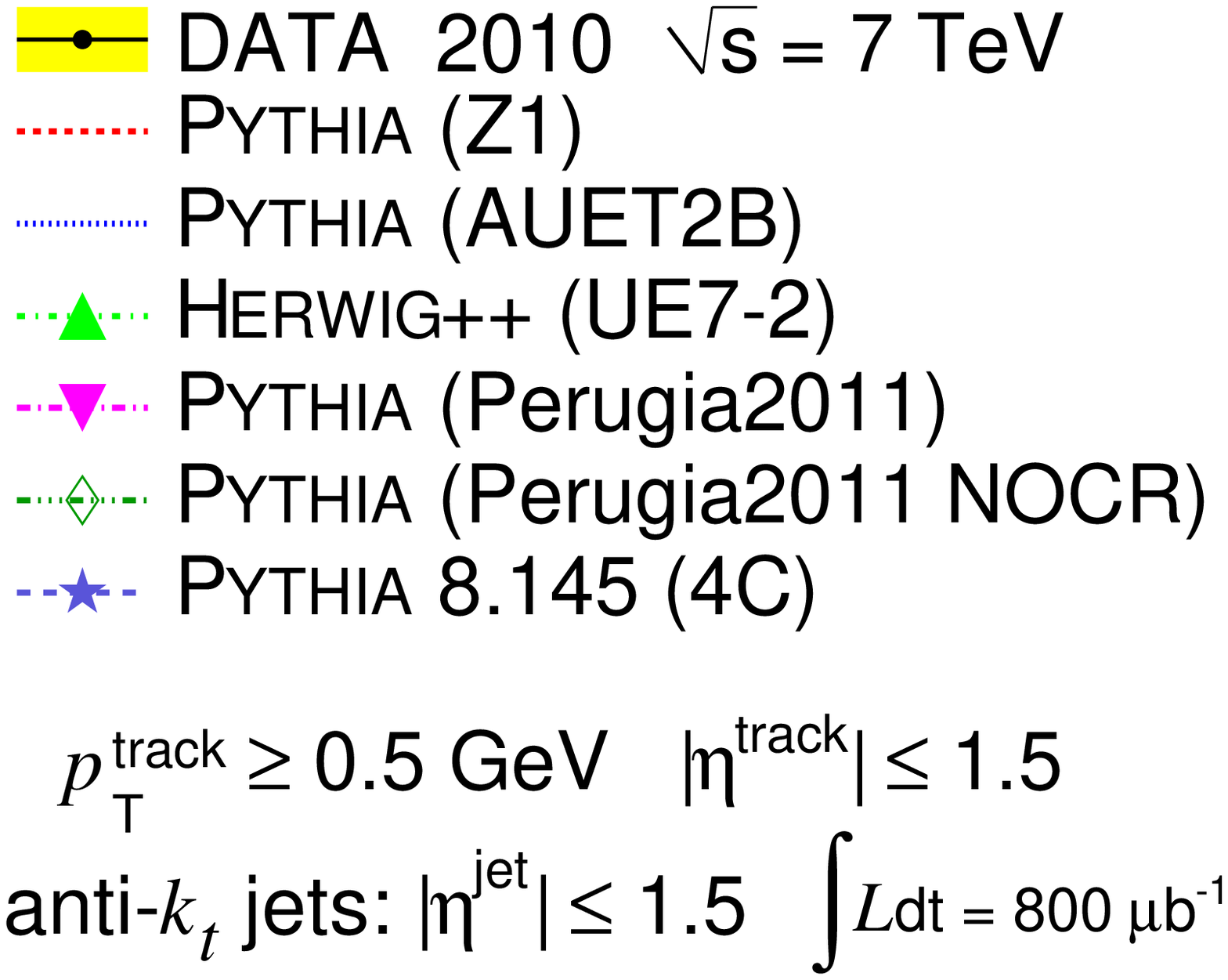}}
\end{center}
\vfil
\caption{The mean value, per unit $ \eta - \phi $, of \sumpt\ in the away region, as a function of
\ptJet\
(a) for $R=0.2$,
(b) for $R=0.4$,
(c) for $R=0.6$,
(d) for $R=0.8$ and
(e) for $R=1.0$.
The shaded band shows the combined statistical and systematic
uncertainty on the data.
The data are compared to the predictions 
obtained with \PythiaS\ using the AUET2B and Z1 tunes.  The bottom
insert shows the ratio of MC predictions to data for several recent MC
tunes.
}
\label{fig:sumptbinsaway}
\end{figure*}

\begin{figure*}[p]
\begin{center}
\subfigure[]{\includegraphics[width=0.44\textwidth]{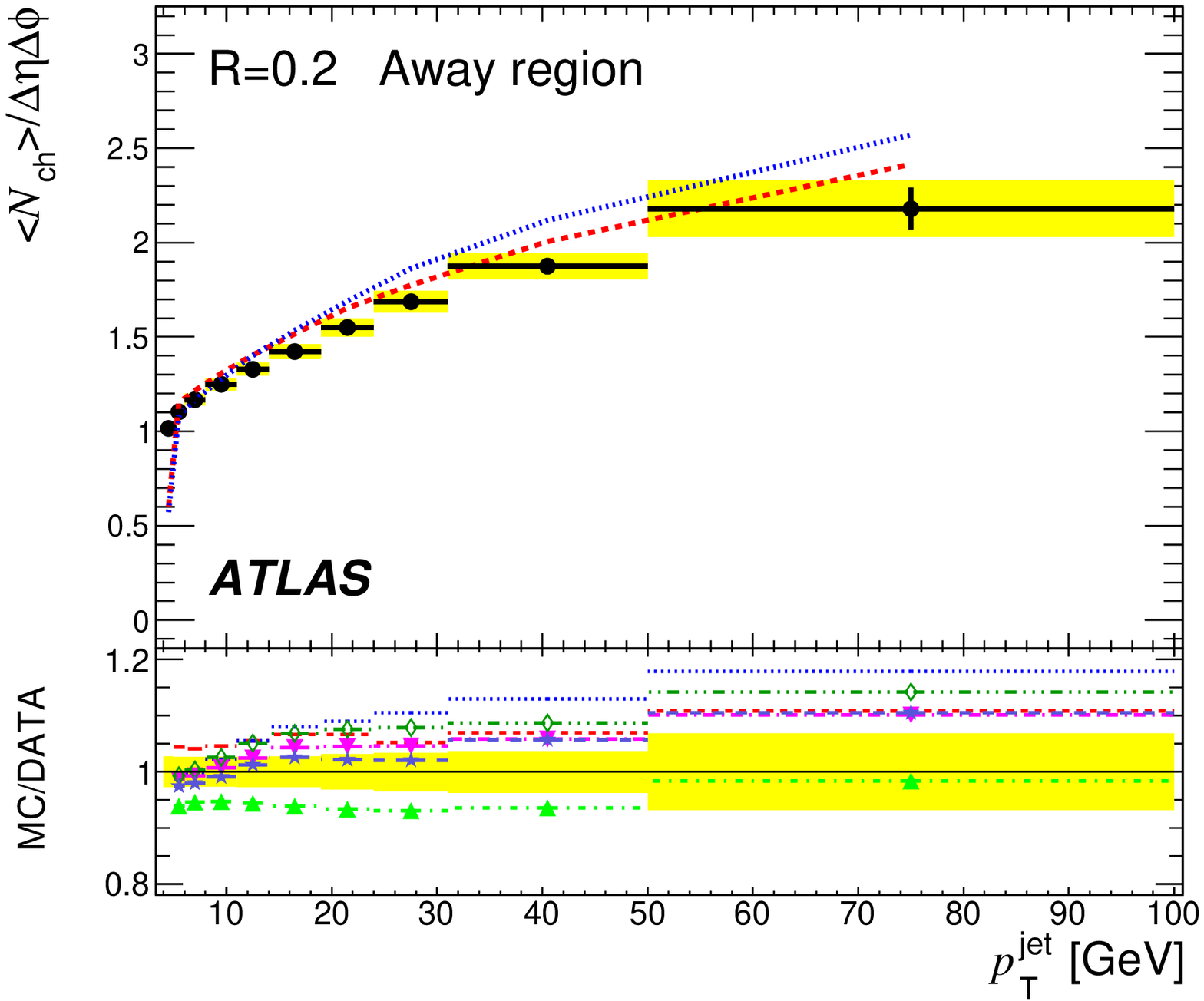}}
\subfigure[]{\includegraphics[width=0.44\textwidth]{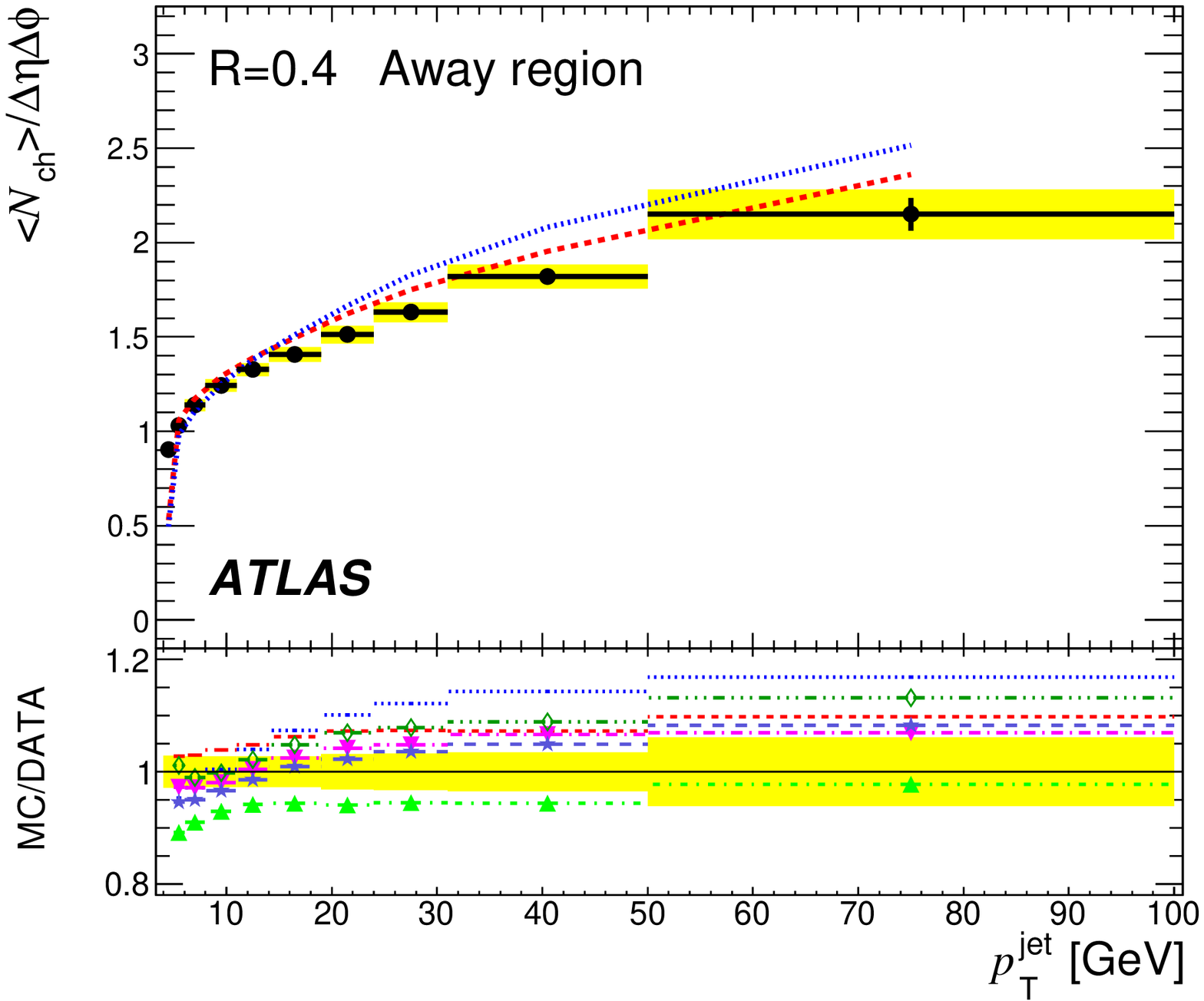}}
\subfigure[]{\includegraphics[width=0.44\textwidth]{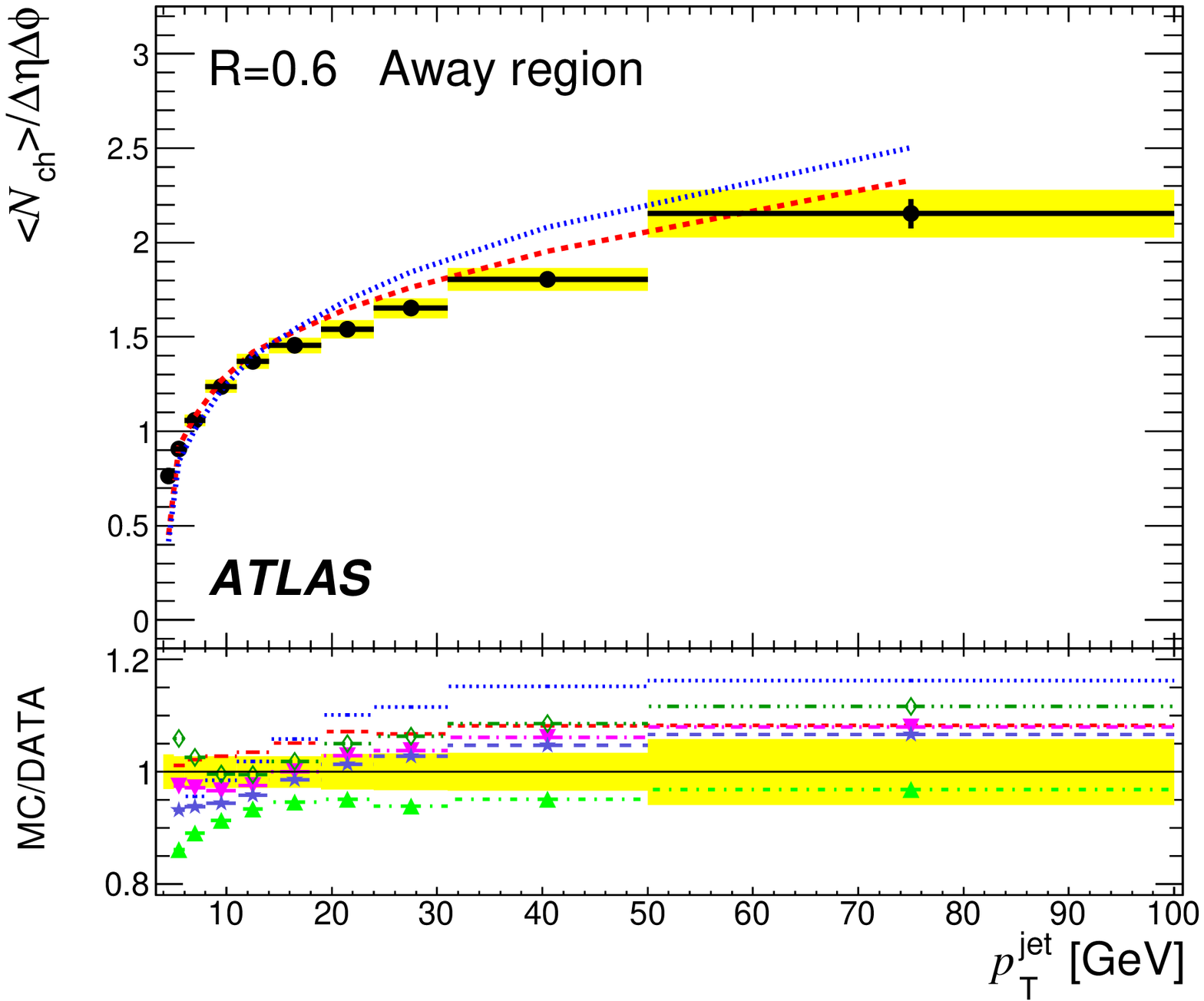}}
\subfigure[]{\includegraphics[width=0.44\textwidth]{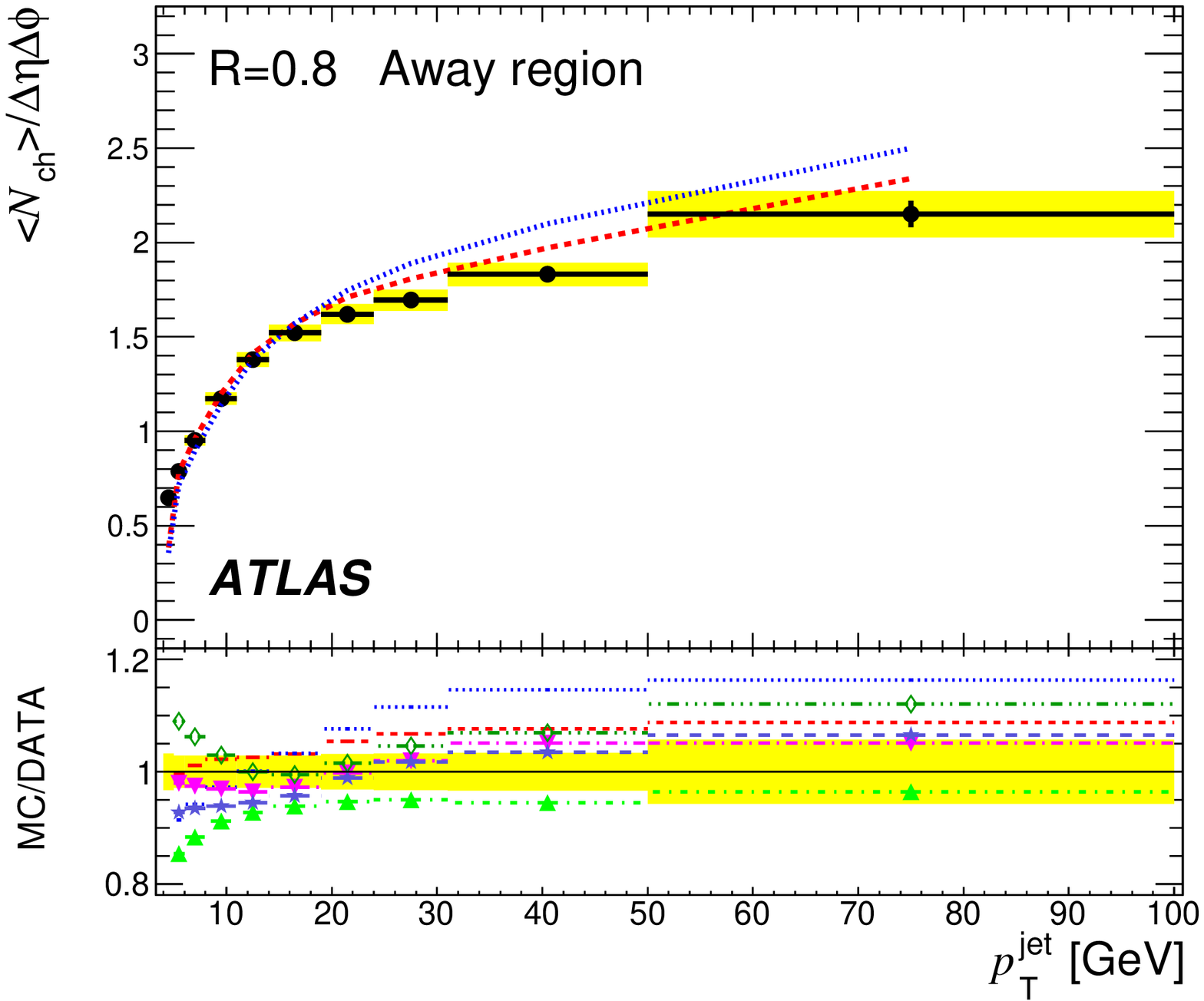}}
\subfigure[]{\includegraphics[width=0.44\textwidth]{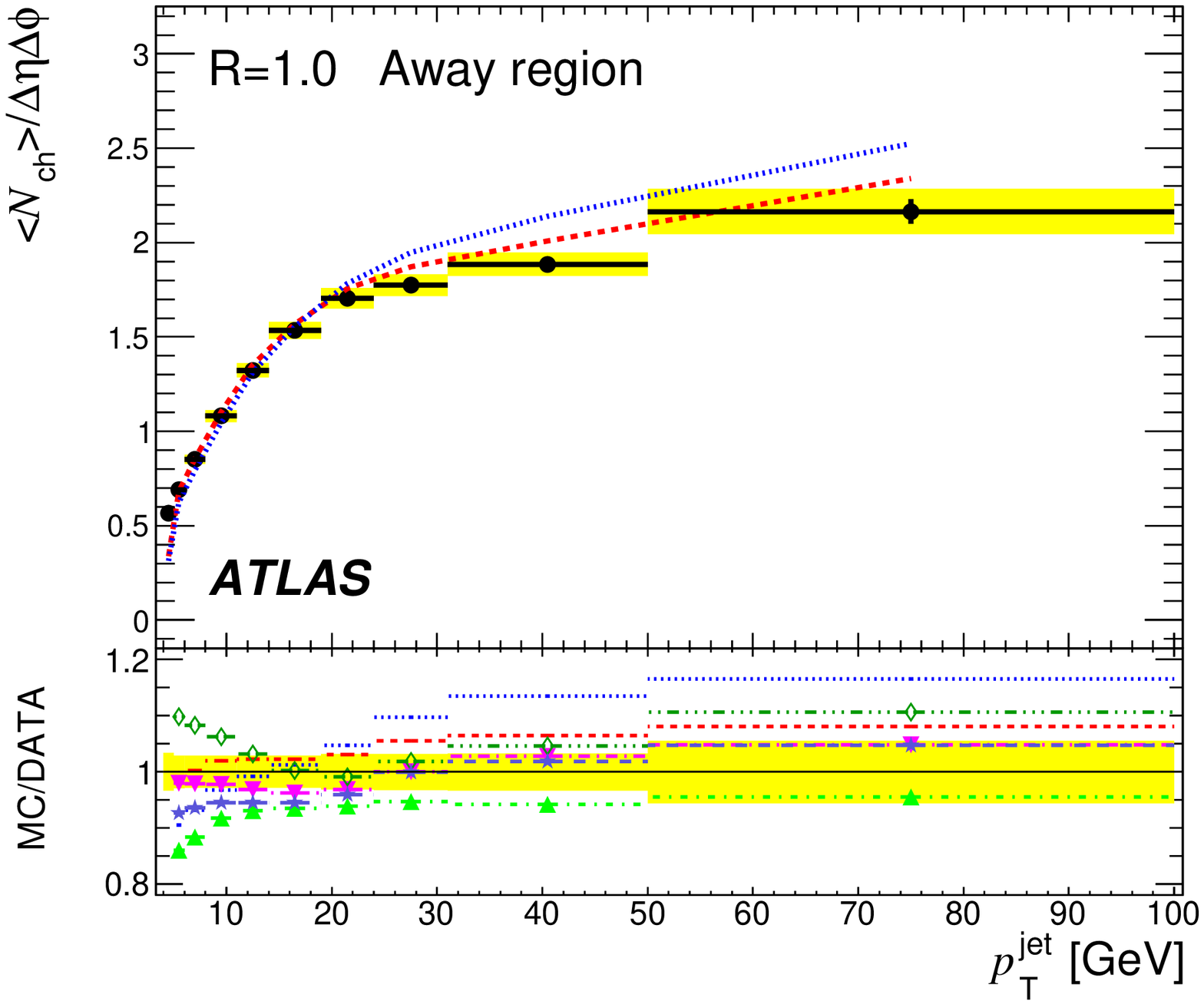}}
{\includegraphics[width=0.44\textwidth]{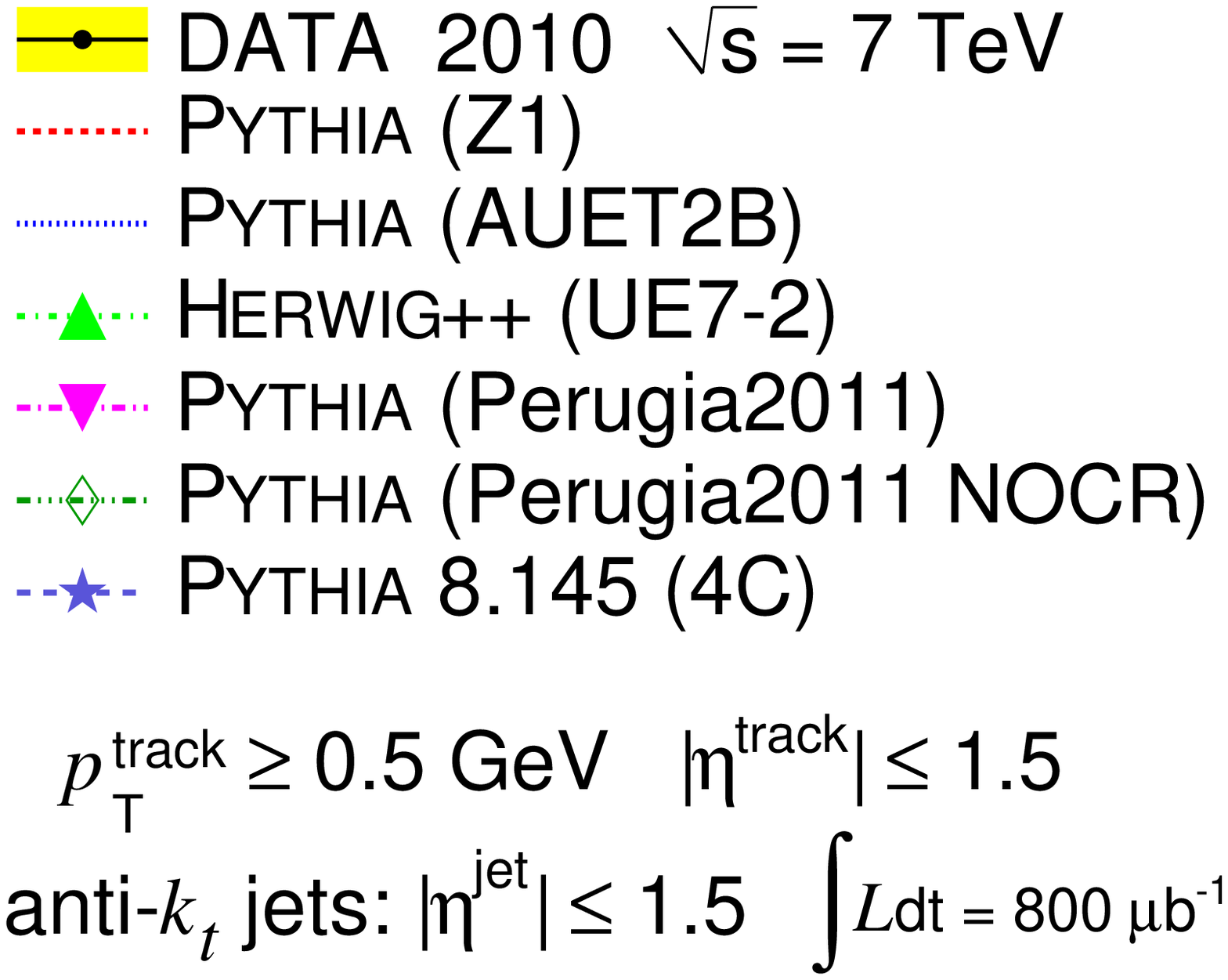}}
\end{center}
\vfil
\caption{The mean value, per unit $ \eta - \phi $, of \ntrks\ in the away region, as a function of
\ptJet\
(a) for $R=0.2$,
(b) for $R=0.4$,
(c) for $R=0.6$,
(d) for $R=0.8$ and
(e) for $R=1.0$.
The shaded band shows the combined statistical and systematic
uncertainty on the data.
The data are compared to the predictions 
obtained with \PythiaS\ using the AUET2B and Z1 tunes.  The bottom
insert shows the ratio of MC predictions to data for several recent MC
tunes.
}
\label{fig:ntrksbinsaway}
\end{figure*}

\begin{figure*}[p]
\begin{center}
\subfigure[]{\includegraphics[width=0.44\textwidth]{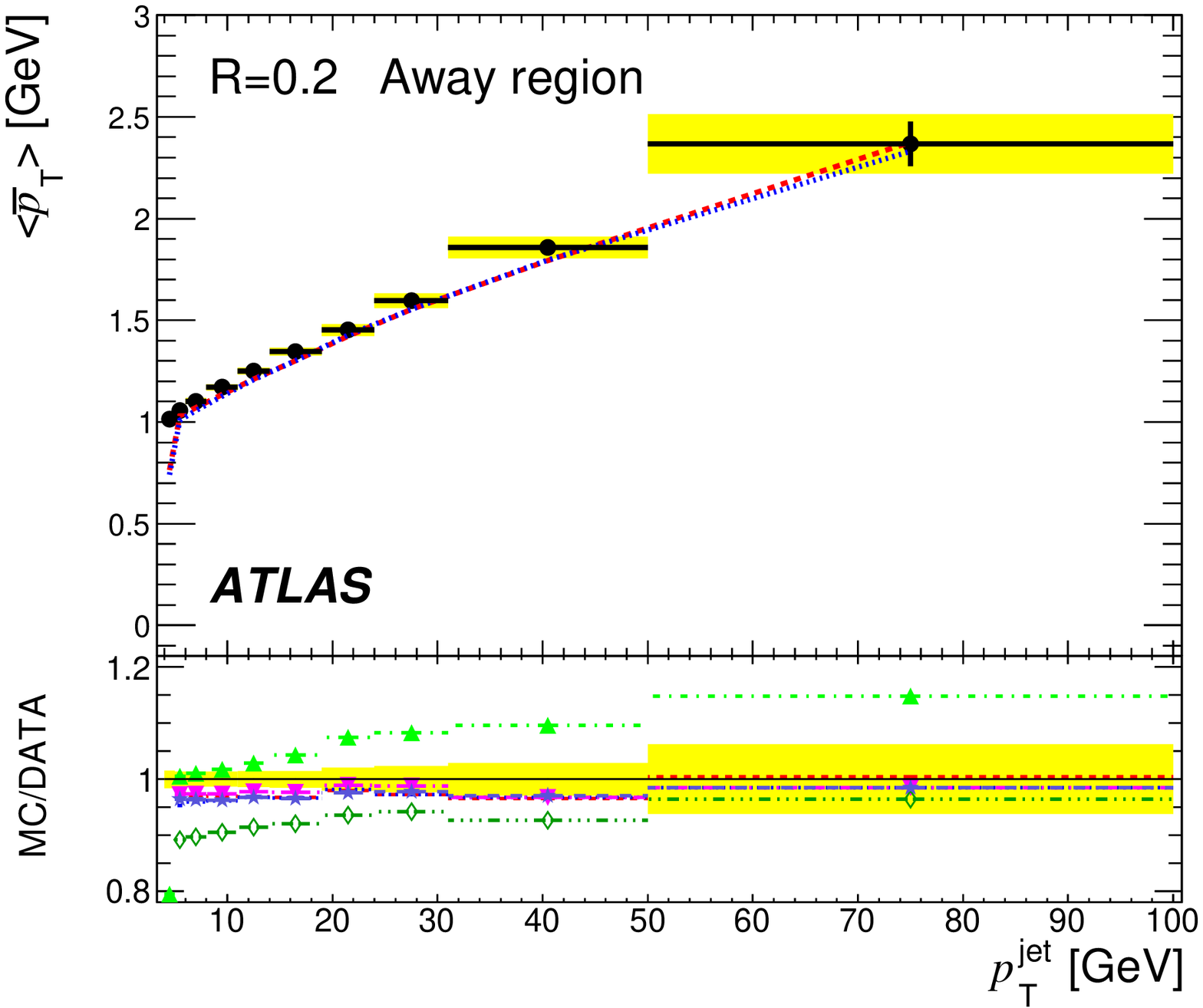}}
\subfigure[]{\includegraphics[width=0.44\textwidth]{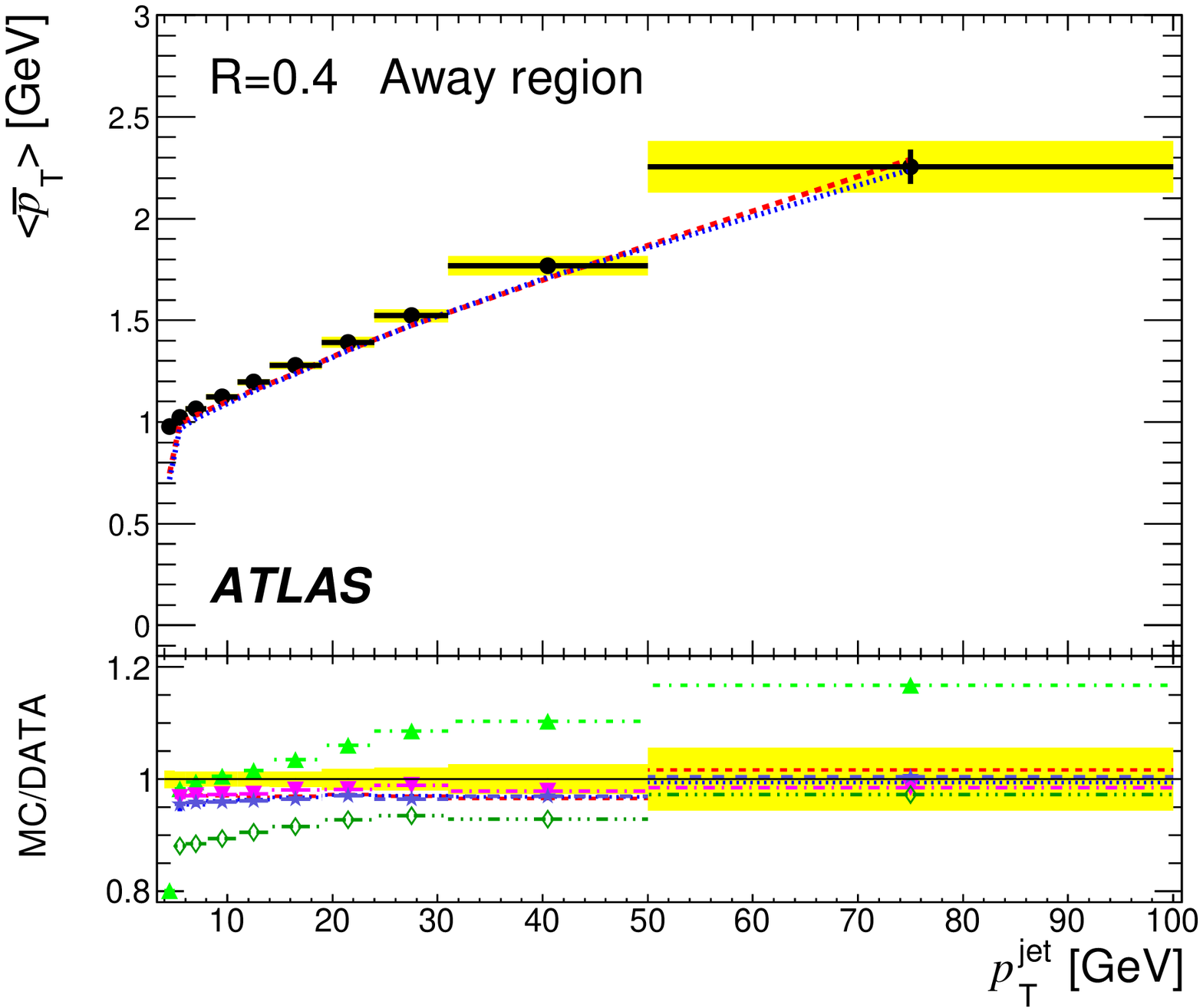}}
\subfigure[]{\includegraphics[width=0.44\textwidth]{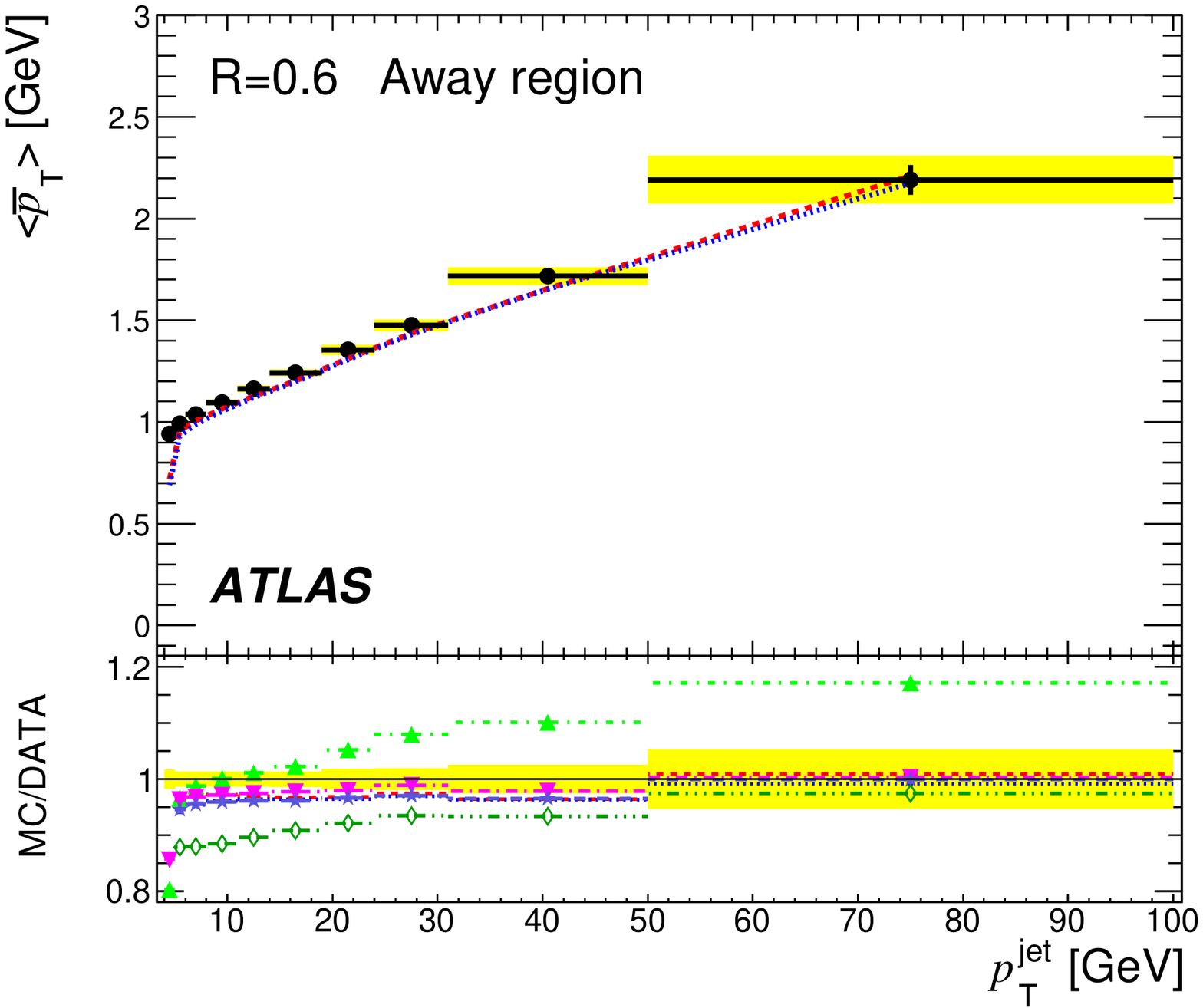}}
\subfigure[]{\includegraphics[width=0.44\textwidth]{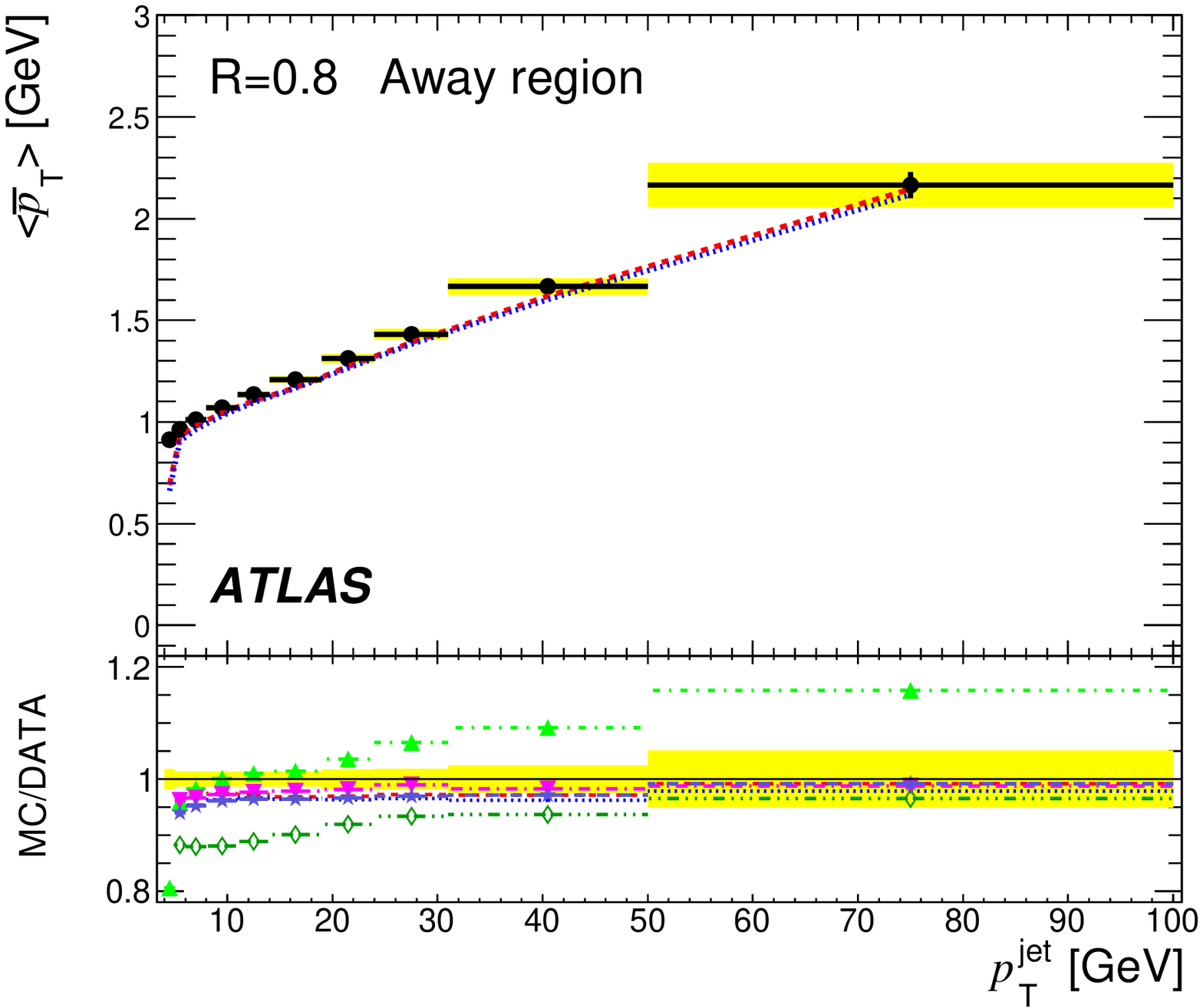}}
\subfigure[]{\includegraphics[width=0.44\textwidth]{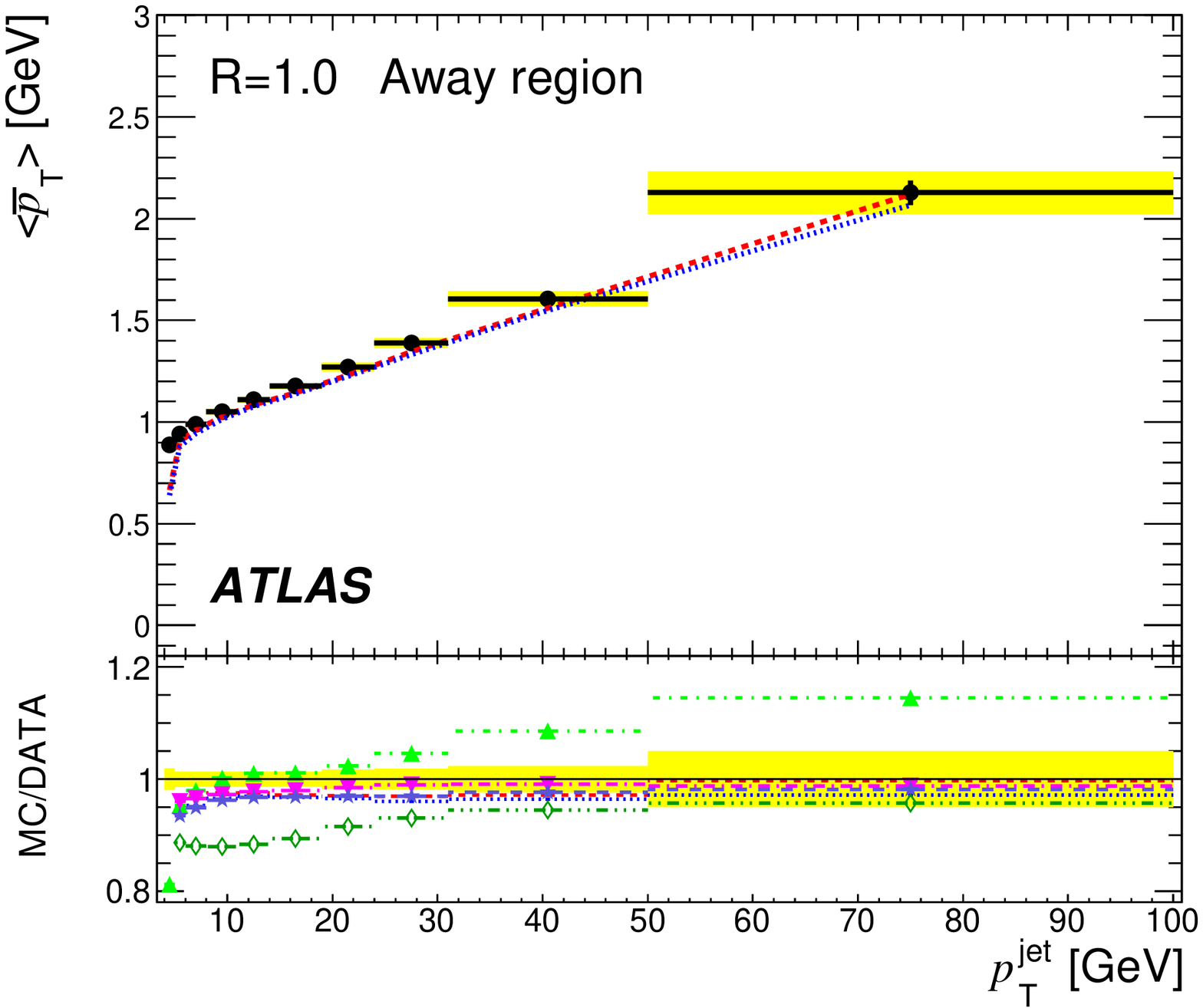}}
{\includegraphics[width=0.44\textwidth]{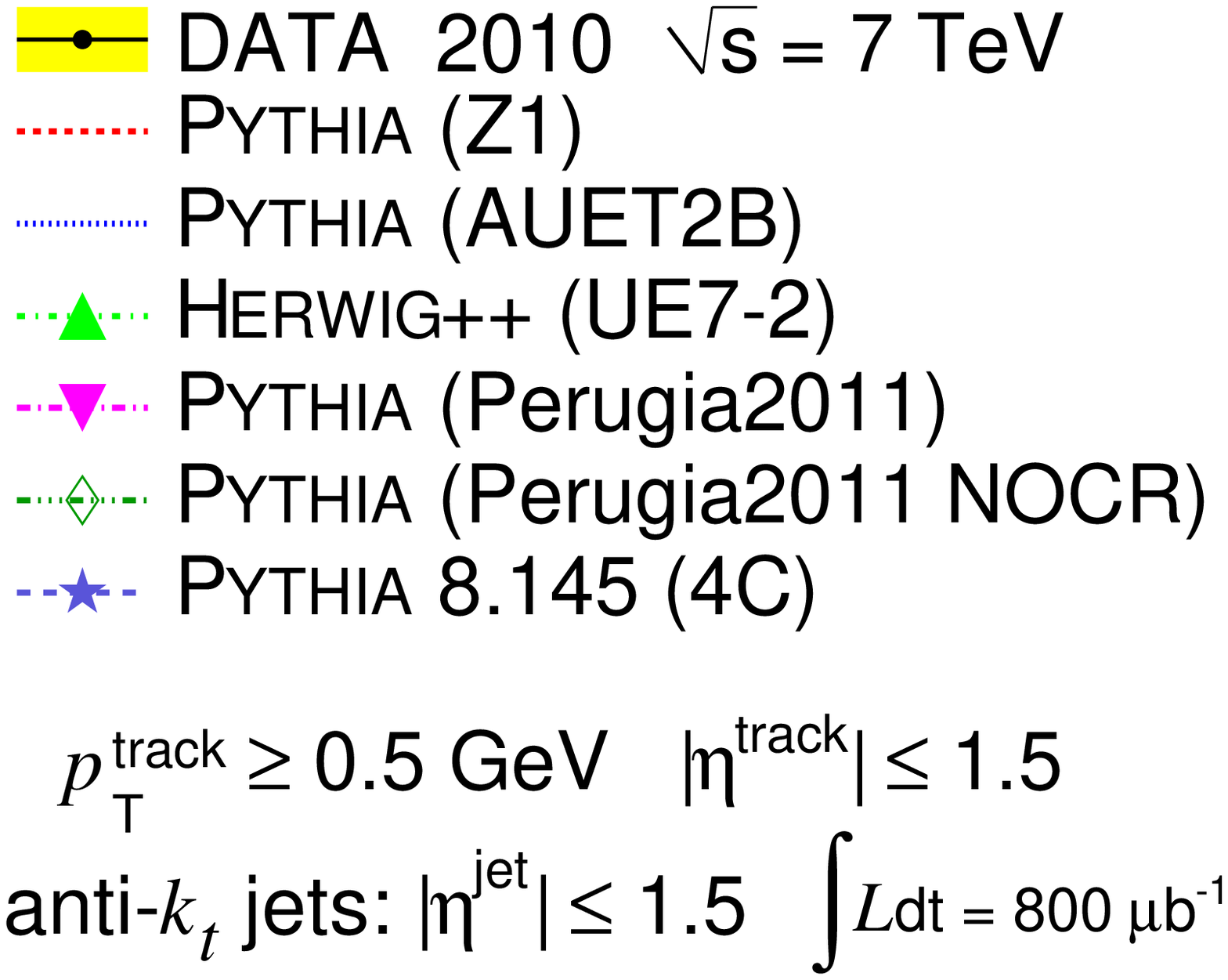}}
\end{center}
\vfil
\caption{The mean value of \ptave\ in the away region, as a function of
\ptJet\
(a) for $R=0.2$,
(b) for $R=0.4$,
(c) for $R=0.6$,
(d) for $R=0.8$ and
(e) for $R=1.0$.
The shaded band shows the combined statistical and systematic
uncertainty on the data.
The data are compared to the predictions 
obtained with \PythiaS\ using the AUET2B and Z1 tunes.  The bottom
insert shows the ratio of MC predictions to data for several recent MC
tunes.
}
\label{fig:ptavebinsaway}
\end{figure*}

The dependence on \ptJet\ of the mean value of \sumpt, \ntrks\ and \ptave\
for all five values of $R$ 
is shown in Figures~\ref{fig:sumptbinsaway} through~~\ref{fig:ptavebinsaway}
for the away region.
The \sumpt, \ntrks\ and \ptave\ values all continue to rise
with \ptJet.  
The away region is expected to have smaller \sumpt\ than the towards
region. This is mainly because the leading jet by definition has the
highest summed  transverse momentum and the distribution of the fraction
of the momentum carried by charged particles is broad.
For $\ptJet\simge 10\ \GeV$, the mean \sumpt\ contained in charged particles
in the away region is typically between 60\%\ (large \ptJet\ and
large $R$) and 100\%\ (small \ptJet\ and small $R$) of the leading jet \pt.
\Herwigpp\ overestimates \ptave\ for $ \ptJet \gtap 10~\gev $.
\PythiaS~Perugia 2011 (without color reconnection) underestimates $ \ptave $ for $ \ptJet \ltap 40~\gev$.

\begin{figure*}[!ptb]
\begin{center}
\subfigure[]{\includegraphics[width=0.44\textwidth]{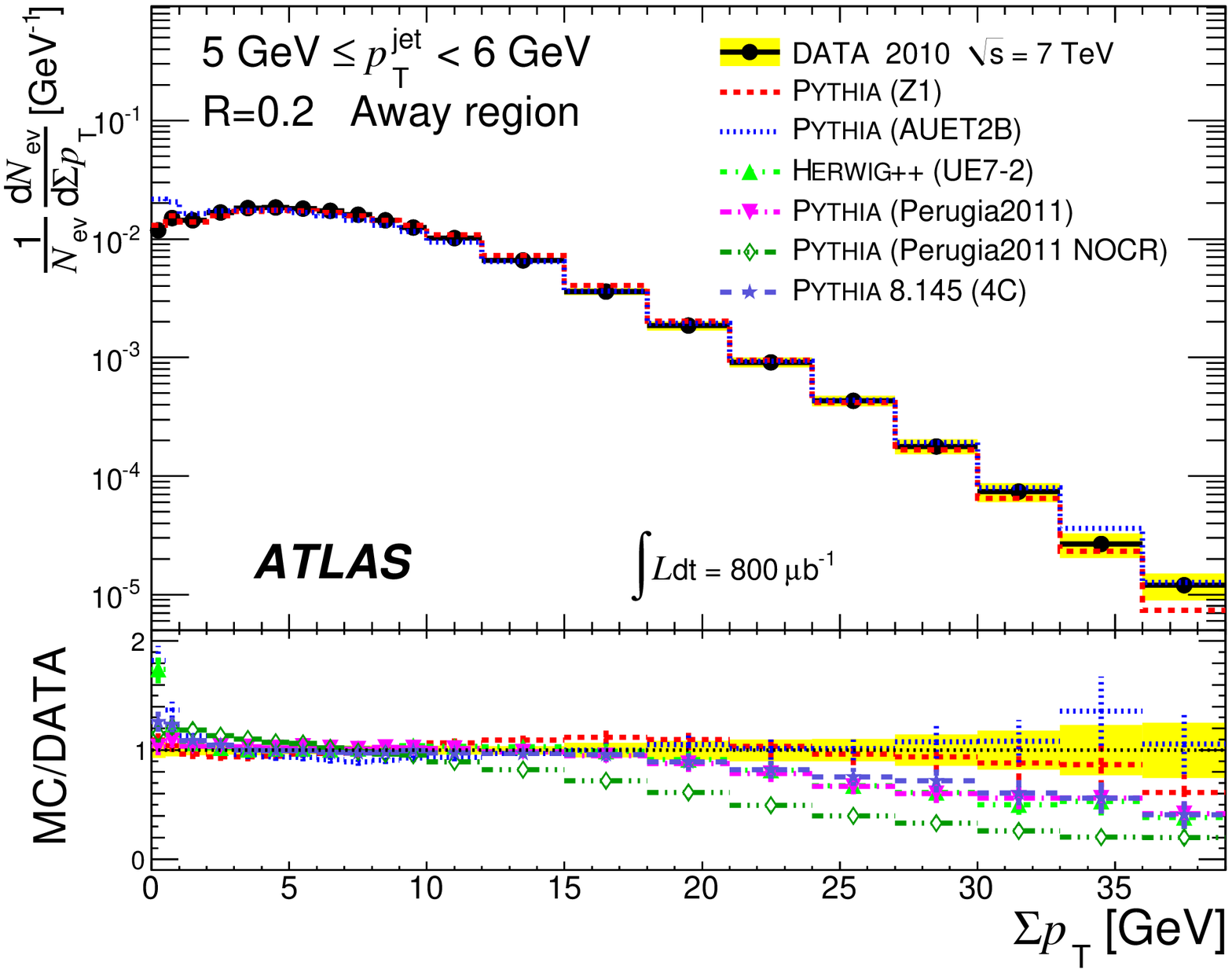}}
\subfigure[]{\includegraphics[width=0.44\textwidth]{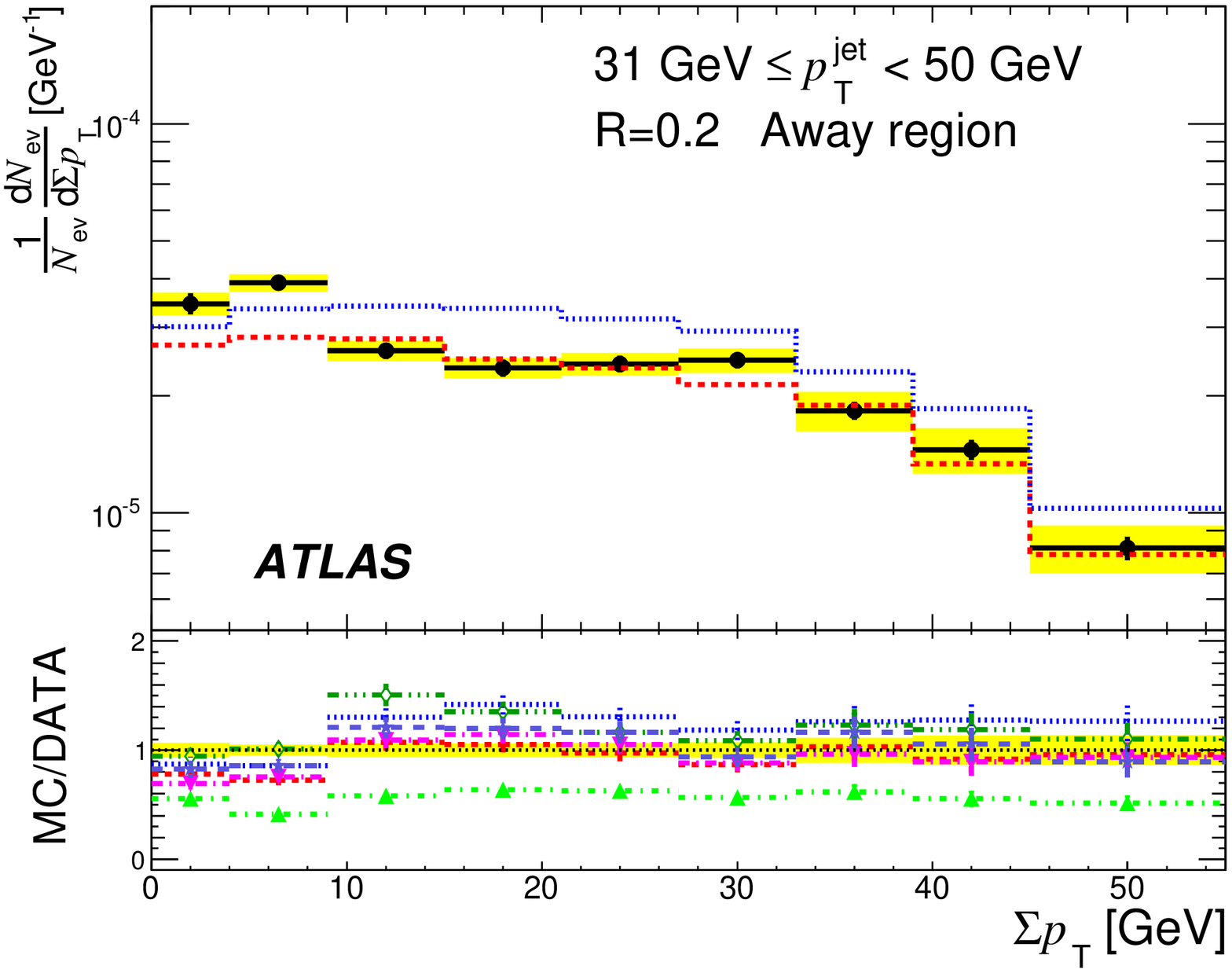}}
\subfigure[]{\includegraphics[width=0.44\textwidth]{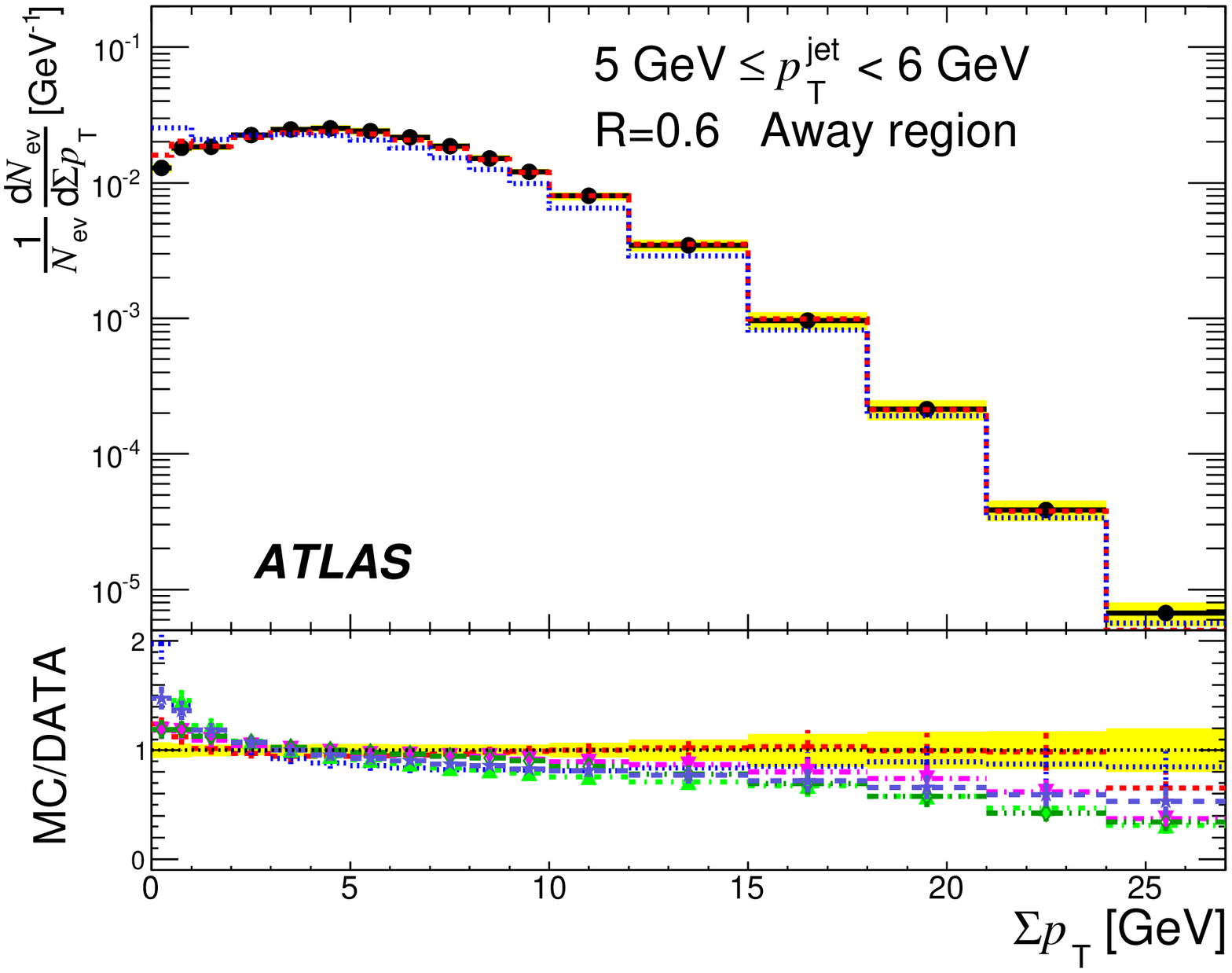}}
\subfigure[]{\includegraphics[width=0.44\textwidth]{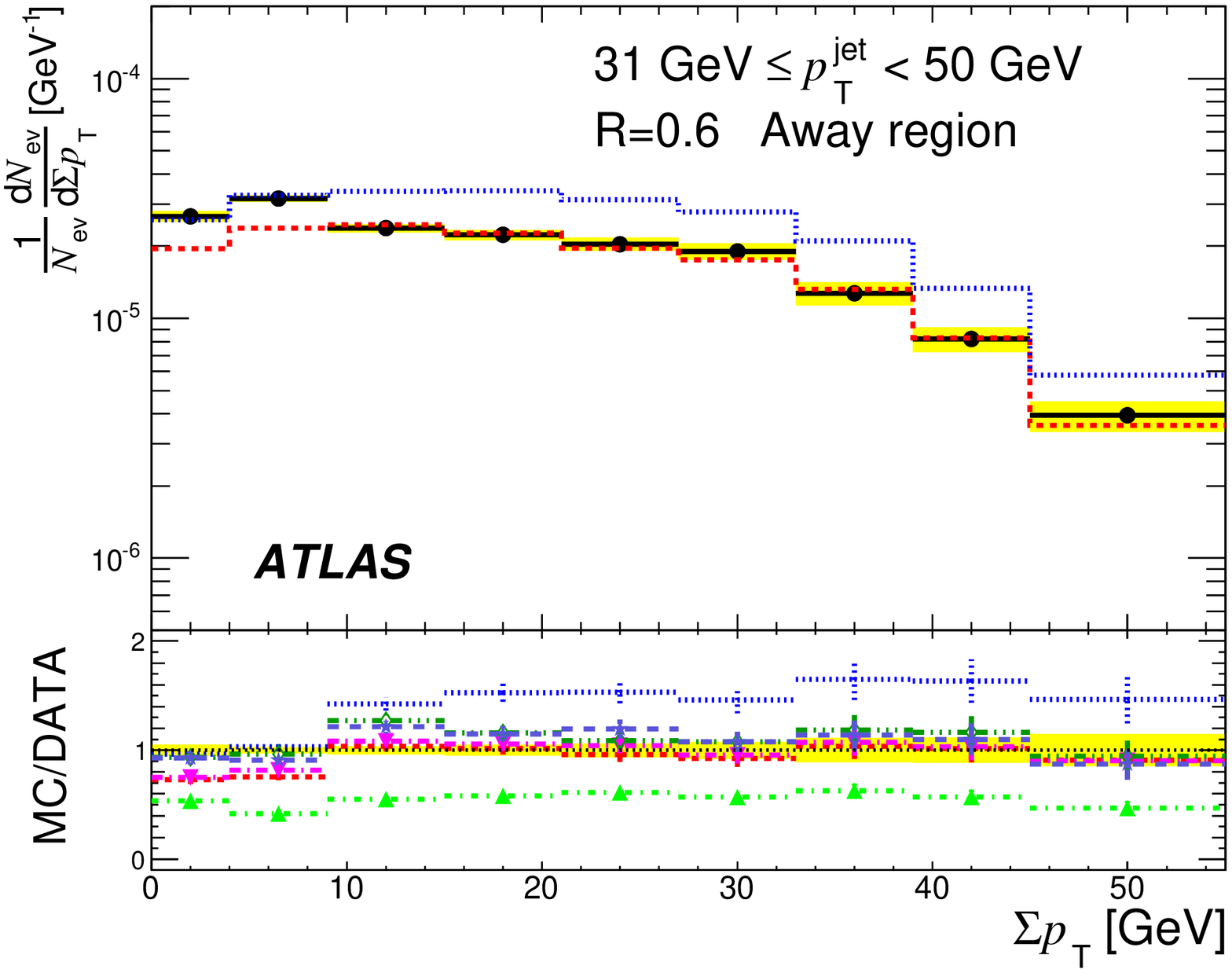}}
\subfigure[]{\includegraphics[width=0.44\textwidth]{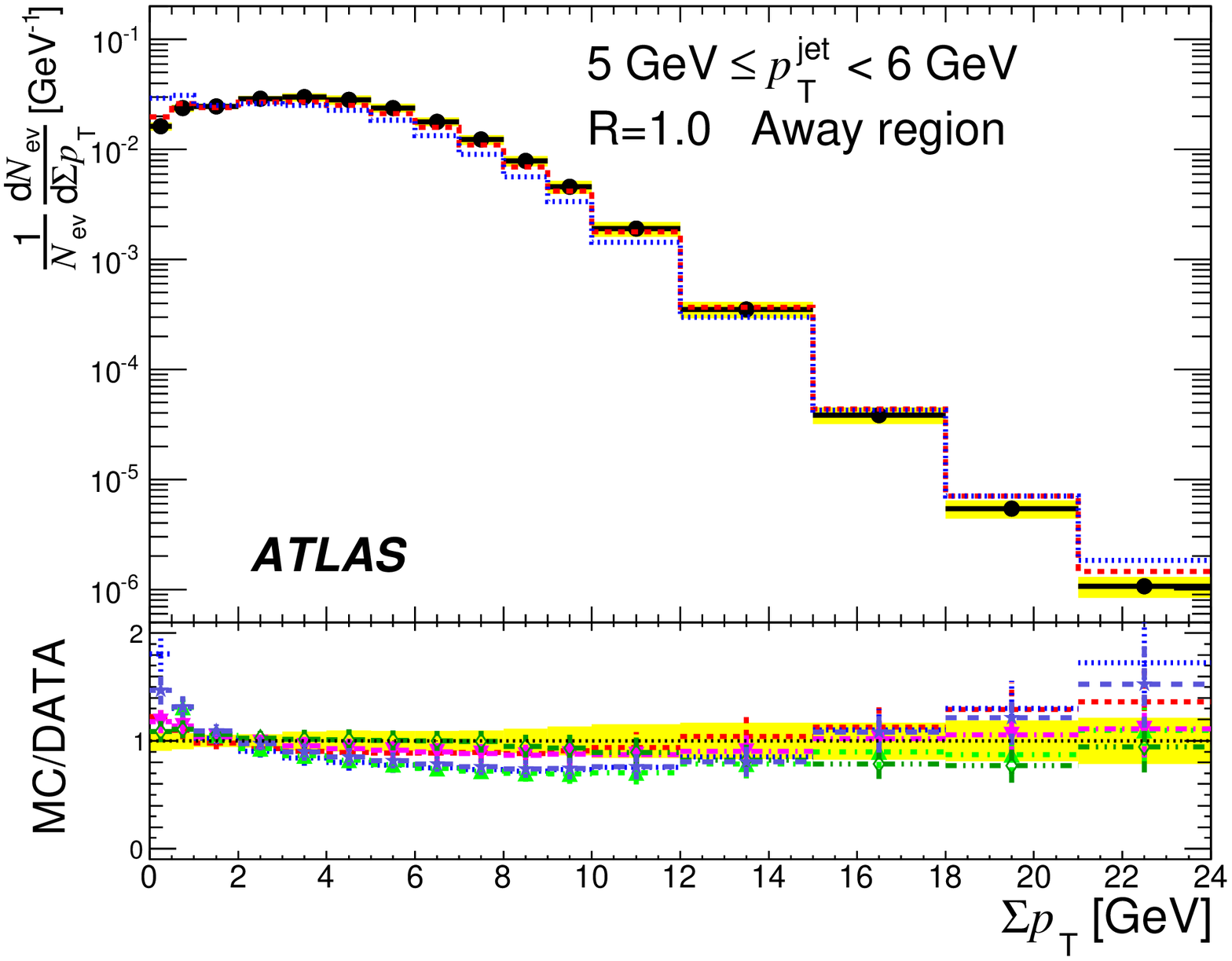}}
\subfigure[]{\includegraphics[width=0.44\textwidth]{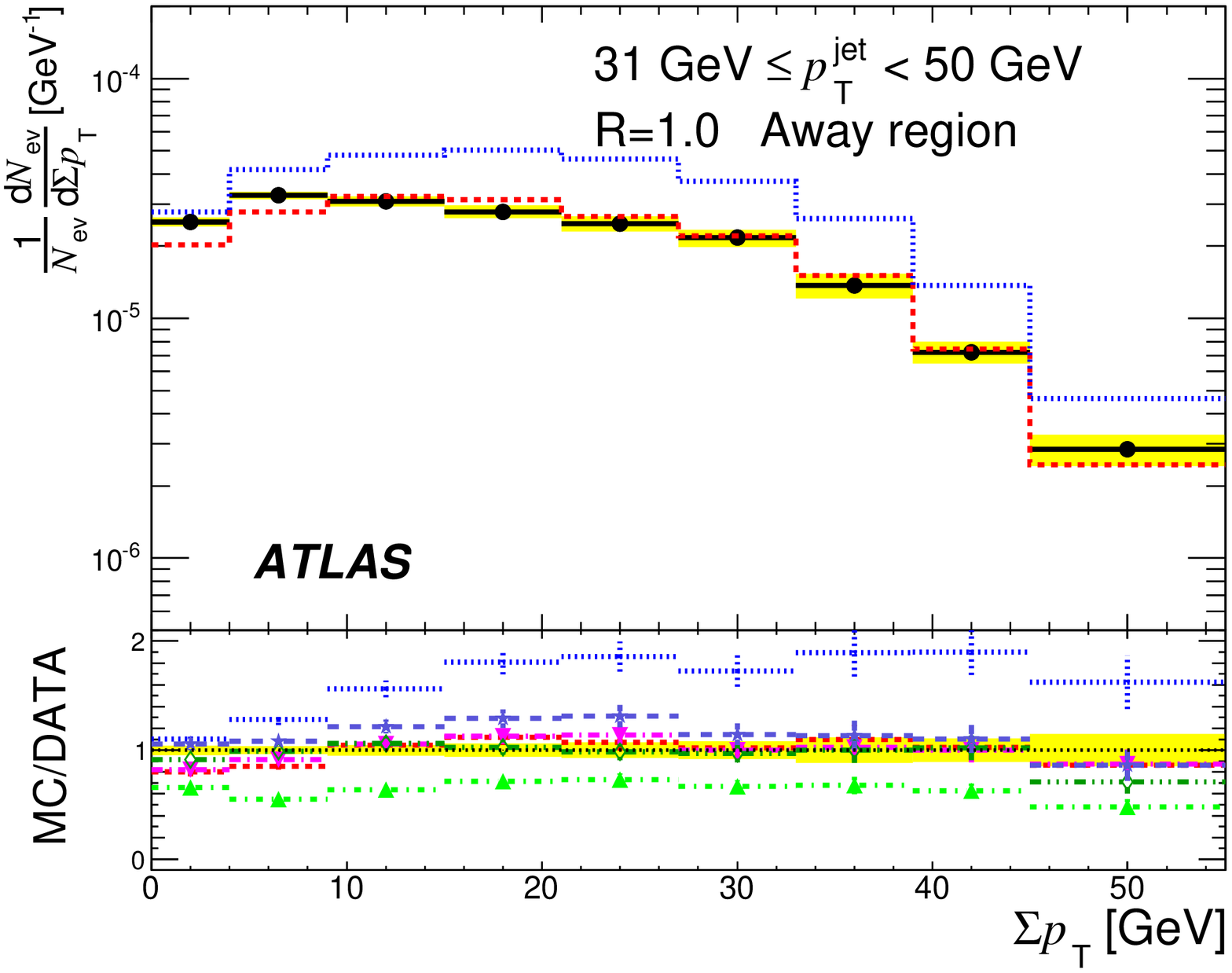}}
\end{center}
\caption{Distributions of \sumpt\ in the 
away region for
(a) $R=0.2$ and $5 \; \GeV \le \ptJet < 6 \;\GeV$,
(b) $R=0.2$ and $31 \; \GeV \le \ptJet < 50 \;\GeV$,
(c) $R=0.6$ and $5 \; \GeV \le \ptJet < 6 \;\GeV$, 
(d) $R=0.6$ and $31 \; \GeV \le \ptJet < 50 \;\GeV$ 
(e) $R=1.0$ and $5 \; \GeV \le \ptJet < 6 \;\GeV$ and 
(f) $R=1.0$ and $31 \; \GeV \le \ptJet < 50 \;\GeV$.
The shaded band shows the combined statistical and systematic 
uncertainty on the data.
The histograms show the predictions
of several MC models.  The bottom insert shows the ratio of
the MC predictions to the data.
}
\label{fig:corr-sumptbinsaway}
\end{figure*}

\begin{figure*}[!ptb]
\subfigure[]{\includegraphics[width=0.44\textwidth]{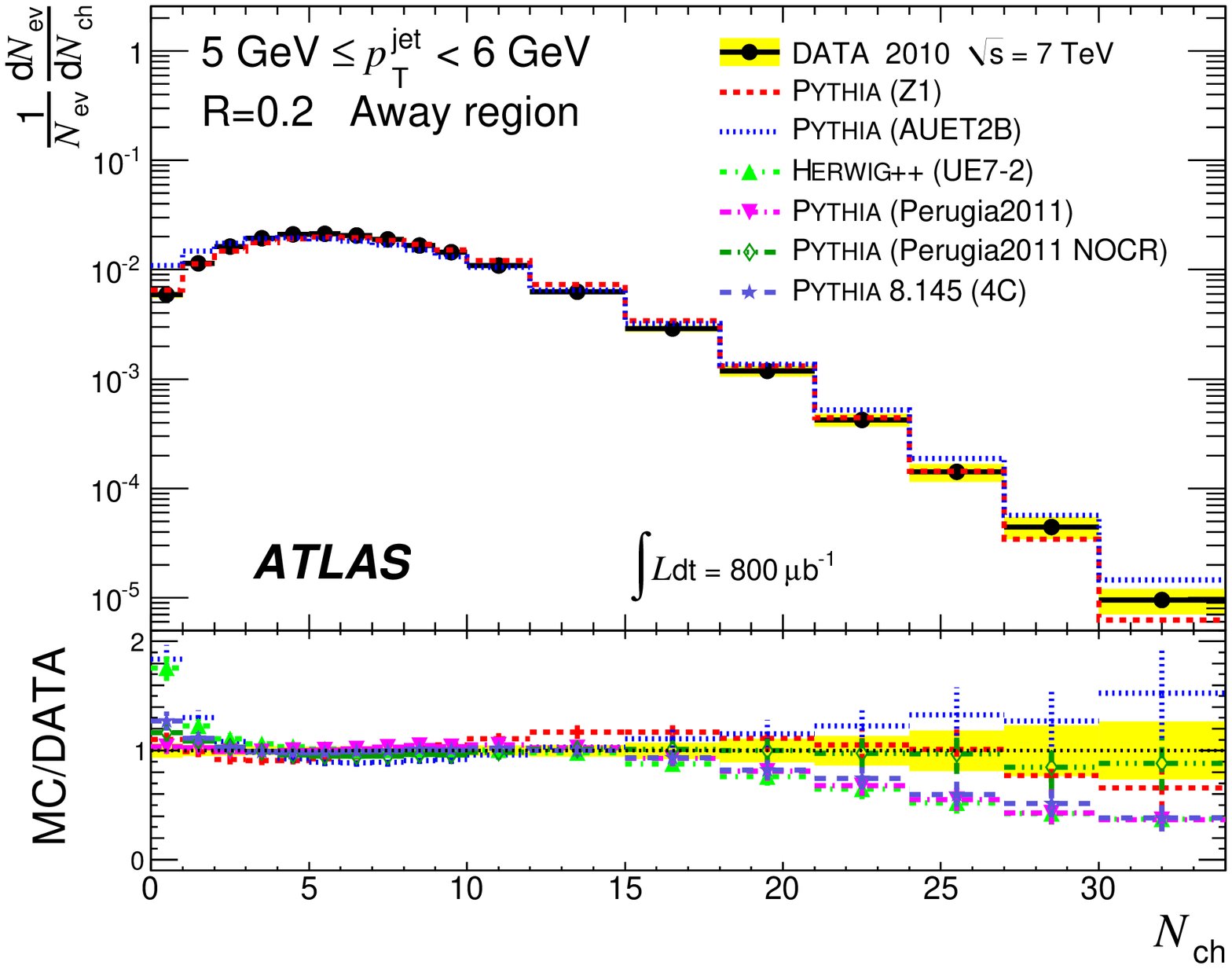}}
\subfigure[]{\includegraphics[width=0.44\textwidth]{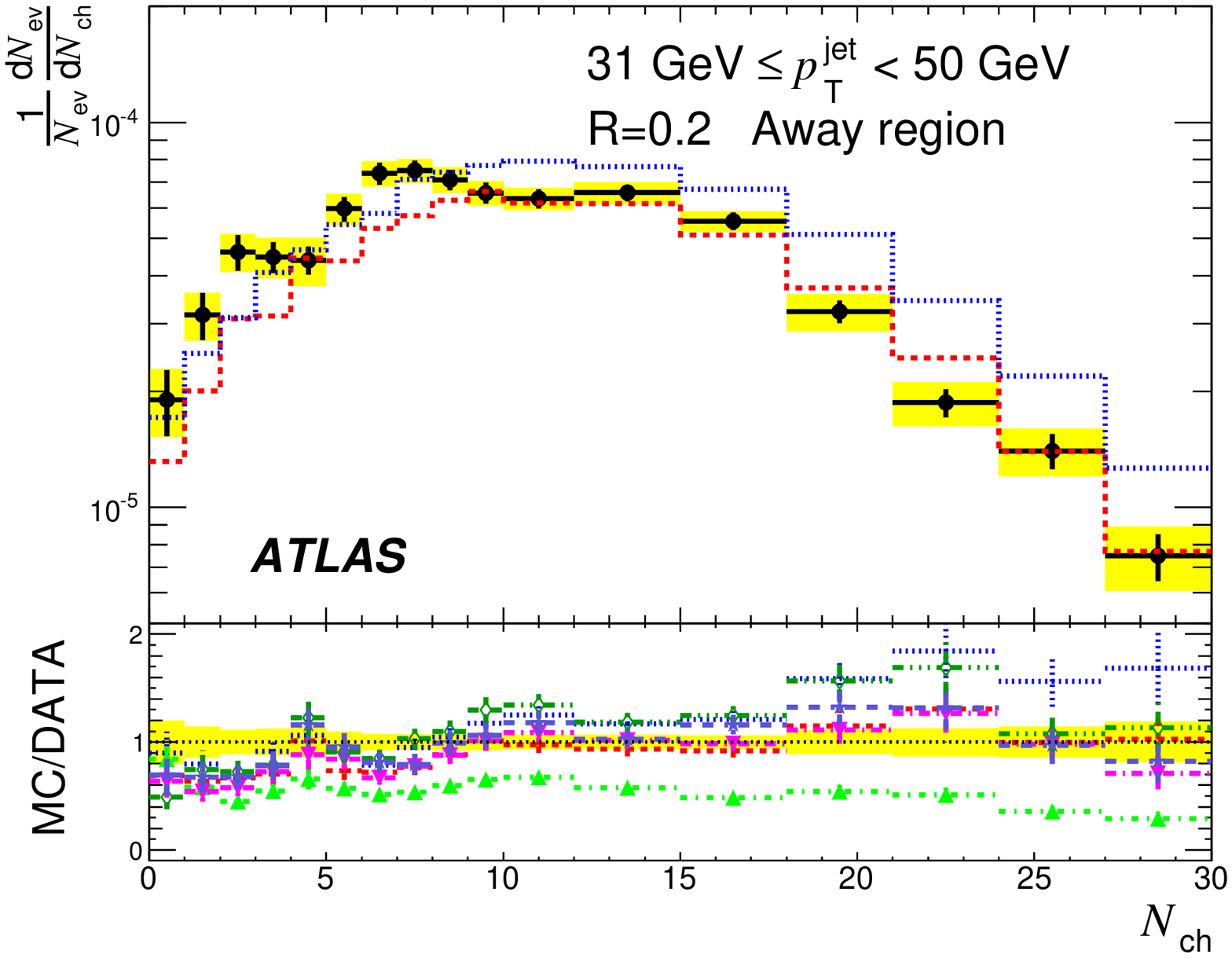}}
\subfigure[]{\includegraphics[width=0.44\textwidth]{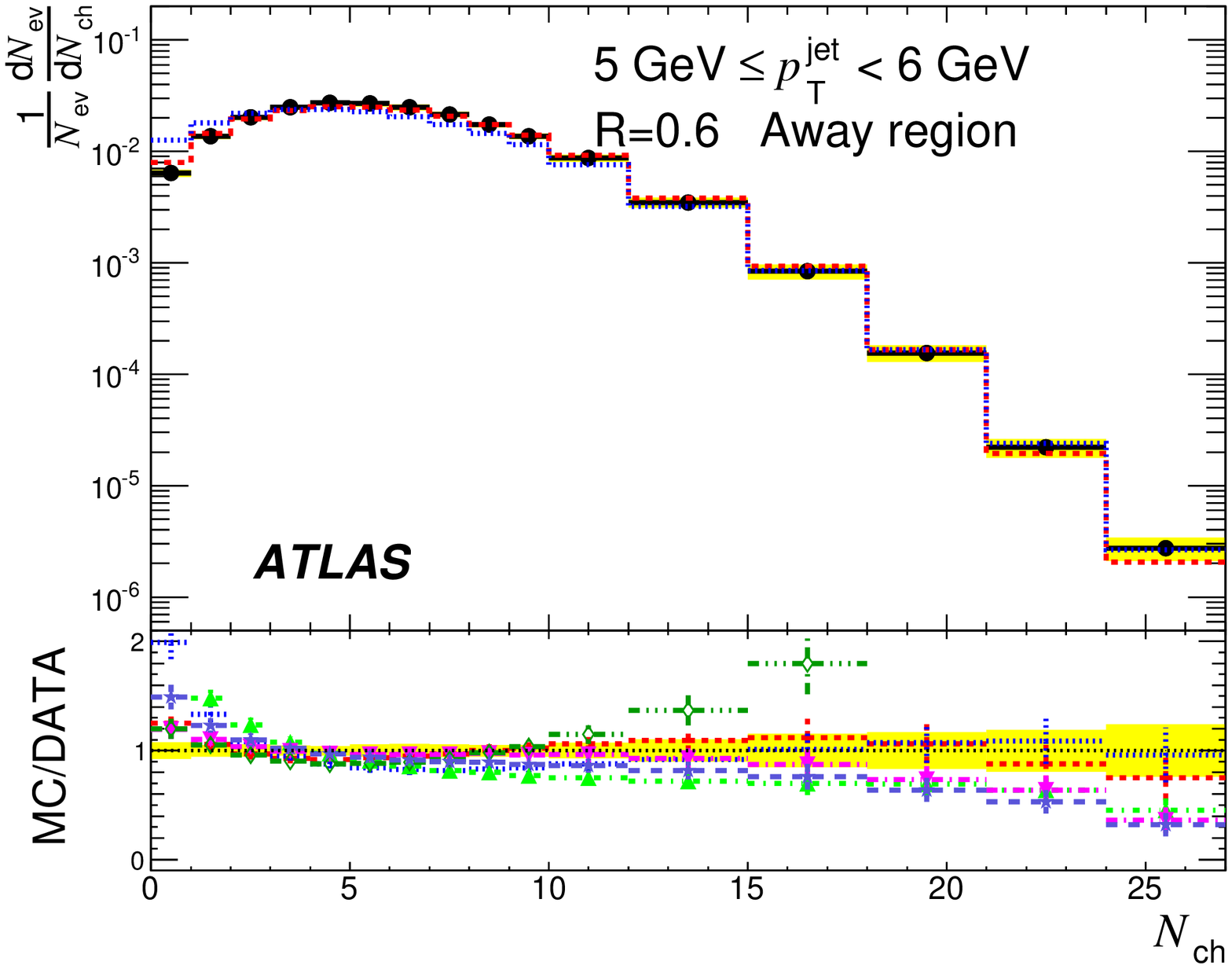}}
\subfigure[]{\includegraphics[width=0.44\textwidth]{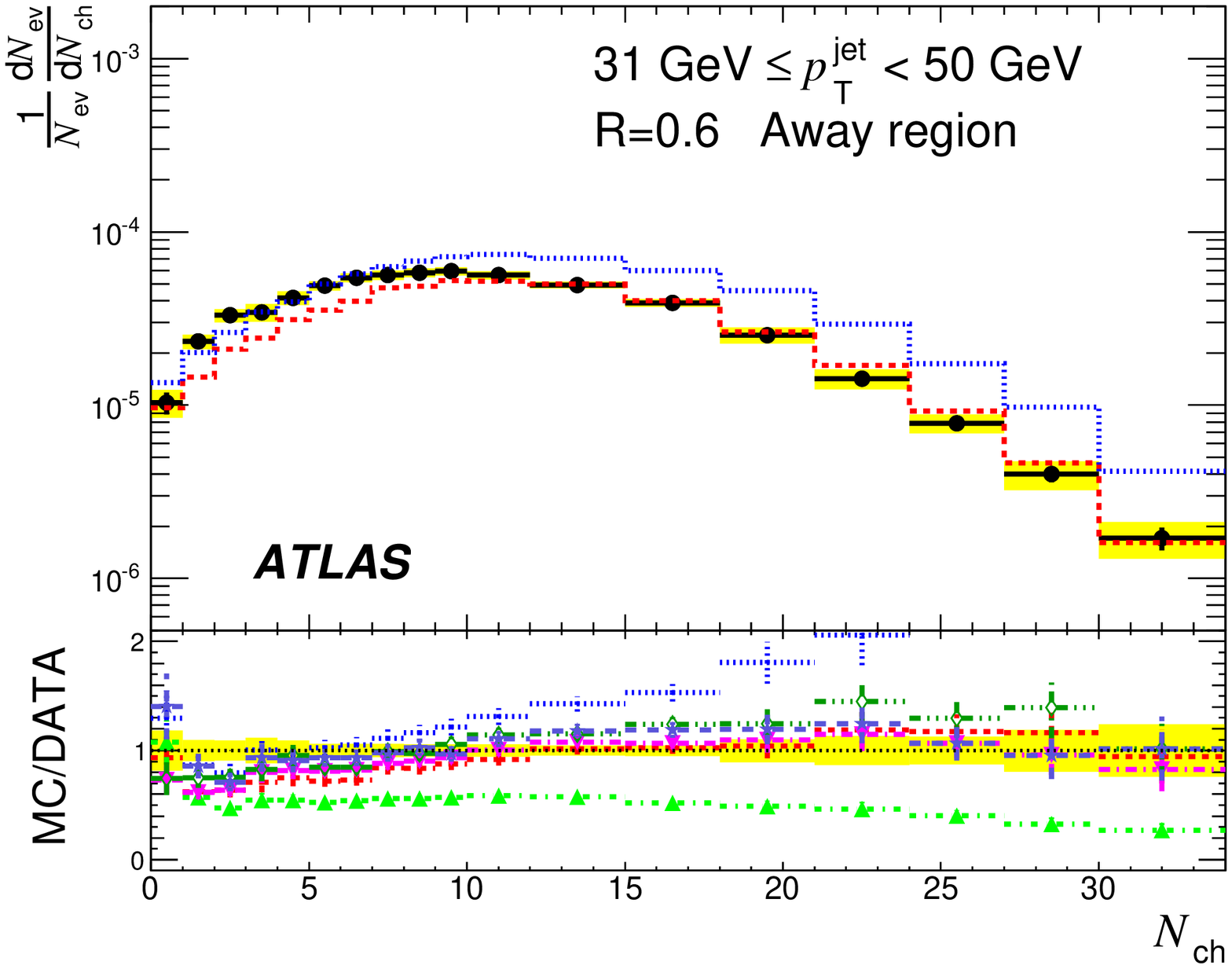}}
\subfigure[]{\includegraphics[width=0.44\textwidth]{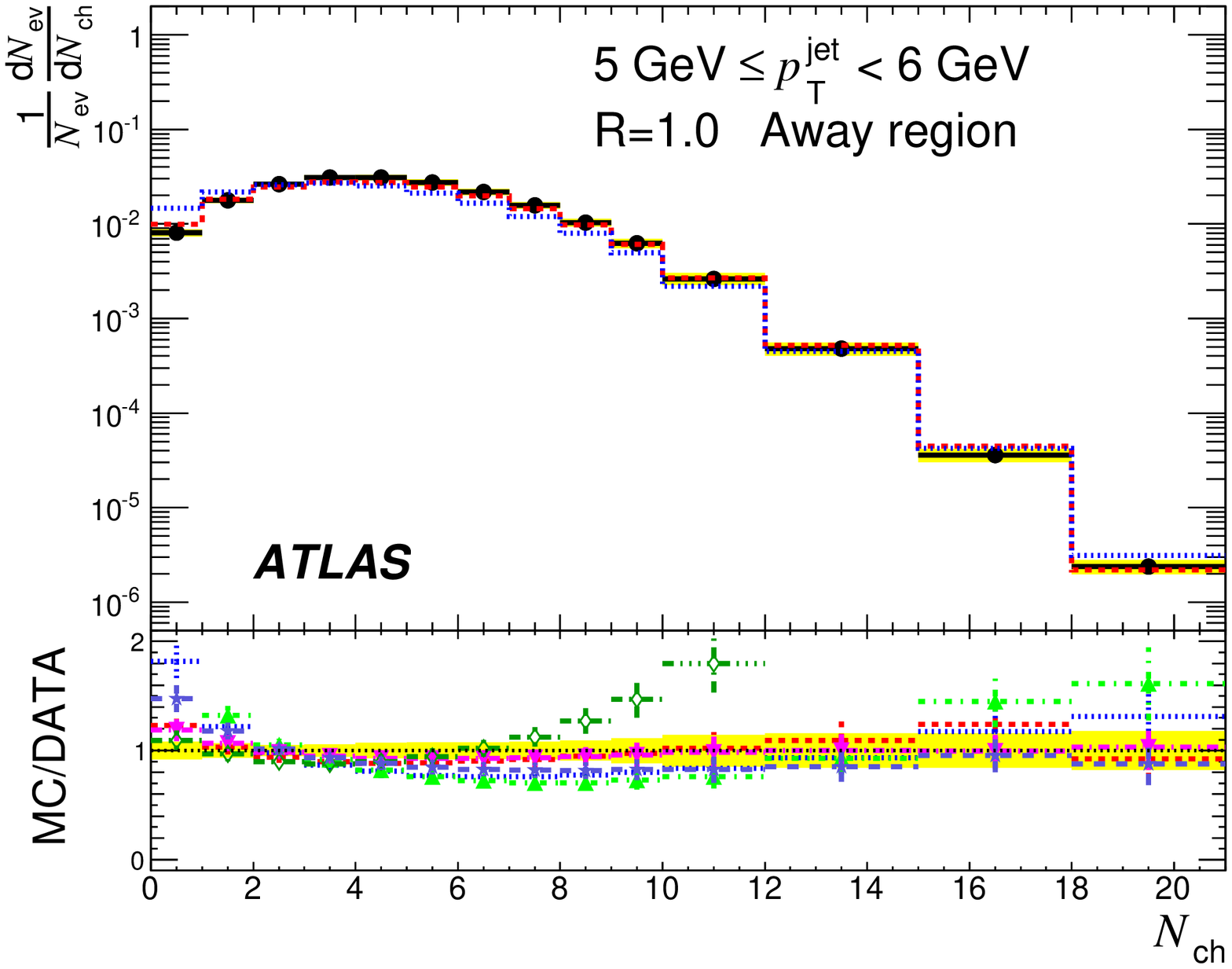}}
\subfigure[]{\includegraphics[width=0.44\textwidth]{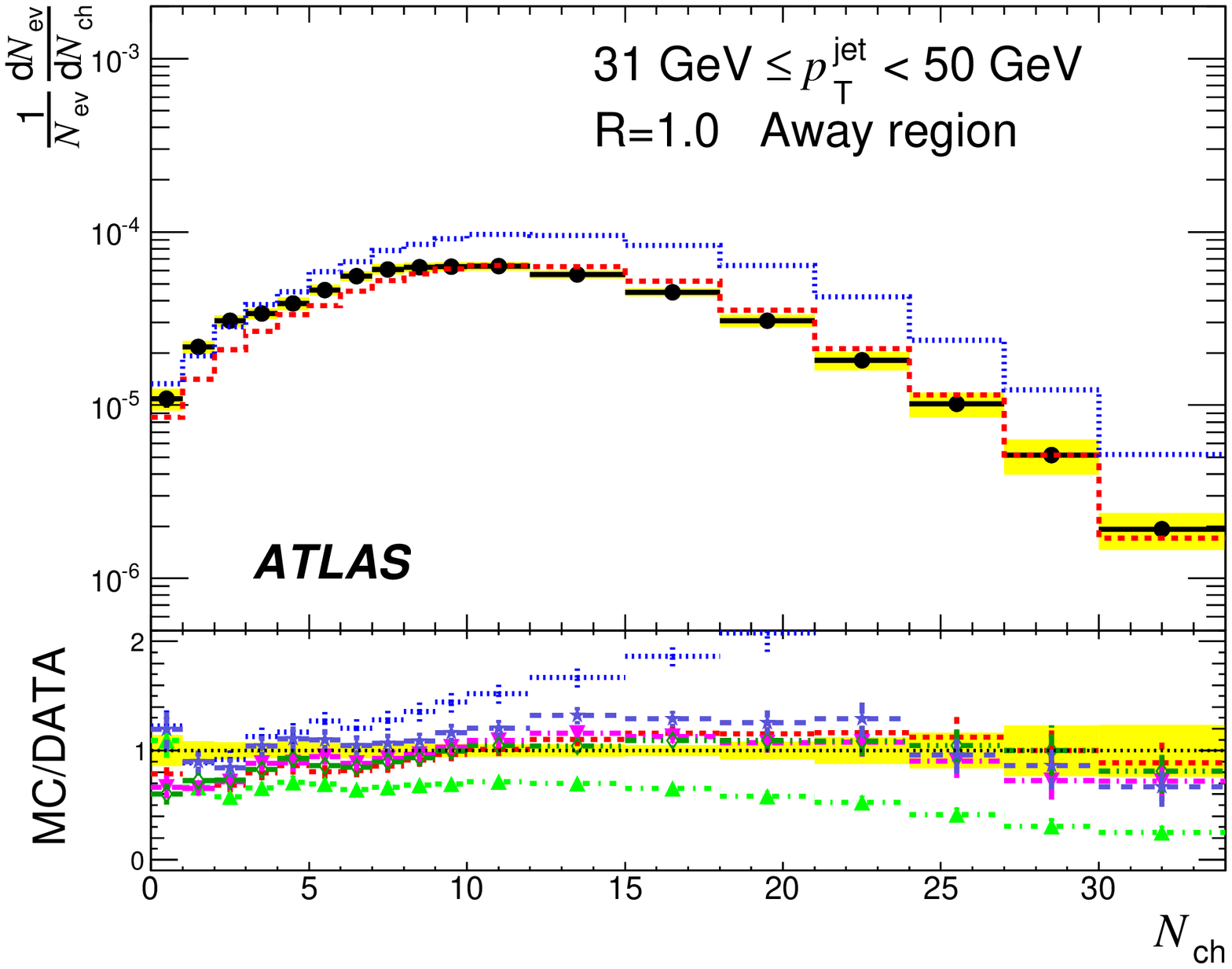}}
\caption{Distributions of \ntrks\ in the 
away region for
(a) $R=0.2$ and $5 \; \GeV \le \ptJet < 6 \;\GeV$,
(b) $R=0.2$ and $31 \; \GeV \le \ptJet < 50 \;\GeV$,
(c) $R=0.6$ and $5 \; \GeV \le \ptJet < 6 \;\GeV$,
(d) $R=0.6$ and $31 \; \GeV \le \ptJet < 50 \;\GeV$ 
(e) $R=1.0$ and $5 \; \GeV \le \ptJet < 6 \;\GeV$ and 
(f) $R=1.0$ and $31 \; \GeV \le \ptJet < 50 \;\GeV$.
The shaded band shows the combined statistical and systematic 
uncertainty on the data.
The histograms show the predictions
of several MC models.  The bottom insert shows the ratio of
the MC predictions to the data.
}
\label{fig:corr-ntrksbinsaway}
\end{figure*}

\begin{figure*}[!ptb]
\subfigure[]{\includegraphics[width=0.44\textwidth]{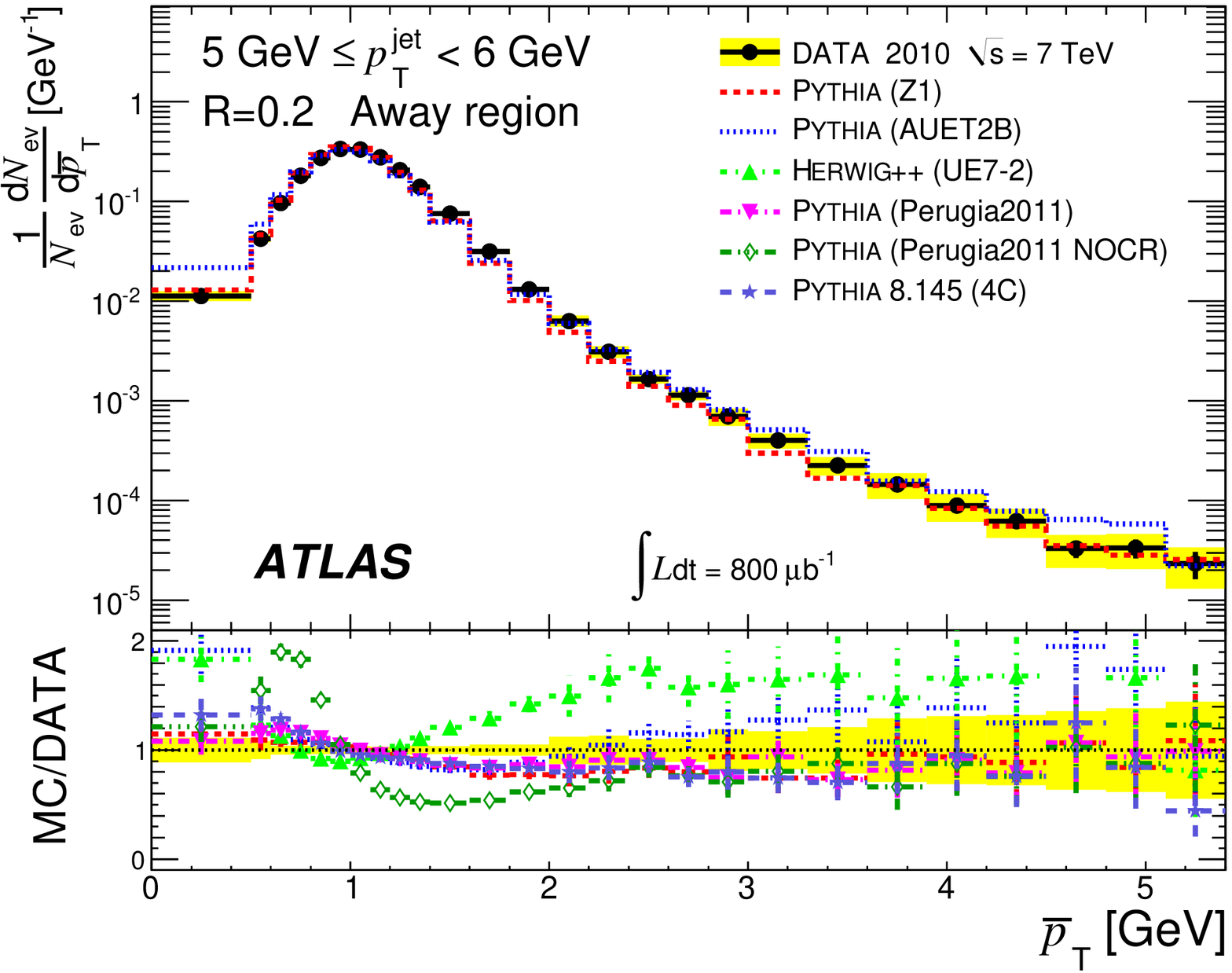}}
\subfigure[]{\includegraphics[width=0.44\textwidth]{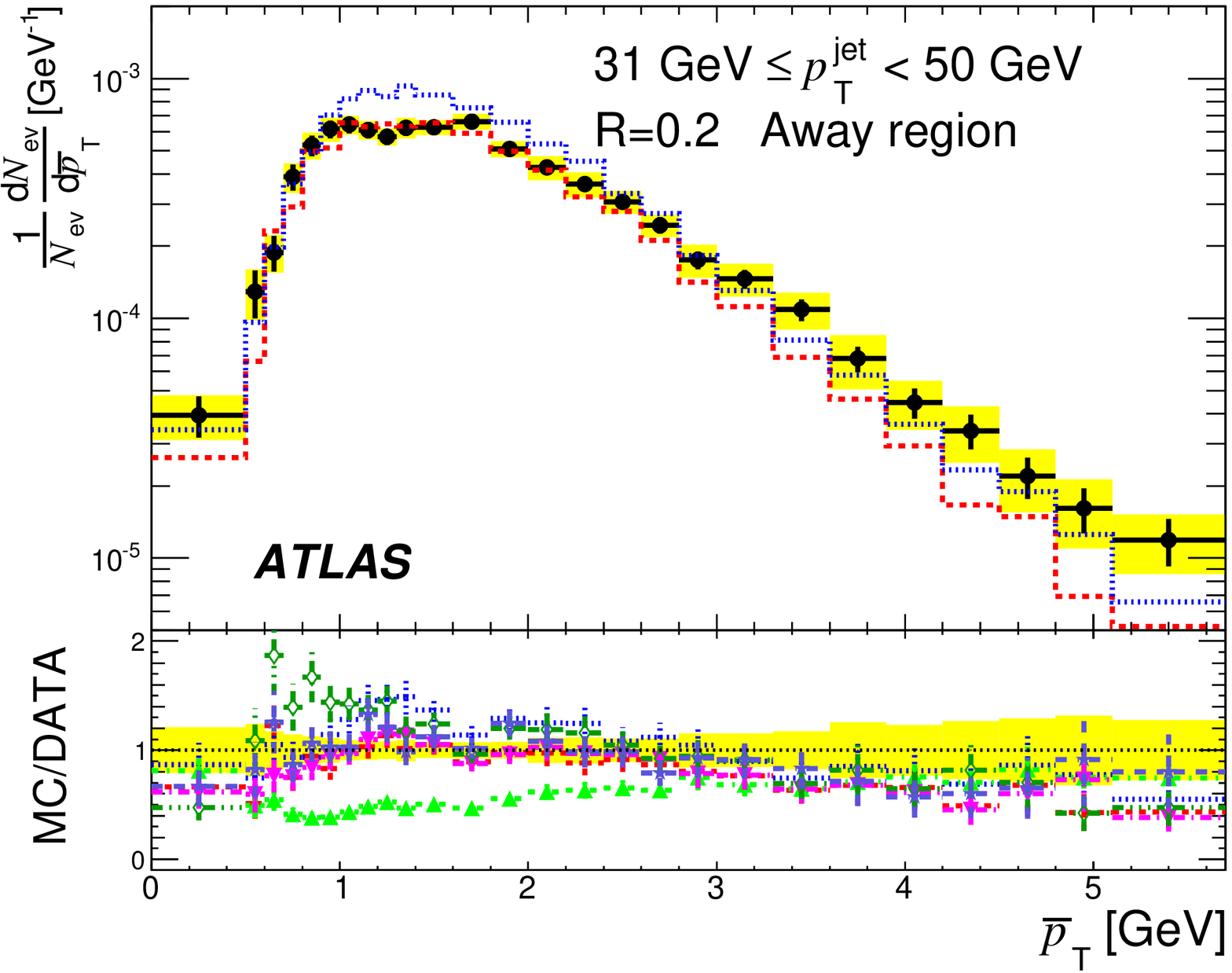}}
\subfigure[]{\includegraphics[width=0.44\textwidth]{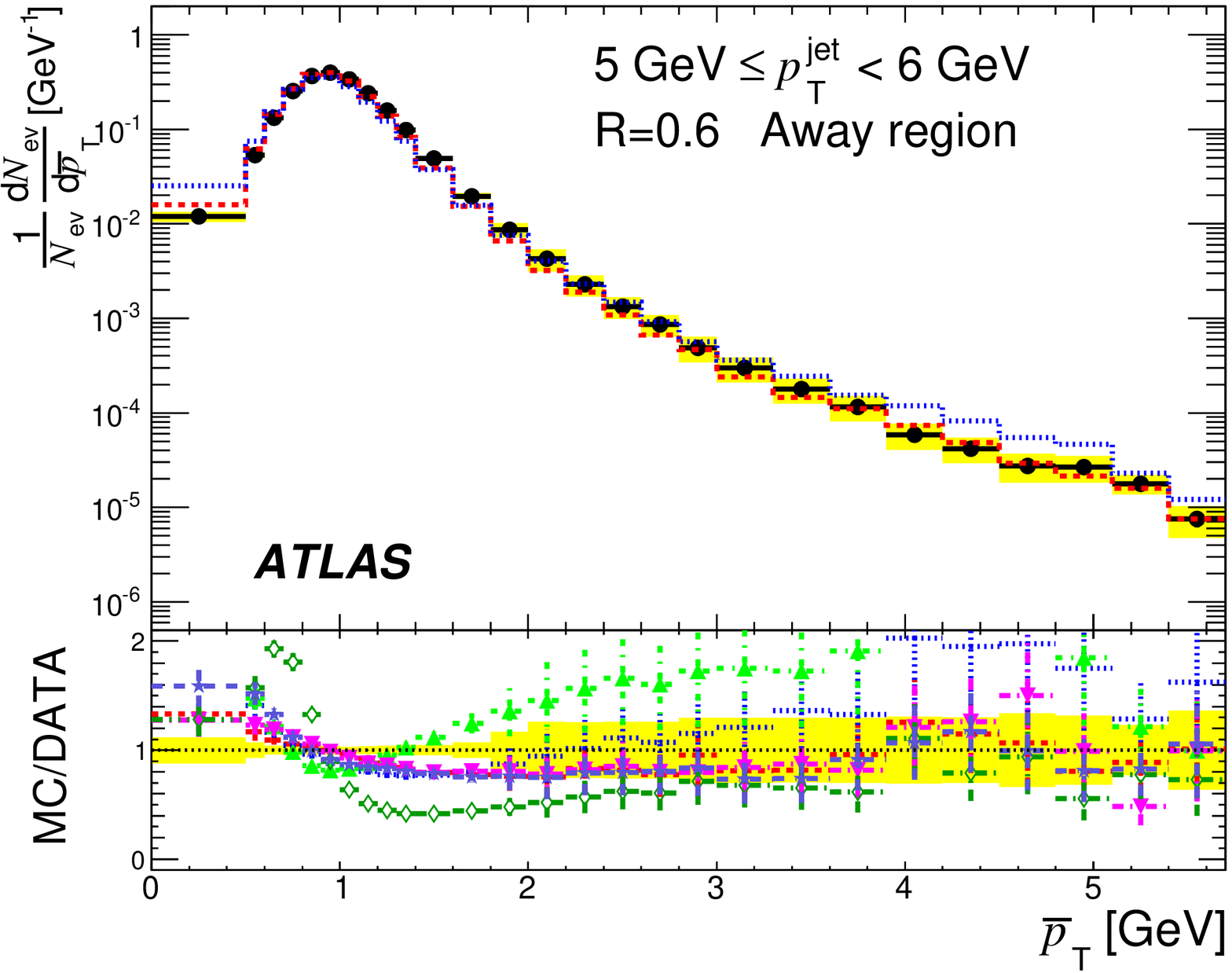}}
\subfigure[]{\includegraphics[width=0.44\textwidth]{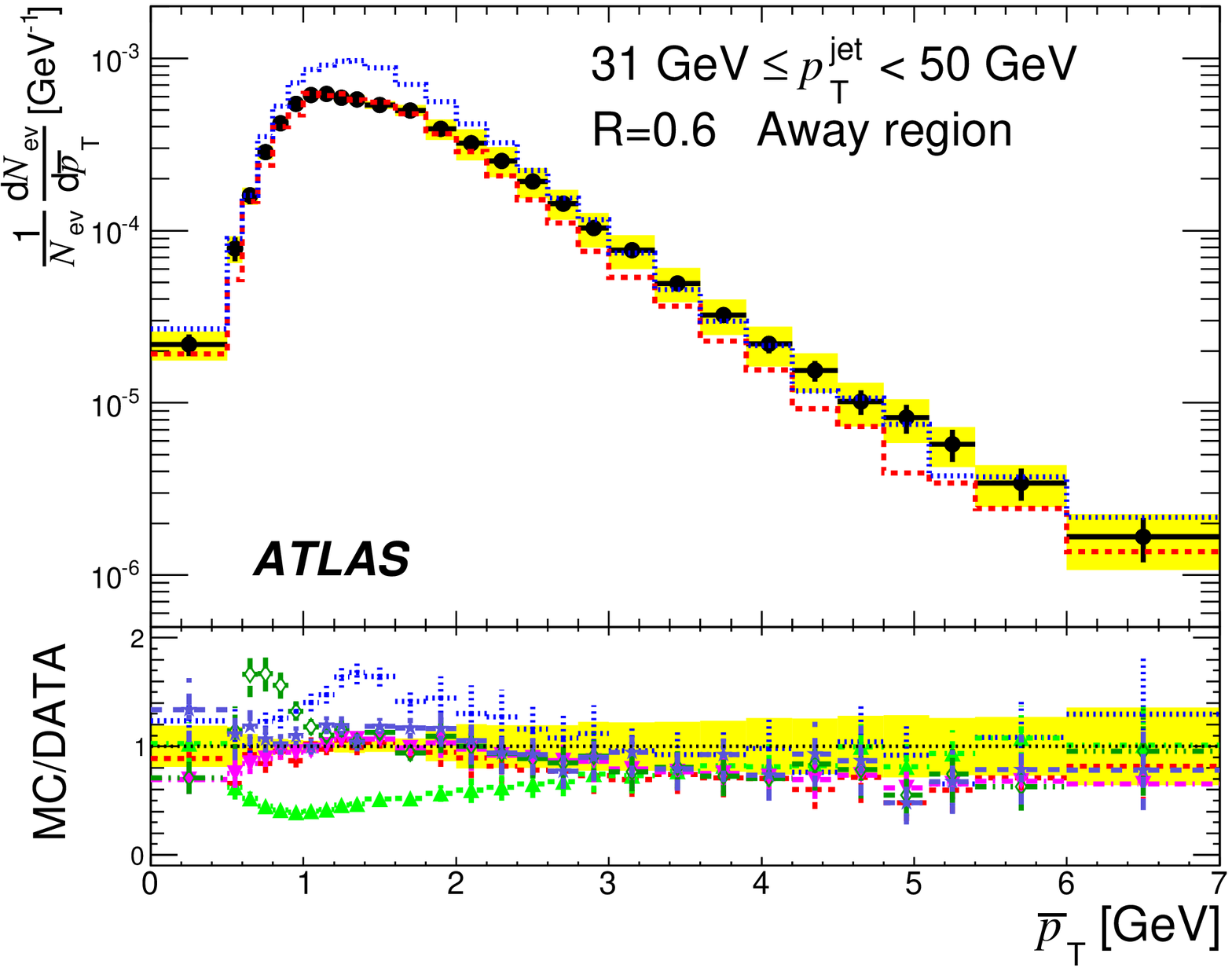}}
\subfigure[]{\includegraphics[width=0.44\textwidth]{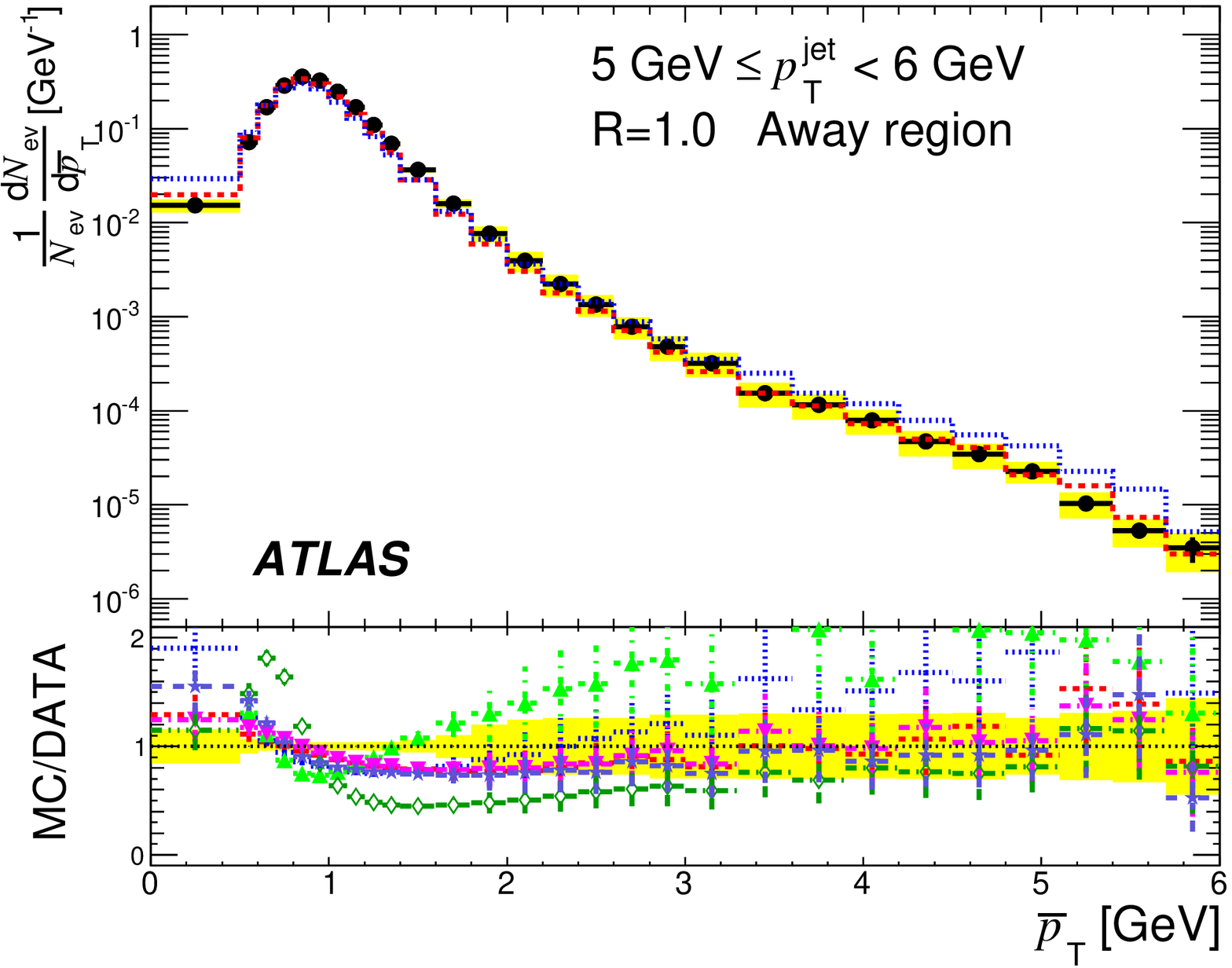}}
\subfigure[]{\includegraphics[width=0.44\textwidth]{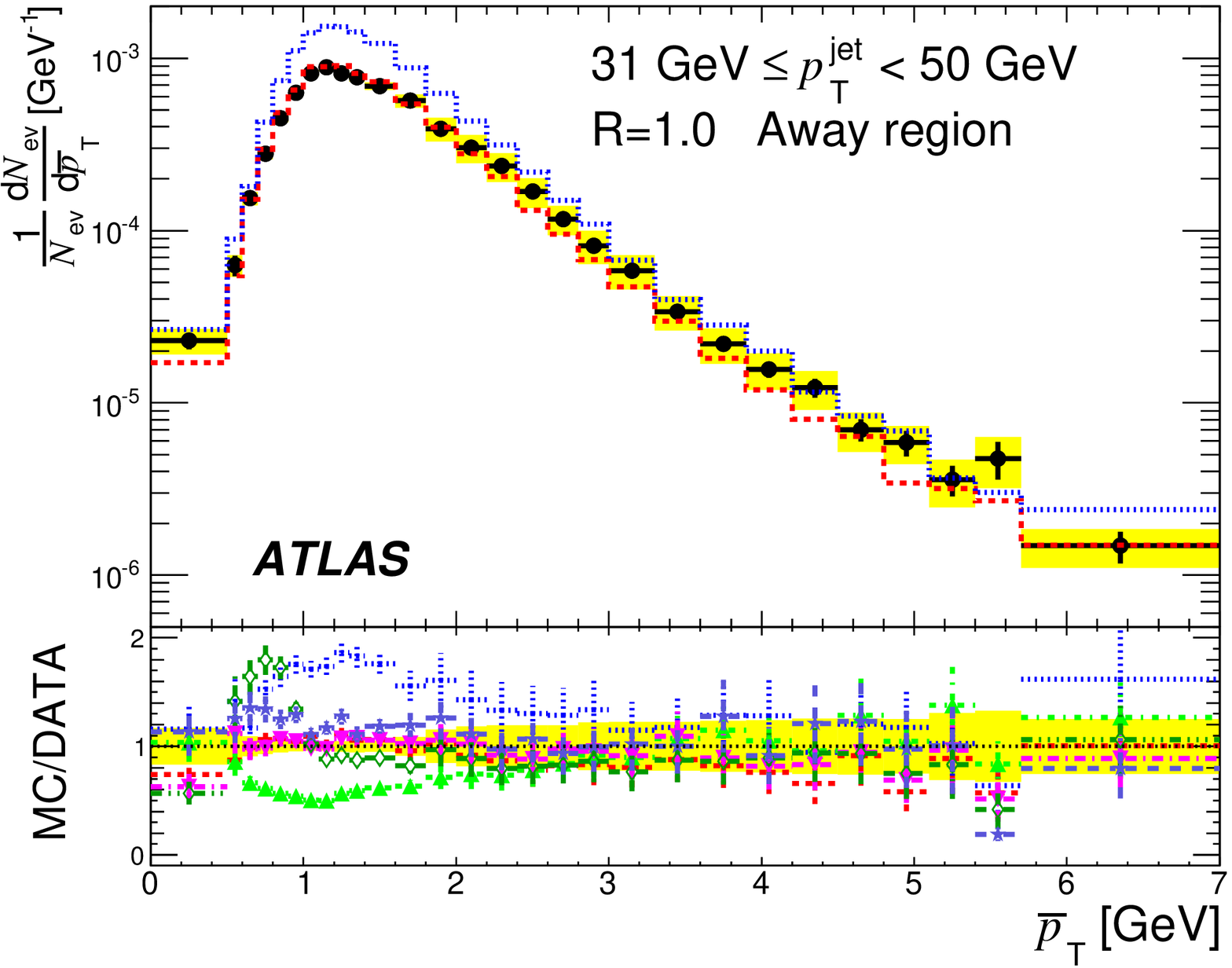}}
\caption{Distributions of \ptave\ in the 
away region for
(a) $R=0.2$ and $5 \; \GeV \le \ptJet < 6 \;\GeV$,
(b) $R=0.2$ and $31 \; \GeV \le \ptJet < 50 \;\GeV$,
(c) $R=0.6$ and $5 \; \GeV \le \ptJet < 6 \;\GeV$, 
(d) $R=0.6$ and $31 \; \GeV \le \ptJet < 50 \;\GeV$ 
(e) $R=1.0$ and $5 \; \GeV \le \ptJet < 6 \;\GeV$ and 
(f) $R=1.0$ and $31 \; \GeV \le \ptJet < 50 \;\GeV$.
The shaded band shows the combined statistical and systematic 
uncertainty on the data.
The histograms show the predictions
of several MC models.  The bottom insert shows the ratio of
the MC predictions to the data.
}
\label{fig:corr-ptavebinsaway}
\end{figure*}

Figures~\ref{fig:corr-sumptbinsaway} through~\ref{fig:corr-ptavebinsaway} show the 
unfolded distributions 
$(1/\Nev) d\Nev/d\sumpt$, $(1/\Nev) d\Nev/d\ntrks$ and $(1/\Nev) d\Nev/d\ptave$ 
in the away region
for three representative values of $R$ and for low (5--6~GeV) and high
(31--50~GeV) bins of \ptJet.    Here \Nev\ is the 
number of events in the sample
with $\ptJet > 4$~GeV and $|\etaJet| < 1.5$. 
Most of the MC models reproduce the shapes of the distributions reasonably, with 
\PythiaS~Z1 and Perugia 2011 providing the best agreement. 
\PythiaS~AUET2B predicts values of \sumpt\ and \ntrks\ that are higher than the data at 
large \ptJet.  


\clearpage
\onecolumngrid
\input{atlas_authlist}

\end{document}

%% file: atlas_authlist.tex
\begin{flushleft}
{\Large The ATLAS Collaboration}

\bigskip

G.~Aad$^{\rm 48}$,
B.~Abbott$^{\rm 111}$,
J.~Abdallah$^{\rm 11}$,
S.~Abdel~Khalek$^{\rm 115}$,
A.A.~Abdelalim$^{\rm 49}$,
O.~Abdinov$^{\rm 10}$,
B.~Abi$^{\rm 112}$,
M.~Abolins$^{\rm 88}$,
O.S.~AbouZeid$^{\rm 158}$,
H.~Abramowicz$^{\rm 153}$
% ...
\bigskip

% ...
$^{10}$ Institute of Physics, Azerbaijan Academy of Sciences, Baku, Azerbaijan\\
$^{11}$ Institut de F\'isica d'Altes Energies and Departament de F\'isica de la Universitat Aut\`onoma  de Barcelona and ICREA, Barcelona, Spain\\
% ...
$^{48}$ Fakult\"{a}t f\"{u}r Mathematik und Physik, Albert-Ludwigs-Universit\"{a}t, Freiburg i.Br., Germany\\
$^{49}$ Section de Physique, Universit\'e de Gen\`eve, Geneva, Switzerland\\
% ...
$^{88}$ Department of Physics and Astronomy, Michigan State University, East Lansing MI, United States of America\\
% ...
$^{111}$ Homer L. Dodge Department of Physics and Astronomy, University of Oklahoma, Norman OK, United States of America\\
$^{112}$ Department of Physics, Oklahoma State University, Stillwater OK, United States of America\\
% ...
$^{115}$ LAL, Universit\'e Paris-Sud and CNRS/IN2P3, Orsay, France\\
% ...
$^{153}$ Raymond and Beverly Sackler School of Physics and Astronomy, Tel Aviv University, Tel Aviv, Israel\\
% ...
$^{158}$ Department of Physics, University of Toronto, Toronto ON, Canada\\
% ...
\end{flushleft}
